\documentclass{aa}
\usepackage{natbib}
\bibpunct{(}{)}{;}{a}{}{,} 
\usepackage[varg]{txfonts}

\usepackage{graphicx}
\usepackage{txfonts}
\begin{document} 

   \title{A Simple Model for Global \ion{H}{i} Profiles of Galaxies.}

   \author{I. M. Stewart
          \inst{1,2},
           S.-L. Blyth \inst{2}
          \and
           W. J. G. de Blok\inst{3,2,4}
          }

   \institute{Argelander Institut f\"ur Astronomie, Universit\"at Bonn,
              Auf dem H\"ugel 71, D-53121 Bonn, Germany\\
              \email{ims@astro.uni-bonn.de}
         \and
             Astrophysics, Cosmology and Gravity Centre,
             Department of Astronomy, University of Cape Town,
              Private Bag X3, Rondebosch 7701, South Africa.
         \and
             Netherlands Institute for Radio Astronomy (ASTRON),
             Postbus 2, NL-7990 AA Dwingeloo, The Netherlands.
         \and
             Kapteyn Astronomical Institute, University of Groningen,
             PO Box 800, NL-9700 AV Groningen, The Netherlands.
             }


 
  \abstract
   {Current and future blind surveys for \ion{H}{i} generate large catalogs of spectral lines for which automated characterisation would be convenient.}
   {A 6-parameter mathematical model for \ion{H}{i} galactic spectral lines is described. The aim of the paper is to show that this model is indeed a useful way to characterise such lines.}
   {The model is fitted to spectral lines extracted for the 34 spiral galaxies of the recent high-definition THINGS survey. Three scenarios with different instrumental characteristics are compared. Quantities obtained from the model fits, most importantly line width and total flux, are compared with analog quantities measured in more standard, non-parametric ways.}
   {The model is shown to be a good fit to nearly all the THINGS profiles. When extra noise is added to the test spectra, the fits remain consistent; the model-fitting approach is also shown to return superior estimates of linewidth and flux under such conditions.}
   {}


   \maketitle

\section{Introduction} \label{s_intro}

Since hydrogen is the most common element in the universe, and makes up most of its mass, the distribution of this element is intimately connected with the dynamics of matter on scales from the cosmic down to the stellar. Although in its ionized or molecular forms hydrogen is problematic to detect, neutral atomic hydrogen (\ion{H}{i}) emits (or absorbs, if it is cooler than the background), due to a hyperfine transition, radio waves at a rest frequency of 1420 MHz. The transition rate is very low, which makes the line faint and relatively difficult to detect, but this has two compensating advantages: firstly, clouds of hydrogen up to the galactic scale are usually optically thin, which makes it relatively straightforward to deduce the mass of the cloud from the intensity of the line; and secondly that the spectral line is intrinsically very narrow, making it a useful way to study the motions, whether thermal, internal or bulk, of the emitting cloud, via Doppler shift and broadening.

In recent years there have been a number of blind surveys for \ion{H}{i} in the local universe, some complete \citep[e.g. HIPASS,][]{barnes_2001}, others ongoing (ALFALFA \citealt{giovanelli_2005}; EBHIS \citealt{winkel_2010, kerp_2011}; CHILES \citealt{fernandez_2013}). Further, deeper blind searches for \ion{H}{i} (LADUMA \citealt{holwerda_2012}; WALLABY \citealt{koribalski_2009}; DINGO \citealt{meyer_2009}; see also \citealt{duffy_2012}) are planned with the upcoming SKA precursor instruments MeerKAT \citep{jonas_2009} and ASKAP \citep{johnston_2008}. Such surveys complement those performed at other wavelengths and avoid potential biases from preferential selection of galaxies which are bright at these wavelengths.

The majority of hydrogen in the universe occurs in galaxies. The width of a galaxy's \ion{H}{i} spectral profile gives its bulk rotation speed, and the area under the profile is proportional to its \ion{H}{i} mass - both quantities of importance in cosmology.

From simple geometrical considerations it is clear that in any survey of objects uniformly distributed in space, the number frequency of objects will increase as their angular size decreases. Since the detectability of objects takes a sharp downturn as their angular sizes become smaller than the instrument resolution, the most common survey object therefore can be expected to be one which is only just brighter than the survey sensitivity cutoff, and which has an angular size no larger than the beam of the instrument. Since detection of \ion{H}{i} sources usually makes use also of spectral information, a galaxy at the limit of detection may be unresolved in any one channel map, but nevertheless kinematically resolved, such that the flux peak moves progressively across channels (see the EBHIS observation of DDO 154 described in section \ref{sss_ebhis_results} for an example of this).

The principal task therefore for post-detection processing of blind \ion{H}{i} surveys is to measure as accurately as possible the width, area and other properties of low signal-to-noise (S/N) spectral lines of unresolved sources: and not only for spatially-integrated spectral lines, but also for the line at each spatial pixel across the extent of the source. This is the aim of the technique discussed in the present paper.

Fitting of a model to a spatially-integrated spectral profile where the source is well-resolved on the other hand should not be expected to yield better values of galaxy parameters than other methods, because with such sources there are problems determining which spatial pixels to include in the sum - significant numbers of pixels with small but non-zero contributions may be excluded, or pixels containing nothing but a chance spike in noise may be mistakenly included. We make use of the well-resolved THINGS observations here for practical, demonstrative reasons. It is unavoidable that pixel masking effects cause flux biases in such spectra, although careful treatment of the data can minimize this.

The question of spatial masking or selection is one that arises even if the source is unresolved, because its flux remains distributed across the imaging beam or point spread function (PSF) of the instrument. Any cutoff criterion will miss some contribution in the wings, but a too generous cutoff will include too much noise from off-source directions. Fitting a model which includes a spatial component shaped like the PSF seems like a simple way around this, but is unsatisfactory because the noise in adjacent spatial pixels is also convolved by the PSF, and is therefore not statistically independent. Statistically speaking, it is better to fit to data which is as unprocessed as practical - i.e., to include instrumental response in the model and fit to raw data, rather than to fit a simpler model to data which has been processed (with accompanying muddying of the statistical waters) in order to remove or at least systematize the instrumental response. But an exploration of such matters is beyond the scope of the present paper. Here we content ourselves with construction of global spectra by spatial masking and summing, and make the associated caveats about the reliability of resulting flux measurements.

The plan of the paper is as follows. The advantages of having a model of the \ion{H}{i} spectral line is discussed in Sect. \ref{ss_theory_intro}, and the model itself is described in Sect. \ref{ss_theory_model}. Various technical matters connected with the process of fitting, including the Bayesian methodology, estimation of uncertainties, and a suitable goodness-of-fit measure, are discussed in Sect. \ref{ss_theory_fitting}. The tests performed on the model are presented in Sect. \ref{s_tests}. \ion{H}{i} spectral lines were obtained under a variety of conditions of noise, spectral resolution and baseline, and were fitted by the model.

We posed three questions about the fits. Firstly, how good a fit is the model in the case of optimum frequency resolution and S/N - does it return good values of bulk properties of the galaxies? This question is addressed in Sect. \ref{ss_tests_orig}, where we fit the model to a spatially-integrated spectrum from each of the 34 THINGS galaxies observed by \citet{walter_2008}.

Secondly, if we add noise to the spectra, and bin the channels more coarsely, does this affect the fitted parameter values? We address this question in Sect. \ref{ss_tests_coarse}.

Thirdly, how does the model perform fitting to a single-dish observation, where it is necessary also to fit a baseline? And can the model return useful kinetic information when the galaxy is at the limit of resolution? In Sect. \ref{ss_tests_ebhis}, data from the EBHIS survey \citep{winkel_2010, kerp_2011} are used to test this.

\section{A model of the \ion{H}{i} spectral line} \label{s_theory}
\subsection{Introduction} \label{ss_theory_intro}

The aim of any model of a physical process is to approximate that process with a small, hence manageable, number of adjustable parameters. A good model will be a good match to the data; will have a small number of parameters; and it is also useful if the model is not just empirical but tied in some way to the underlying physics of the object. Since these desirables can be in opposition, a model is usually then also a compromise.

The shape of an \ion{H}{i} spectral line represents the distribution and motions of neutral hydrogen within a galaxy. On top of the bulk rotation which prevents gravitational collapse there can be innumerable variations in the pattern of local \ion{H}{i} flow from galaxy to galaxy, which will be reflected in differences between the shapes of the respective spectral lines. It might therefore seem difficult to formulate a simple model of the \ion{H}{i} spectral line. However as shown in Sect. \ref{sss_orig_results}, in practice the 6-parameter model presented here works fairly well. Local deviations in flux density between the data and the fitted model don't exceed about 10\%, and tend not to affect either the total flux or the linewidth.

In all the \ion{H}{i} surveys the authors are aware of, total flux and linewidth have been estimated from the data directly, without fitting a model to the line profile.\footnote{\citet{saintonge_2007} described a model constructed from Hermite polynomials which is used in the ALFALFA source-\emph{detection} procedure \citep{haynes_2011}, but these authors do not, so far as we are aware, make further use of the fit parameters.} So why use one? Firstly because it is more systematic: there is no need for either human intervention or ad-hoc prescriptions for the number of channels to consider. Secondly, as is shown in Sect. \ref{ss_tests_coarse}, the bias in linewidth measurement is much reduced through model fitting. Thirdly, as shown in Sect. \ref{ss_tests_ebhis}, the modelling approach very naturally accommodates a modelling of the spectral background or baseline. It's no longer necessary to decide where the line profile `ends' before fitting the baseline, since both can be considered together. Fourthly, fitting of a model arguably lends itself more easily to automated processing, which becomes an ever more pressing consideration as survey datasets grow in size.

Lastly, a parametrized model opens the door to the use of Bayesian techniques, which are becoming increasingly accepted as useful tools in astronomy. Advantages here fall under three main heads. The first of these is that a Bayesian formulation is the formally correct (and therefore optimum) procedure for estimating the parameters of interest and for incorporating prior knowledge. This method is applied in the present paper. The second head or category is the use of Bayes' theorem to assess the relative suitability of differing models. This is also used in the present paper, to determine the best order of baseline model (Sect. \ref{ss_app_B_bkg}). Note that the same formulation can be used to estimate detection probability directly, although this topic is not explored in depth in the present paper: the Bayesian approach automatically takes into account both prior knowledge of the expected range of line shapes, as well as the total bandwidth and area of the survey. Such an approach to detection is more fundamental and rigorous than relying on either the 5-sigma rule or ad hoc prescriptions for calculating signal-to-noise ratio (S/N). The third advantage of the Bayesian formalism is in the extraction of statistical results from large ensembles of low-quality data via hierarchical modelling \citep{loredo_2012u}. This technique is however not used here.

\citet{roberts_1978} discussed the \ion{H}{i} spectral line and explained why it often has a double-horned shape with a relatively steep rise and fall. Roberts explored several semi-physical models of the line profile, and in fact his model C is a special case of ours. Roberts was concerned rather to emphasize and explore the connection between galactic linewidth and luminosity now more usually associated with \citet{tully_1977}, and his profile models don't have the flexibility to fit a wide variety of \ion{H}{i} line shapes.

In the second of two papers which presented a comprehensive simulation of gas in the local universe, \citet{obreshkow_2009b} described a profile model (which they apply to both \ion{H}{i} and CO spectral lines) which uses 5 parameters. These parameters, labelled by the authors $k_1$ to $k_5$, are purely empirical in themselves, but can be derived from more fundamental properties of the line profile via a set of formulae (equations A2 to A6, appendix A of their paper). This second tier of parameters includes the flux density at profile centre, the maximum flux density, and the velocity widths at the 50\% and 20\% levels. In an earlier paper by some of the same authors \citep{obreshkow_2009a}, the second-tier parameters are derived by constructing each profile from an appropriate projection and integration of a comprehensive model of the distribution of \ion{H}{i} gas density and velocity as a function of radius from the galaxy centre. This radial model in turn depends on several third-tier fundamental parameters, such as masses and characteristic radii for disk, halo and bulge components. It might be possible to find a reasonably simple way to connect these fundamental parameters with the eventual $k$ values, but these authors did not attempt this themselves. Their intent was to generate the second-tier parameters for a large number of simulated galaxies and make these available in a database. The $k$-prescription was provided simply as a convenience for the user who wished to construct approximate profiles from these data.

One could make use of the $k$-model of \citeauthor{obreshkow_2009b} for fitting to real data, but there are a number of desirable improvements:
\begin{enumerate}
  \item The $k$ parameters provide no direct physical insight.
  \item Many real \ion{H}{i} spectral lines exhibit a noticeable asymmetry \citep[see e.g.][]{richter_1994}. The $k$ model does not cater for this.
  \item Since the prescribed functions of the $k$ model are discontinuous at sharply-defined velocity bounds, not aligned \emph{a priori} with velocity channel boundaries, integrating the model over a finite channel width is a little fiddly.
  \item There is no way to simulate any artefacts arising from the autocorrelation process which generates spectra from radio signals; nor can the $k$-model simulate applied filters, such as Hanning or Tukey filters.
\end{enumerate}
The model described in the present paper addresses all these issues.

\subsection{The model} \label{ss_theory_model}

In order to derive a model we need to arrive at a reasonable approximation to the motion of \ion{H}{i} in a generic galaxy. We will begin by breaking the motions into two categories: bulk vs. random.

\subsubsection{Bulk motions} \label{sss_model_bulk}

We approximate bulk motions firstly by assuming that all gas in a (late-type, therefore gas-rich) galaxy rotates about a common centre and in a common plane. As shown for example in \citet{deblok_2008}, this is a reasonable assumption for the majority of spiral galaxies. We don't assume that the density of gas is azimuthally symmetric: departures from same are catered for, albeit in a crude manner, via the asymmetry parameter of our model.

As is well known, for most galaxies with a total \ion{H}{i} mass larger than about $10^9$ solar masses, the curve of rotation speed as a function of radius from the centre is seen to be remarkably flat outside the core. We assume here that the rotation curve is always exactly flat outside a certain radius. Within the core itself, the velocity usually rises steadily from (nominally) zero at the galactic centre, then turns over smoothly when it reaches the `flat' value of velocity. In the present model we approximate this rise by a straight line, such as would be observed for example in a rotating solid body; we also assume that the density of gas in the core is uniform. This simplified rotation curve is similar in shape to that labelled `C' in Fig. 3 of \citet{roberts_1978}.

   \begin{figure}
   \centering
      \includegraphics[width=\hsize]{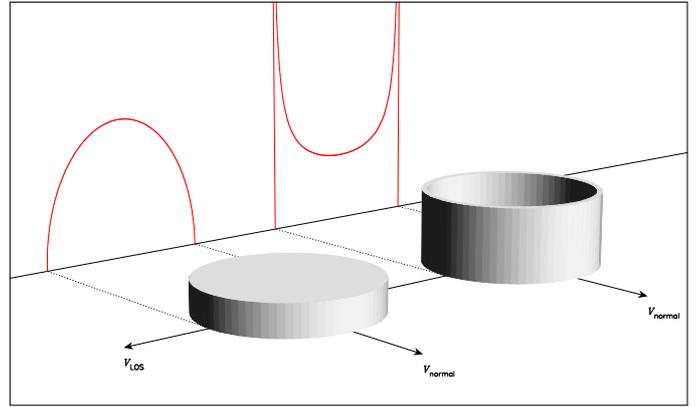}
      \caption{A schematic showing the model distribution of \ion{H}{i} in phase space for an edge-on galaxy. The disk and the ring represent respectively the inner and outer components of the model. The height of each body represents the density of \ion{H}{i} as a function of velocity. $v_\mathrm{LOS}$ is the velocity in the line of sight and $v_\mathrm{normal}$ is the velocity normal to that, but in the plane of the galaxy. The figures in red represent the \ion{H}{i} densities due to the two components, as projected onto the line of sight. Since the densities are displayed as functions of velocity, these red projections in fact give directly the spectral line shapes of the two components. Notes: (i) The ring formally speaking ought to be infinitely thin, but some visible width has been given to it for the sake of easier interpretation. (ii) No attempt has been made to make the height scales of the projection graphs consistent with those of the density figures.
              }
         \label{fig_U}
   \end{figure}

Understanding of the way a profile model represents the bulk gas motions is assisted by mapping the gas motions and densities in phase space - that is, presenting the gas density not as a function of spatial location but of velocity components. The line profile may then be obtained by projecting the phase space gas distribution onto a line directed towards the observer. This is diagrammed in Fig. \ref{fig_U}. In phase space, the outer mass of gas with a constant rotation speed appears as a ring of infinitesimal thickness, whereas the `solid-rotating' inner part appears as a uniform disk.

The ring gives in projection the characteristic double-horned profile so often observed in \ion{H}{i} spectral lines:
\begin{equation} \label{equ_model_outer}
  s_\mathrm{outer}(v) = \frac{2 S}{\pi \Delta v} \ \rho^{-1} \! \left( \frac{v - v_\mathrm{ctr}}{\Delta v/ 2} \right)
\end{equation}
where
{\setlength\arraycolsep{2pt}
  \begin{eqnarray*}
	\rho^2(u) & = & 1 - u^2 \textrm{ for } |u|<1,\nonumber \\
	          & = & 0 \textrm{ else.}
  \end{eqnarray*}
}
Here $S$ is the total flux from the \ion{H}{i} in this section, $\Delta v$ is the range between maximum and minimum gas velocities in the line of sight, and $v_\mathrm{ctr}$ is the mean line-of-sight velocity. The disk in projection yields a half ellipse:
\begin{displaymath}
  s_\mathrm{inner}(v) = \frac{4 S}{\pi \Delta v} \, \rho \left( \frac{v - v_\mathrm{ctr}}{\Delta v / 2} \right).
\end{displaymath}
These are the two fundamental components of our model. Asymmetry in the line profile is accommodated by multiplying both components by
\begin{displaymath}
	1 + 2 \alpha (v - v_\mathrm{ctr})/\Delta v
\end{displaymath}

To describe the model so far we require 5 parameters: the line centre $v_\mathrm{ctr}$; the so-called intrinsic line width $\Delta v$; the total flux $S$; the fraction $f$ of the gas which is found in the `solid-rotating' part which is associated with the inner part of the galaxy; and the asymmetry parameter $\alpha$. The model so far is thus represented by
{\setlength\arraycolsep{2pt}
  \begin{eqnarray} \label{equ_model_a}
	s_\mathrm{intrinsic}(v) & = & \frac{2 S}{\pi \Delta v} \, \left[1 + \frac{2 \alpha (v - v_\mathrm{ctr})}{\Delta v} \right] \times\nonumber\\
	                        & \times & \left[(1-f) \rho^{-1} \! \left( \frac{v - v_\mathrm{ctr}}{\Delta v /2} \right) + 2 f \rho \left( \frac{v - v_\mathrm{ctr}}{\Delta v / 2} \right) \right].
  \end{eqnarray}
}

\subsubsection{Random motions} \label{sss_model_random}

Random motions include both thermal motions of the atoms and turbulent motions. The latter may occur on length scales much larger than the atomic, but provided they are much smaller than the scale of bulk, i.e. galaxy-scale motions, we can lump them together with thermal motions. In the present study we assume that the distribution of random velocities is the same throughout any galaxy. Maps of the local linewidth in \citet{walter_2008} indicate that this is a reasonable assumption for the THINGS galaxies. In this approximation, random motions may be treated mathematically as a 3-dimensional convolution of the bulk-motion line profile. Further, if we approximate the distribution of random velocities by a 3D Gaussian, then the 1D projection of this in any direction is also a Gaussian:
\begin{equation} \label{equ_dispersion}
	g(v) = \frac{1}{\Delta v_\mathrm{rand} \sqrt{2 \pi}} \exp \left( \frac{-v^2}{2 [\Delta v_\mathrm{rand}]^2} \right).
\end{equation}
The characteristic width $\Delta v_\mathrm{rand}$ is the sixth and final model parameter.

It should be emphasized that representation of random gas motions by a single Gaussian is only an approximation. In reality there may be several co-located components exhibiting different degrees of dispersion \citep[see for example][]{braun_1997}.

The full equation for the profile model is
\begin{equation} \label{equ_model_full}
	s(v) = s_\mathrm{intrinsic}(v) \star g(v)
\end{equation}
where $\star$ indicates convolution. The full list of 6 model parameters is $v_\mathrm{ctr}$, $\Delta v$, $S$, $\Delta v_\mathrm{rand}$, $f$ and $\alpha$. These have natural ranges, that is, ranges imposed by physical reasonableness, as follows:
\begin{itemize}
  \item $v_\mathrm{ctr}$: constrained in practice by the ends of the spectrum.
  \item $\Delta v>=0$
  \item $S>0$
  \item $\Delta v_\mathrm{rand}>0$
  \item $0<=f<=1$
  \item $-1<=\alpha<=1$
\end{itemize}

Note however that, for some of the fits reported in the present paper, up to 6 additional parameters were used for fitting the baseline.

The asymmetry and fraction-solid parameters can be expected to be correlated respectively with the third and fourth moments of the spectral line, as described by \citet{andersen_2009}. These authors make a comprehensive study of the statistics of \ion{H}{i} and \ion{H}{ii} spectral line shapes and their relation to galaxy morphology. We don't here retrace that analysis, but only observe that calculation of the appropriate moments of our spectral line model is easy and avoids the necessity, mentioned by \citeauthor{andersen_2009}, of \emph{a priori} human selection of a velocity range, when extracting moments from data. Other measures of asymmetry, such as that of \citet{tifft_1988}, are equally easy to calculate from our model.

\subsection{Fourier-space formulation} \label{ss_theory_method}

Although the model can be calculated directly using equation \ref{equ_model_full} and the preceding formalism, there are several advantages to calculating the profile first in Fourier space, then transforming to velocity space. Firstly, the convolution in equation \ref{equ_model_full} turns into a product; secondly, the singularities in equation \ref{equ_model_outer} are avoided; and thirdly, the process mimics the processing of real signals in an XF-type correlator, and thus allows inclusion of some of the artefacts which result from same.

This Fourier-space model construction was followed in all the calculations described in the present paper. To make it easy for readers to do this themselves, the Fourier transform of equation \ref{equ_model_full} is given in appendix \ref{s_app_A}.

\subsection{Fitting considerations} \label{ss_theory_fitting}

A frequent use for such a model is to fit it to data. For this purpose one needs an objective function describing the goodness of the fit, and one must choose an algorithm for minimizing this function. These considerations are discussed in Sects. \ref{sss_fitting_bayes} through \ref{sss_fitting_goodness}.

\subsubsection{Bayesian formulation} \label{sss_fitting_bayes}

According to Bayes' theorem, given a set of measurements of flux density $\mathbf{y}$, the posterior probability density function of the model parameters $p(\mathbf{q}|\mathbf{y})$, where we use $\mathbf{q}$ as shorthand for the six model parameters, is given by \citep[see e.g.][]{dagostini_2003}
\begin{equation} \label{equ_bayes}
	p(\mathbf{q}|\mathbf{y}) = \frac{1}{E} \, p(\mathbf{q}) \, p(\mathbf{y}|\mathbf{q}),
\end{equation}
where $E$, known as the evidence, is just a normalizing constant:
\begin{equation} \label{equ_evidence}
	E = \int d\mathbf{q} \, p(\mathbf{q}) \, p(\mathbf{y}|\mathbf{q}).
\end{equation}
The first function in the integrand is the prior probability distribution which represents our prior knowledge of the parameter values; the second is the likelihood. For $N$ data values $y_j$ which include Gaussian-distributed noise, the likelihood is given by
{\setlength\arraycolsep{2pt}
  \begin{eqnarray} \label{equ_like}
	p(\mathbf{y}|\mathbf{q}) & = & \prod_{j=1}^N \frac{1}{\sigma_j \sqrt{2\pi}} \exp \left\{ \frac{-[y_j - s(v_j,\mathbf{q})]^2}{2\sigma_j^2} \right\}\nonumber \\
	                    & = & (2\pi)^{-N/2} \exp \left( \frac{-\chi^2}{2} \right) \prod_{j=1}^N \frac{1}{\sigma_j}
  \end{eqnarray}
}
where $s(v_j,\mathbf{q})$ is the profile model evaluated for velocity channel $j$, $\sigma_j$ is the standard deviation of the noise in channel $j$, and $\chi^2$ has its usual formulation.

A Bayesian fit (that is, a fit procedure which optimizes the Bayesian posterior probability) may tend towards being data-dominated, or it may be prior-dominated. The first case occurs if the data has high signal-to-noise (S/N) and if there has not been much earlier fitting experience. The original THINGS dataset observed with the VLA matches this criterion for us. The EBHIS observations of the THINGS galaxies however generally have much lower S/N values, as do the semi-simulated profiles described in Sect. \ref{ss_tests_coarse}. For the latter two cases we therefore thought it appropriate to set some cautious priors, derived from the fits to the VLA THINGS profiles. Some of the model parameters are poorly constrained by the data in these lower-S/N fits and these are thus prior-dominated.

The low-S/N priors are described in detail in appendix \ref{ss_app_B_justification}.

\subsubsection{Fitting algorithms} \label{sss_fitting_algorithms}

Three fitting techniques have been used in the present study: the Levenberg-Marquardt (LM) method \citep[chapter 15.5]{press_1992}, simplex optimization (\citealt{nelder_1965}, see also \citealt{press_1992} chapter 10.4) and Markov-chain Monte Carlo (MCMC), specifically using the Metropolis algorithm (\citealt{metropolis_1953}, see also \citealt{bhanot_1988} for other useful references). For fitting to the THINGS profiles, both from the original VLA observations (Sect. \ref{ss_tests_orig}) and as observed as part of EBHIS (Sect. \ref{ss_tests_ebhis}), the LM procedure was used to obtain initial values of the approximate centre and width of the posterior, then the exact form of the posterior was explored via MCMC. For the semi-artificial profiles described in Sect. \ref{ss_tests_coarse}, the simplex procedure alone was used.

The simplex algorithm is robust but slow. In practice it was `fast enough', requiring only about 10 to 20 seconds on a standard laptop to fit a 6-parameter model to 450 data points.

Technical details of the MCMC are given in appendix \ref{ss_app_B_mcmc}.

\subsubsection{Uncertainties} \label{sss_fitting_uncerts}

The parameter uncertainties quoted in Table \ref{tab_O} are the square roots of the diagonal elements of a covariance matrix estimated from the ensemble of MCMC points. For the LM and simplex fits, the covariance matrix was obtained where necessary by inverting the matrix of second derivatives (known as the Hessian) of the posterior with respect to the parameters. As can be seen in the figures in appendix \ref{s_app_C}, the MCMC values are often about 50\% larger than the LM ones. The reason for this is not known.

All flavours of uncertainty scale with the uncertainties in the original values of flux density. The calculation of these is described in appendix \ref{ss_app_B_uncerts}.

\subsubsection{Goodness of fit} \label{sss_fitting_goodness}

The value of $\chi^2$ is commonly used to assess how well a model fits the data. In effect, reduced $\chi^2$ is a measure of the ratio between the data-minus-model residuals and the amplitude of the measurement noise. Noise plays the vital role in this because it affects the probability that the observed residuals would occur by chance with a perfect fit. In the present case however this is not quite what we want. We expect from the start that the model will not be a perfect fit to the data, so in a sense this question is already decided in the negative: what we want is rather to measure the size of the imperfections. Noise plays no role in this, provided only that it is not so large as to swamp the deviations; hence the bare value of $\chi^2$ will not return the information we want.

The line shape of NGC 925 in Fig. \ref{fig_H} provides a good example of the issue. Clearly seen are many local `bumps' or `wiggles' in the \ion{H}{i} distribution which the model is not able to track. It is the relative size of these bumps and dips compared to the average height of the profile which it would be most useful to know. For example, from Fig. \ref{fig_H} one can clearly see that the model is a cleaner fit to NGC 3184 than to NGC 925: what we want is some formula to quantify this difference.

The formula we devised to estimate the wiggle fraction $J$ is
\begin{equation} \label{equ_bump}
	J^2 = \frac{1}{S^3} \left( \frac{\Delta v}{\Delta v_\mathrm{chan}} \right)^2 \sum_j y_j \left[ \left( y_j - s_j \right)^2 - \sigma_j^2 \right].
\end{equation}
where $S$ and $\Delta v$ are respectively the total flux and line width model parameters as described in Sect. \ref{ss_theory_model}, and $\Delta v_\mathrm{chan}$ is the width of the spectral channels. The reasoning behind this formula is as follows. Firstly, we want the square of the residual in each channel. We then want to subtract the noise from this in quadrature. The resulting term should be weighted by the flux density $y_j$ of the line profile. This is normalized by dividing by the sum $S$ of the model profile values. The result so far is the square of the average residual, within the line profile, due solely to model/data mismatch. Finally the square root is taken and that result divided by the average flux density of the model, which is
\begin{displaymath}
	\langle s \rangle = \frac{\Delta v_\mathrm{chan}}{\Delta v} S.
\end{displaymath}
The result is the average fractional residual due to `wiggles'. This value is given in Col. 8 of Table \ref{tab_O}.

Local fluctuations in \ion{H}{i} can be visually deceptive. A good example is the global spectrum of NGC 4214, which is shown together with a model fit in Fig. \ref{fig_H}. At first inspection it is not clear why the model has not better fitted the seemingly regular double-horned profile. However, the linewidth of this galaxy is, at 56 km s$^{-1}$, relatively small: less than 5 times the `sigma' width of the turbulent broadening, which here is 12 km s$^{-1}$. Further consultation of Table \ref{tab_O} shows that the fraction of solid rotation fitted (which tends to reduce the depth of the valley between the horns) is small - consistent with zero. In fact the fitting procedure has chosen the sharpest and deepest possible double-horned profile which, after smoothing by the turbulent-broadening Gaussian, is consistent with the line slopes. With this amount of smoothing it is not possible for the model to follow deviations over velocity scales less than 10 km s$^{-1}$, as occurs with NGC 4214. These fluctuations only fool the eye into assuming a regularity which isn't in fact there, because NGC 4214 is too narrow to make obvious the actually random nature of its wiggles; in contrast to the spectrum of NGC 925, for example (shown in the same figure).

It is worth observing too that, although the wiggles in NGC 4214 do rather offend the eye, the $J$ factor for this galaxy is only 2.3\%, well below the average seen in Fig. \ref{fig_T}.

\section{Tests of the model} \label{s_tests}
\subsection{Spectra from the original THINGS observations} \label{ss_tests_orig}
\subsubsection{Introduction} \label{sss_orig_desc}

\citet{walter_2008} used the VLA to observe the \ion{H}{i} distribution in 34 nearby late-type galaxies. Spatial resolution, frequency resolution and S/N were all relatively high, certainly when compared to the most common objects detected in blind \ion{H}{i} surveys. The set of observations is known as THINGS (The \ion{H}{i} Nearby Galaxy Survey). We fitted our profile model to a spectrum extracted from THINGS observations of each of the 34 targets.

\citet{walter_2008} provide these spectra in the online version of their paper. According to their description, the global spectral profile for each galaxy is obtained by adding together a subset of pixels for each channel of the data cube. Pixels where there was no measurable emission from \ion{H}{i} were not included in the sum. This has the effect of making the noise in a channel proportional to the square root of the number of unmasked pixels. The mask cubes are not available on the THINGS web site but were kindly provided on request by F. Walter. This allowed us to estimate the noise per channel for each global profile. The exact procedure for doing so is described in appendix \ref{ss_app_B_uncerts}.

Some of the THINGS galaxies have a velocity range which straddles zero. For some of these, interference from Milky-Way \ion{H}{i} is evident. A few channels where this effect was obvious have been omitted when fitting to the spectra for galaxies DDO 53, NGC 2976, NGC 3077 and NGC 6946. Note that the total flux values in Col. 4 of Table \ref{tab_O} are the values of the respective parameter of the fitted model: thus these represent an interpolation over missing channels for the four galaxies mentioned. One can however also obtain the total flux under the fitted profile by adding together the samples of the profile model obtained at each channel. For Fig. \ref{fig_A}, in which the flux under the data spectrum is compared to the model, the relevant channels have been omitted from these sums for both the data and the model for the four galaxies mentioned. The percentage differences in model flux from the values in the Table are respectively 7, 13, 4 and 3.

The question discussed in the remainder of Sect. \ref{ss_tests_orig} is: how good a fit is the model in the case of optimum frequency resolution and S/N - does it return good values of bulk properties of the galaxies?

\subsubsection{Graphical and tabular display of fit results} \label{sss_orig_results}

\begin{figure}
   \centering
      \includegraphics[width=\hsize]{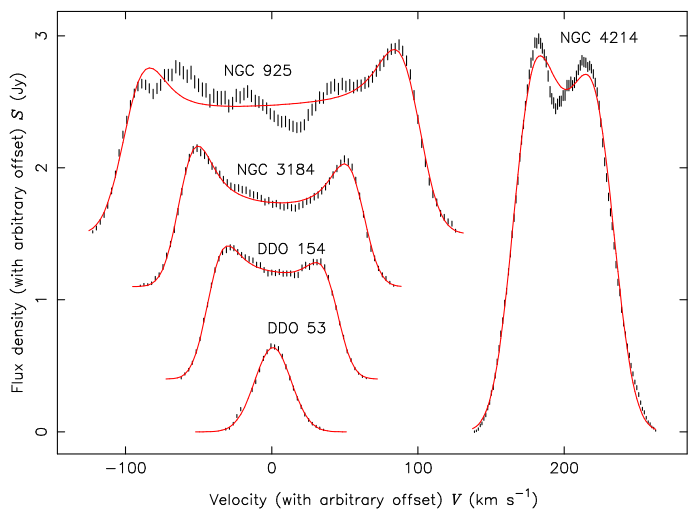}
      \caption{Fitted profiles (red curves) compared to raw data (black error bars) for 5 of the THINGS galaxies. Widths and heights of the profiles have not been altered but arbitrary offsets to both flux density and velocity have been added for clarity of plotting.
              }
         \label{fig_H}
   \end{figure}

Five examples of profiles fitted to THINGS spectra are shown in Fig. \ref{fig_H}. These are the five galaxies chosen as `simulation inputs' in Sect. \ref{ss_tests_coarse}. Figures showing all 34 individual fit results are given in appendix \ref{s_app_C}.

The results of the fits are given in Table \ref{tab_O}. Shown in Cols. 2 to 7 are the mean values with uncertainties for each of the six model parameters. Also given in Col. 8 is the `wiggle fraction' $J$ calculated from equation \ref{equ_bump}. A low value indicates a good match between model and data. Note that it is possible for $J^2$ to be negative. In this case, formally speaking, the root is imaginary, as listed in the table. This has no physical meaning, it just indicates that any local deviations in \ion{H}{i} from the model are insignificant compared to the measurement noise.

\begin{figure}
   \centering
      \includegraphics[width=\hsize]{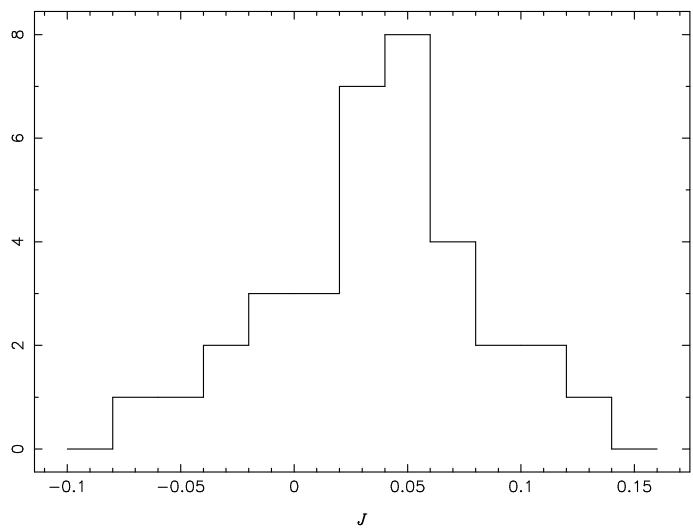}
      \caption{A histogram showing the distribution of `wiggle fraction' $J$ among the THINGS galaxies, as specified by equation \ref{equ_bump}. This quantity indicates the approximate extent of local fluctuations of \ion{H}{i} density away from the fitted model. Values shown here as less than zero are technically imaginary (see the respective values in Col. 8 of Table \ref{tab_O}), since they are square roots of differences in quadrature which turned out to be negative. Such values simply indicate that fluctuations away from the model fit are, for that galaxy, dominated by measurement noise.
              }
         \label{fig_T}
   \end{figure}

A histogram showing the distribution of wiggle fraction is given in Fig. \ref{fig_T}.

Column 10 of the table gives the velocity separation between points on the fitted profile where the flux density decreases to 20\% of its maximum value. The remaining columns, 9 and 11, are described in the following section.

\subsubsection{Numerical comparisons between model and data} \label{sss_orig_comparison}

\begin{figure}
   \centering
      \includegraphics[width=\hsize]{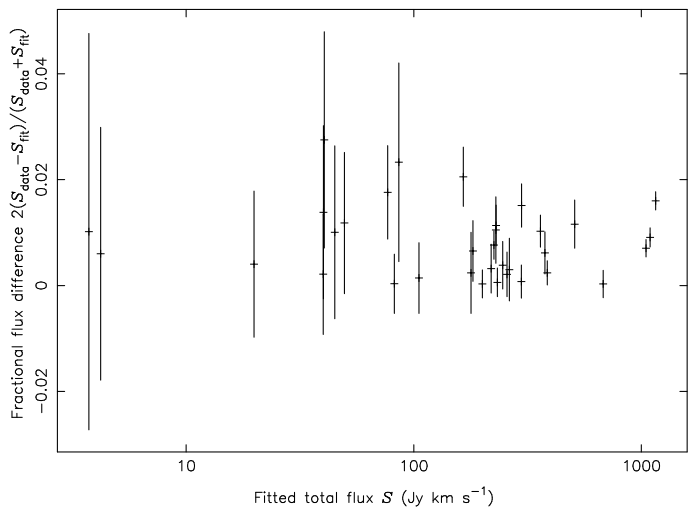}
      \caption{Fractional difference between the flux $S_\mathrm{data}$, derived from summing the flux densities in each data channel, and the value $S_\mathrm{fit}$, the equivalent sum of flux densities from the fitted line-profile model. This fraction was calculated from the expression $2(S_\mathrm{data}-S_\mathrm{fit})/(S_\mathrm{data}+S_\mathrm{fit})$. The $x$ coordinate is the fitted flux $S$ given in Col. 4 of Table \ref{tab_O}, whereas the fraction itself is given in Col. 9. $S_\mathrm{fit}$ is identical to $S$ except in the cases of DDO 53, NGC 2976, NGC 3077 and NGC 6946. For these galaxies, some channels around velocity zero were omitted from the calculation of both $S_\mathrm{data}$ and $S_\mathrm{fit}$. Uncertainties were calculated from the two independent contributions using the standard propagation formula.
              }
         \label{fig_A}
   \end{figure}

\begin{figure}
   \centering
      \includegraphics[width=\hsize]{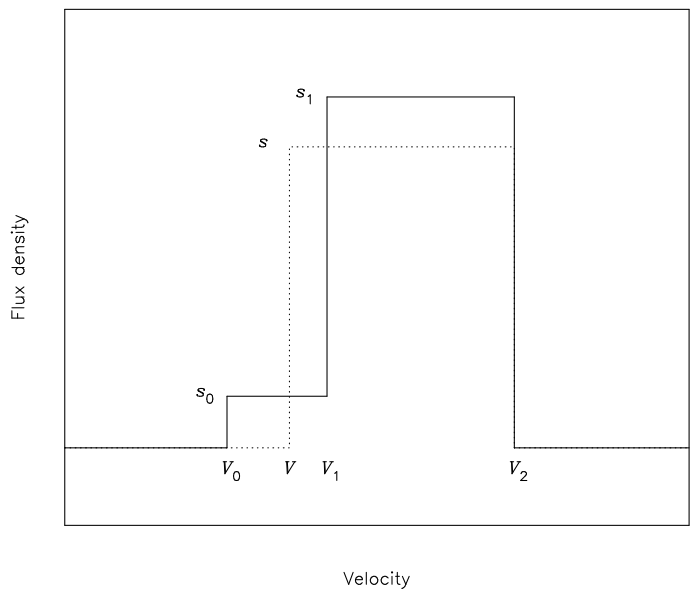}
      \caption{Demonstrates a mechanism by which the flux under a spectral line may be underestimated. The solid line shows a simple model of a continuous spectrum which has a pedestal extending some way to the left. The model, indicated by the dotted line, has itself no matching pedestal. $\chi^2$ is minimized if $v \to v_1$ and $s \to s_1$.
              }
         \label{fig_S}
   \end{figure}

\begin{figure}
   \centering
      \includegraphics[width=\hsize]{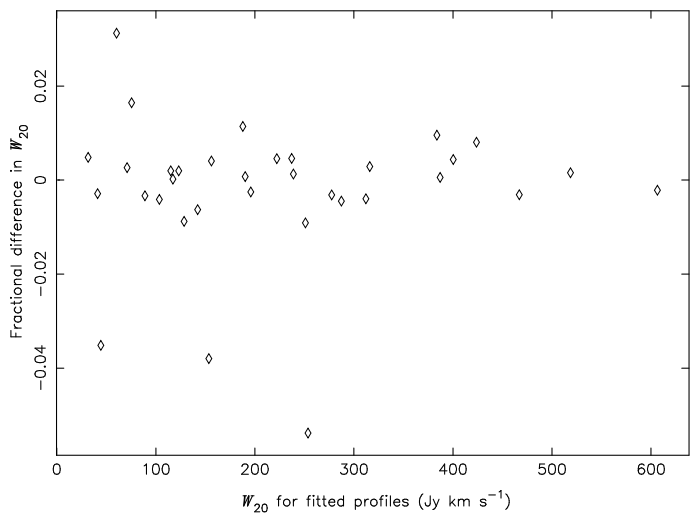}
      \caption{This plot compares the linewidth $W_{20,\mathrm{data}}$ at the 20\% height estimated from the observed THINGS spectral line to $W_{20,\mathrm{fit}}$, the same value for the profile fitted in the present paper to that line. The fractional linewidth difference was calculated from the expression $2(W_{20,\mathrm{fit}}-W_{20,\mathrm{data}})/(W_{20,\mathrm{fit}}+W_{20,\mathrm{data}})$. Values for $W_{20,\mathrm{fit}}$ are given in Col. 10 of Table \ref{tab_O}, and the fractional difference itself is given in Col. 11.
              }
         \label{fig_E}
   \end{figure}

\begin{figure}
   \centering
      \includegraphics[width=\hsize]{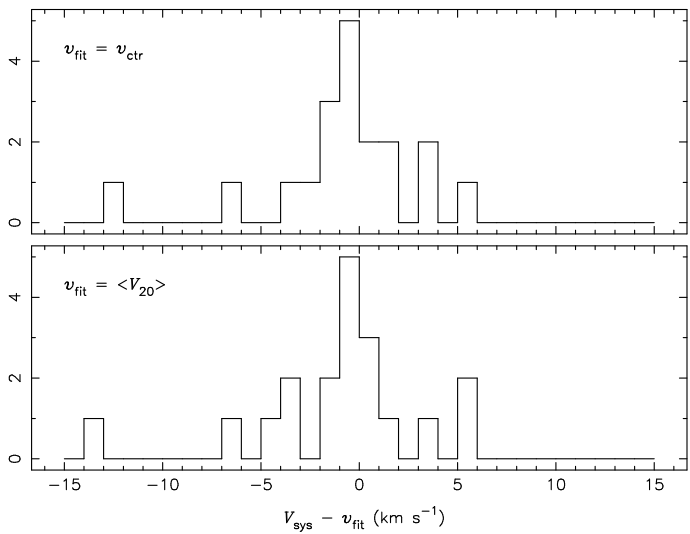}
      \caption{Frequency histograms of the difference between values of systemic velocity from two different sources. The first value, $V_\mathrm{sys}$, is that given in Col. 6 of Table 2 of \citet{deblok_2008}. These values were derived from tilted-ring fits to rotation curves of a 19-member subset of THINGS galaxies. The second value, $v_\mathrm{fit}$, was calculated in the present work. In the upper plot, the line centre parameter $v_\mathrm{ctr}$ of the fitted profile was used for $v_\mathrm{fit}$; in the lower plot, $v_\mathrm{fit}$ is the mean of the low and high velocity values at the 20\% height level on the fitted profile.
              }
         \label{fig_D}
   \end{figure}

The reason for doing these fits is to see how well the model matches a variety of real spectra. One can gain a qualitative impression from looking at the profiles but it is more informative to perform some numerical comparisons between the fitted model parameters and equivalent values from other sources. Some examples are shown in Figs. \ref{fig_A} to \ref{fig_D}, which are described individually below.

Figure \ref{fig_A} compares the total flux $S_\mathrm{fit}$ from summing valid values of the fitted flux density to a similar sum $S_\mathrm{data}$ over the raw data values. What is displayed in the figure is the fractional difference between these `fit' and `data' flux values for each galaxy.

\begin{sidewaystable*}
\caption{Parameter values for profile models fitted to the original THINGS global spectra.}             
\label{tab_O}      
\centering                          
\begin{tabular}{l r @{$\pm$} l r @{$\pm$} l r @{$\pm$} l r @{$\pm$} l r @{$\pm$} l r @{$\pm$} l c r @{$\pm$} l c c}        
\hline\hline                 
$\ \ \ 1$ & \multicolumn{2}{c}{2} & \multicolumn{2}{c}{3} & \multicolumn{2}{c}{4} & \multicolumn{2}{c}{5} & \multicolumn{2}{c}{6} & \multicolumn{2}{c}{7} & 8 & \multicolumn{2}{c}{9} & 10 & 11\\
Name & \multicolumn{2}{c}{$v_\mathrm{ctr}$} & \multicolumn{2}{c}{$\Delta v$} & \multicolumn{2}{c}{$S$} & \multicolumn{2}{c}{$\Delta v_\mathrm{rand}$} & \multicolumn{2}{c}{$f$} & \multicolumn{2}{c}{$\alpha$} & $J$ & \multicolumn{2}{c}{$\Delta S/S$} & $W_{20}$ & $\Delta W_{20}/W_{20}$\\
 & \multicolumn{2}{c}{km s$^{-1}$} & \multicolumn{2}{c}{km s$^{-1}$} & \multicolumn{2}{c}{Jy km s$^{-1}$} & \multicolumn{2}{c}{km s$^{-1}$} & \multicolumn{2}{c}{} & \multicolumn{2}{c}{} & \% & \multicolumn{2}{c}{\%} & km s$^{-1}$ & \%\\
\hline
DDO 154 & 375.69 & 0.12 & 81.24 & 0.46 & 82.10 & 0.33 & 9.04 & 0.15 & 0.2600 & 0.0164 & -0.0901 & 0.0081 & $\phantom{0}1.0\mathrm{i}$ &  0.0 &  0.6 & $103.5$ & $-0.4$\\
DDO 53 & 18.83 & 0.20 & 0.08 & 0.18 & 19.88 & 0.20 & 12.37 & 0.14 & 0.5029 & 0.2654 & -0.0005 & 0.5241 & $\phantom{0}0.0\mathrm{i}$ &  0.4 &  1.4 & $\phantom{0}44.4$ & $-3.5$\\
Ho II & 157.27 & 0.10 & 48.97 & 0.97 & 218.86 & 0.73 & 9.19 & 0.22 & 0.2449 & 0.0486 & -0.0663 & 0.0090 & $\phantom{0}0.5\phantom{\mathrm{i}}$ &  0.3 &  0.5 & $\phantom{0}71.0$ & $\phantom{-}0.3$\\
Ho I & 143.11 & 0.86 & 25.70 & 3.68 & 40.00 & 0.33 & 8.83 & 0.42 & 0.5369 & 0.2501 & -0.6686 & 0.1627 & $\phantom{0}1.6\mathrm{i}$ &  0.2 &  1.1 & $\phantom{0}41.3$ & $-0.3$\\
IC 2574 & 50.06 & 0.07 & 97.84 & 0.38 & 385.97 & 0.64 & 13.08 & 0.10 & 0.3410 & 0.0099 & -0.0788 & 0.0038 & $\phantom{0}3.4\phantom{\mathrm{i}}$ &  0.2 &  0.2 & $128.6$ & $-0.9$\\
M81 dwA & 112.84 & 0.22 & 0.08 & 0.18 & 4.21 & 0.08 & 8.83 & 0.19 & 0.5001 & 0.2662 & 0.0124 & 0.5316 & $\phantom{0}0.0\phantom{\mathrm{i}}$ &  0.6 &  2.4 & $\phantom{0}31.7$ & $\phantom{-}0.5$\\
M81 dwB & 345.86 & 0.96 & 42.27 & 4.75 & 3.74 & 0.10 & 8.33 & 1.08 & 0.3803 & 0.2460 & 0.1226 & 0.1059 & $\phantom{0}6.9\mathrm{i}$ &  1.0 &  3.7 & $\phantom{0}60.4$ & $\phantom{-}3.1$\\
NGC 628 & 659.36 & 0.12 & 45.47 & 0.20 & 297.88 & 0.89 & 11.88 & 0.10 & 0.0080 & 0.0068 & -0.1235 & 0.0091 & $\phantom{0}2.6\phantom{\mathrm{i}}$ &  1.5 &  0.4 & $\phantom{0}75.6$ & $\phantom{-}1.6$\\
NGC 925 & 552.93 & 0.17 & 191.24 & 0.56 & 229.62 & 1.05 & 12.06 & 0.24 & 0.2731 & 0.0114 & 0.0603 & 0.0074 & $\phantom{0}6.9\phantom{\mathrm{i}}$ &  1.0 &  0.6 & $222.3$ & $\phantom{-}0.5$\\
NGC 1569 & -71.38 & 3.77 & 73.00 & 9.28 & 86.04 & 1.21 & 27.65 & 1.25 & 0.1929 & 0.1735 & -0.8676 & 0.0963 & $\phantom{0}2.2\phantom{\mathrm{i}}$ &  2.3 &  1.9 & $123.2$ & $\phantom{-}0.2$\\
NGC 2366 & 100.08 & 0.07 & 94.84 & 0.42 & 232.91 & 0.46 & 9.43 & 0.12 & 0.5250 & 0.0107 & 0.0767 & 0.0047 & $\phantom{0}2.7\phantom{\mathrm{i}}$ &  0.1 &  0.3 & $115.1$ & $\phantom{-}0.2$\\
NGC 2403 & 133.44 & 0.05 & 228.75 & 0.12 & 1047.36 & 1.23 & 8.91 & 0.05 & 0.3580 & 0.0022 & 0.1191 & 0.0021 & $\phantom{0}4.4\phantom{\mathrm{i}}$ &  0.7 &  0.2 & $251.1$ & $-0.9$\\
NGC 2841 & 632.49 & 0.19 & 566.35 & 0.44 & 181.87 & 0.74 & 14.62 & 0.17 & 0.1228 & 0.0060 & 0.0065 & 0.0049 & $12.9\phantom{\mathrm{i}}$ &  0.7 &  0.6 & $606.4$ & $-0.2$\\
NGC 2903 & 555.39 & 0.12 & 353.80 & 0.28 & 229.88 & 0.63 & 12.71 & 0.14 & 0.1224 & 0.0043 & -0.1072 & 0.0044 & $\phantom{0}5.6\phantom{\mathrm{i}}$ &  1.1 &  0.4 & $386.9$ & $\phantom{-}0.1$\\
NGC 2976 & 4.96 & 0.47 & 133.88 & 2.33 & 44.99 & 0.54 & 10.80 & 0.75 & 0.5288 & 0.0486 & -0.2219 & 0.0232 & $\phantom{0}3.6\phantom{\mathrm{i}}$ &  1.0 &  1.6 & $156.2$ & $\phantom{-}0.4$\\
NGC 3031 & -38.67 & 0.09 & 366.28 & 0.23 & 1155.72 & 1.45 & 23.62 & 0.10 & 0.0702 & 0.0024 & 0.2966 & 0.0020 & $11.2\phantom{\mathrm{i}}$ &  1.6 &  0.2 & $423.8$ & $\phantom{-}0.8$\\
NGC 3077 & -23.58 & 0.36 & 117.12 & 0.78 & 257.06 & 0.79 & 17.13 & 0.10 & 0.0009 & 0.0008 & 0.9943 & 0.0039 & $\phantom{0}7.3\phantom{\mathrm{i}}$ &  0.2 &  0.4 & $117.3$ & $\phantom{-}0.0$\\
NGC 3184 & 593.79 & 0.14 & 117.17 & 0.37 & 105.37 & 0.51 & 9.60 & 0.17 & 0.1042 & 0.0119 & -0.0798 & 0.0074 & $\phantom{0}0.8\phantom{\mathrm{i}}$ &  0.1 &  0.7 & $142.2$ & $-0.6$\\
NGC 3198 & 661.11 & 0.07 & 283.65 & 0.16 & 224.91 & 0.44 & 10.58 & 0.08 & 0.1201 & 0.0031 & 0.0353 & 0.0030 & $\phantom{0}4.1\phantom{\mathrm{i}}$ &  0.8 &  0.3 & $312.2$ & $-0.4$\\
NGC 3351 & 778.93 & 0.24 & 255.37 & 0.50 & 49.54 & 0.48 & 8.05 & 0.27 & 0.0051 & 0.0040 & -0.0001 & 0.0141 & $\phantom{0}3.8\mathrm{i}$ &  1.2 &  1.3 & $277.6$ & $-0.3$\\
NGC 3521 & 798.22 & 0.10 & 415.67 & 0.29 & 297.23 & 0.66 & 19.17 & 0.11 & 0.0975 & 0.0039 & -0.0461 & 0.0034 & $\phantom{0}7.9\phantom{\mathrm{i}}$ &  0.1 &  0.3 & $467.1$ & $-0.3$\\
NGC 3621 & 729.07 & 0.08 & 257.31 & 0.20 & 679.58 & 1.26 & 11.34 & 0.09 & 0.2857 & 0.0034 & -0.0272 & 0.0031 & $\phantom{0}3.1\phantom{\mathrm{i}}$ &  0.0 &  0.3 & $287.3$ & $-0.4$\\
NGC 3627 & 720.58 & 0.73 & 327.58 & 2.15 & 40.07 & 0.47 & 22.07 & 0.93 & 0.1764 & 0.0249 & 0.1190 & 0.0189 & $\phantom{0}5.3\mathrm{i}$ &  1.4 &  1.6 & $383.8$ & $\phantom{-}1.0$\\
NGC 4214 & 292.74 & 0.07 & 56.34 & 0.12 & 200.23 & 0.39 & 12.40 & 0.05 & 0.0043 & 0.0035 & -0.0458 & 0.0046 & $\phantom{0}2.3\phantom{\mathrm{i}}$ &  0.0 &  0.3 & $\phantom{0}88.9$ & $-0.3$\\
NGC 4449 & 200.43 & 0.56 & 133.64 & 3.68 & 263.17 & 1.15 & 20.29 & 0.53 & 0.9261 & 0.0487 & -0.0222 & 0.0271 & $\phantom{0}8.5\phantom{\mathrm{i}}$ &  0.3 &  0.6 & $153.5$ & $-3.8$\\
NGC 4736 & 309.36 & 0.27 & 195.56 & 0.50 & 76.78 & 0.49 & 15.47 & 0.25 & 0.0016 & 0.0013 & -0.0500 & 0.0104 & $10.7\phantom{\mathrm{i}}$ &  1.8 &  0.9 & $237.2$ & $\phantom{-}0.5$\\
NGC 4826 & 408.87 & 0.54 & 288.68 & 1.22 & 40.44 & 0.59 & 11.79 & 0.56 & 0.0110 & 0.0087 & 0.3199 & 0.0223 & $\phantom{0}5.5\phantom{\mathrm{i}}$ &  2.8 &  2.0 & $316.1$ & $\phantom{-}0.3$\\
NGC 5055 & 497.74 & 0.16 & 357.96 & 0.58 & 376.67 & 1.09 & 17.09 & 0.22 & 0.4021 & 0.0065 & 0.1099 & 0.0050 & $\phantom{0}7.0\phantom{\mathrm{i}}$ &  0.6 &  0.4 & $400.2$ & $\phantom{-}0.4$\\
NGC 5194 & 455.88 & 0.28 & 148.64 & 0.81 & 165.24 & 0.67 & 17.36 & 0.31 & 0.1259 & 0.0145 & 0.3317 & 0.0081 & $\phantom{0}5.3\phantom{\mathrm{i}}$ &  2.1 &  0.6 & $187.8$ & $\phantom{-}1.1$\\
NGC 5236 & 506.81 & 0.17 & 172.90 & 0.33 & 360.42 & 0.79 & 31.73 & 0.14 & 0.0030 & 0.0024 & 0.1650 & 0.0045 & $\phantom{0}4.4\phantom{\mathrm{i}}$ &  1.0 &  0.3 & $253.6$ & $-5.4$\\
NGC 5457 & 228.74 & 0.09 & 138.68 & 0.41 & 1093.04 & 1.39 & 23.15 & 0.11 & 0.0134 & 0.0083 & 0.2260 & 0.0026 & $\phantom{0}5.0\phantom{\mathrm{i}}$ &  0.9 &  0.2 & $195.8$ & $-0.3$\\
NGC 6946 & 45.03 & 0.13 & 210.84 & 0.36 & 508.91 & 1.71 & 11.09 & 0.14 & 0.3078 & 0.0068 & 0.1004 & 0.0057 & $\phantom{0}9.3\phantom{\mathrm{i}}$ &  1.2 &  0.5 & $238.9$ & $\phantom{-}0.1$\\
NGC 7331 & 816.52 & 0.26 & 479.34 & 0.57 & 178.52 & 0.98 & 15.07 & 0.27 & 0.0414 & 0.0087 & 0.1150 & 0.0080 & $\phantom{0}2.1\mathrm{i}$ &  0.2 &  0.8 & $518.8$ & $\phantom{-}0.2$\\
NGC 7793 & 226.75 & 0.11 & 157.62 & 0.40 & 245.89 & 0.80 & 12.43 & 0.15 & 0.1312 & 0.0092 & -0.0601 & 0.0054 & $\phantom{0}5.7\phantom{\mathrm{i}}$ &  0.4 &  0.5 & $190.3$ & $\phantom{-}0.1$\\
\hline
\end{tabular}
\tablefoot{
  Columns are as follows. 1: source name. 2-7: the 6 model parameters, being mean values of the MCMC distribution. 8: wiggle fraction $J$ as defined in equation \ref{equ_bump}. 9: fractional difference between data and fitted values of total flux, also plotted in Fig. \ref{fig_A}. 10: width at 20\% of peak flux density of the fitted profile. 11: fractional difference between data and fit values of $W_{20}$, also plotted in Fig. \ref{fig_E}.
}
\end{sidewaystable*}

For about half of the galaxies, the difference between fluxes $S_\mathrm{data}$ and $S_\mathrm{fit}$ as displayed in Fig. \ref{fig_A} is consistent with zero. Almost all of the galaxies have flux differences less than about 1\%. Where there is a perceptible difference, the total flux values for the data ($S_\mathrm{data}$) are always larger than for the fitted profile.

No detailed explanation for this flux anomaly is known at present, but any explanation ought to start with the observation that the model is nonlinear in some of its parameters, and that it doesn't have support across the whole spectrum. Worth particular attention are the line wings, which in the model become closer to Gaussian in shape the further away from line centre one goes. This Gaussian component is intended to model the distribution of random gas velocities. However, it is known that the true distribution of velocities is better described by a sum of at least 2 Gaussians \citep{ianja_2012}. It is not hard to see that a Gaussian which is the best fit to the steep line edges might nevertheless miss flux present in broad, non-Gaussian wings for example.

As a toy model, with a non-linear parameter and limited support, to demonstrate how a pedestal can lead to flux underestimation in such circumstances, consider a simplified situation in which we wish to fit the two-step profile shown in Fig. \ref{fig_S} with a simple top-hat model with three free parameters: its left and right edges, and its height $s$.

Obviously there is nothing to be gained by moving the right edge of the model away from $v_2$; and given any two values for the edge locations, determining the best-fit height is trivial. The only open question is where within the range $[v_0,v_1]$ should we place the left edge $v$ of the profile such that $\chi^2$ is minimized. We define a kind of non-discrete, noiseless analog of $\chi^2$ as
\begin{displaymath}
  \chi^2 = (v_1 - v)(s - s_0)^2 + (v_2 - v_1)(s_1 - s)^2.
\end{displaymath}
After optimizing the model height $s$ this becomes
\begin{displaymath}
  \chi^2 = (s_1 - s_0)^2 (v_2 - v_1) \frac{v_1 - v}{v_2 - v}.
\end{displaymath}
Clearly this has its lowest allowed value when $v = v_1$. But this means that the profile is not fitting the pedestal at all, and the data then has more flux than the model, which is consistent with the characteristic upward deviations seen in Fig. \ref{fig_A}. Pedestals are noticeable in 5 of the galaxy profiles shown in Figs. \ref{fig_gal1} to \ref{fig_gal34}, namely for Ho II, NGC 628, NGC 5194, NGC 5236 and NGC 5457. Of these, all but Ho II are among the group of 6 worst cases of flux underestimation.

Figure \ref{fig_E} compares the velocity widths at the 20\% height of the fitted profiles ($W_{20}$) to the same figure calculated from the raw data values. For both model and data, the 20\% flux density level was calculated with respect to the maximum value for that spectrum; the velocity at this level on either wing of the profile was calculated by linear interpolation between the most distal pair of velocities which straddled the 20\% level. This simple technique is only accurate for the data because of the relatively high S/N of the THINGS observations. Some discussion of difficulties which arise in linewidth estimation when the spectrum is noisy is given in Sect. \ref{sss_coarse_mc}.

The agreement between the fitted width and the widths from the data is seen to be very good. In few cases is the difference larger than the 1\% level. The anomalously high value ($W_{20,\mathrm{fit}} > W_{20,\mathrm{data}}$) belongs to the M81 dwarf B; the three low values ($W_{20,\mathrm{data}} > W_{20,\mathrm{fit}}$) to DDO 53, NGC 4449 and NGC 5236. For M81 dwB and DDO 53, the width of the spectrum channels is several percent of the $W_{20}$ width, and therefore entirely accounts for the error. In NGC 4449 and 5236 the profile model is clearly seen not to be a good fit. However, we note that NGC 4449 is an interacting galaxy with a significant amount of \ion{H}{i} outside the disk. And for NGC 5236 the THINGS observations have many missing spacings, which may explain the curious steps in its line wings, which are probably the cause of the inflated value of $W_{20,\mathrm{data}}$.

De Blok et al. (2008) fitted a tilted-ring model to a 19-member subset of the THINGS galaxies and obtained high-precision \ion{H}{i} rotation curves for these. In Fig. \ref{fig_D} we compare their systemic velocities with two measures of line centre from the fitted profile, namely the fitted line centre parameter $v_\mathrm{ctr}$, and the velocity $\langle V_{20} \rangle$ obtained from the mean of the 20\% velocities. The systemic velocities $V_\mathrm{sys}$ are taken from Col. 6 of Table 2 of \citeauthor{deblok_2008}. With exception of the outlier at about -14 km s$^{-1}$, which is NGC 3627, the spread in differences is centred on zero with a standard deviation on the same order as the channel width for these galaxies, which was about 5.2 km s$^{-1}$ for 12 out of the 19, half that for the rest.

De Blok et al. found that a $V_\mathrm{sys}$ calculated from the global line wings of NGC 3627 returned values in the range 717 to 720 km s$^{-1}$, which fall much closer to our value of 720.6. NGC 3627 appears to have kinematic asymmetries, such as have been described and discussed by \citet{swaters_1999}. However, in the two galaxies studied by \citeauthor{swaters_1999}, the disturbances seem to be confined to the inner velocity regions. Disturbances in the inner-disk kinematics don't affect the rise and fall of the global profile, and thus should have only a small effect on the $V_\mathrm{sys}$ derived via fitting the present model. In contrast to this, for NGC 3627, Fig. 79 of \citeauthor{deblok_2008} shows a significant excess of gas at velocities 10 km s$^{-1}$ or more greater than their tilted-ring fit, right at the trailing edge of the velocity distribution. The present model is sensitive to such distortions, so this is arguably the reason for the inconsistency between our value of $V_\mathrm{sys}$ for this galaxy and that of \citet{deblok_2008}.

\begin{figure}
   \centering
      \includegraphics[width=\hsize]{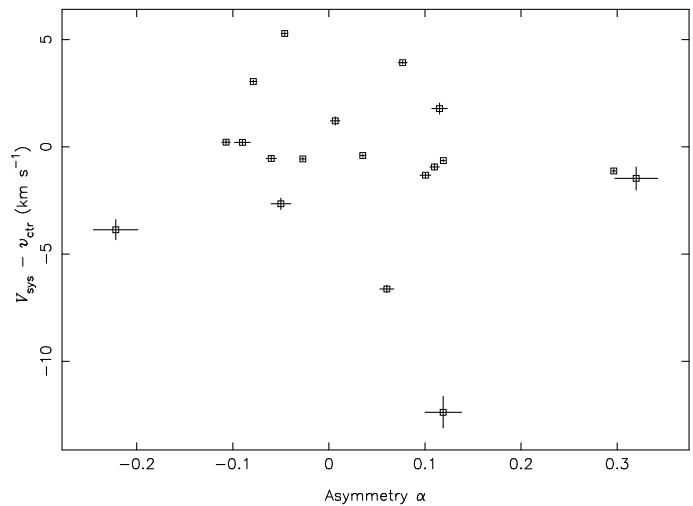}
      \caption{Scatter plot comparing the asymmetry parameter $\alpha$ of the fitted profile model to differences between the present fitted centre velocity $v_\mathrm{ctr}$ and the systemic velocities $V_\mathrm{sys}$ obtained by \citet{deblok_2008}. The velocity difference is the same as that histogrammed in the upper panel of Fig. \ref{fig_D}. Only the 19-member subset of THINGS galaxies examined by \citeauthor{deblok_2008} are shown.
              }
         \label{fig_V}
   \end{figure}

\citet{andersen_2009} found that line centres obtained via a moment analysis were biased in proportion to the asymmetry as measured by the 3rd moment. Since asymmetry is built in to our model from the start, we would not expect it to suffer from a similar problem. Figure \ref{fig_V} confirms that this is the case. This figure is analogous to Fig. 8 of \citeauthor{andersen_2009}. Despite that both our velocity scatter and uncertainties are smaller, no correlation is apparent in our Fig. \ref{fig_V}.

\subsection{Fitting to coarsened and noisified data} \label{ss_tests_coarse}
\subsubsection{Introduction} \label{sss_coarse_intro}

We argue in the introduction that the main use for this profile model is to extract information in a systematic way from the kinds of \ion{H}{i} spectra most commonly encountered in blind surveys, namely with low S/N, and a channel width of about 10 km s$^{-1}$. In order to test how well the model serves this purpose, data is needed with \emph{a priori} known values of line width, total flux etc. to allow comparison with the values obtained from fitting the model. Also a large number of test spectra is desirable to reduce or at least ascertain the uncertainties in the results. The raw THINGS spectra will not serve for this: they are too few in number, have unknown \emph{a priori} values, have too high S/N, and significantly higher spectral resolution than we normally expect to encounter in blind \ion{H}{i} surveys.

An alternative would be simply to simulate the required numbers of noisy, coarsely-binned profiles. However such a simulation always carries with it some uncertainty about how applicable its results are to real measurements. Simulations are also vulnerable to the criticism that one only gets out only what one puts in.

We can however make use of the THINGS profiles to generate realistic data simply by binning them into wider velocity channels and adding a lot more noise. \citet{lewis_1983} made use of a similar method in investigating line width measurement biases. This is also the approach which is taken in the present section.

What we do here is run several Monte Carlos, in each of which an ensemble of test spectra is generated. The line-profile model is fitted to each spectrum in an ensemble, and the width, centre and total flux of the line are also estimated using traditional direct methods. For each ensemble, and for each property of interest (line width etc.), an average is formed from the values obtained from the fitted profiles on the one hand and the direct measurement on the other. The input values of the properties are known and so are available for comparison. The aim is to demonstrate that the average value of each property is better estimated via model fitting than by direct measurements from the spectrum.

\subsubsection{The Simulation Monte Carlo} \label{sss_coarse_mc}

Five of the THINGS galaxies were selected, the sample being chosen so as to cover a range of shapes and sizes. These five are the ones shown in Fig. \ref{fig_H}. They all had channel widths of about 2.6 km s$^{-1}$, except for NGC 4214, for which the width is half this. For each of the five, a Monte Carlo ensemble of 100 spectra was generated for each of a set of twelve S/N values, the S/N figure being calculated according to the ALFALFA formula, which is discussed in Sect. \ref{sss_coarse_snr}. The twelve values were evenly spaced in a logarithmic sense over a range between about 2 and 100. Different amounts of S/N offset were applied to the five galaxies to prevent overlay of graph points when results from several galaxies appear on the same figure.

As already mentioned, the desired channel width in this Monte Carlo was about 10 km s$^{-1}$. This was most simply achieved by averaging an integer number of the channels of the original spectra, that number being 4 for all but NGC 4214, for which a factor of 8 was used. This scheme also permitted dithering of individual spectra by random offsets in the range from zero to one less than the rebinning factor. This dithering was to avoid systematic effects from the centre velocity of the input spectrum always having the same relation to the channel boundaries in the rebinned scheme.

\subsubsection{Calculating the S/N} \label{sss_coarse_snr}

Signal-to-noise ratio can be defined most simply as the ratio between the maximum or peak height of the input spectrum and the noise standard deviation or RMS. However this does not correlate well with the detectability of \ion{H}{i} lines, because a broader line is usually easier to detect than a narrow one having the same peak-to-RMS S/N. For this reason it is usual to define some alternative measure which takes account of the line width. The ALFALFA survey makes use of the following formula \citep[equation 2 of][]{haynes_2011}:
\begin{displaymath}
	S/N_\mathrm{ALFALFA} = \frac{S}{\sigma W_\mathrm{50}} \sqrt{\frac{\min(400, W_\mathrm{50})}{2 \; \Delta v_\mathrm{chan}}}
\end{displaymath}
We used the same to calculate the S/N values for our Monte Carlo.

Note firstly that, for any given profile shape, $S/N_\mathrm{peak} \propto S/N_\mathrm{ALFALFA}$, but the proportionality constant differs from profile to profile. Note secondly that ALFALFA take $S/N_\mathrm{ALFALFA} > 6$ as their detection criterion \citep{haynes_2011}.

\begin{figure}
   \centering
      \includegraphics[width=\hsize]{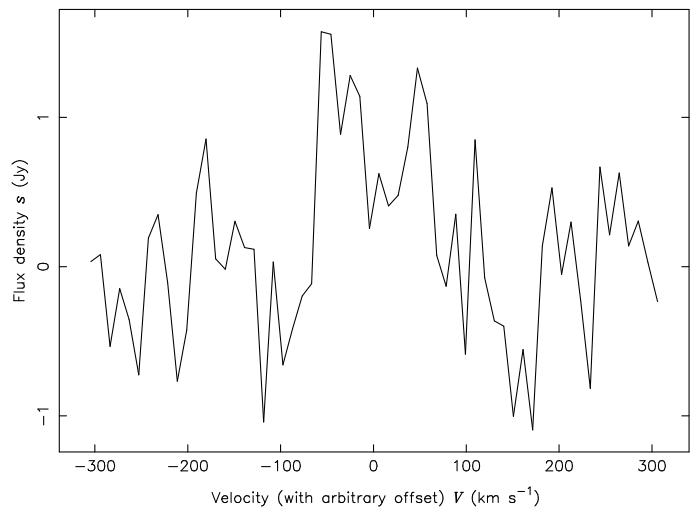}
      \caption{An example of a rebinned and noisified profile such as described in Sect. \ref{ss_tests_coarse}. The template spectrum was the THINGS observation of NGC 3184. The template data have been rebinned by a factor of 4 to give a channel width of 10.35 km/s; noise has then been added to bring the signal-to-noise ratio to a figure of 2.0 in peak-to-RMS terms, which for this galaxy is equivalent to about 3.9 according to the ALFALFA formula, as described in Sect. \ref{sss_coarse_snr}.
              }
         \label{fig_I}
   \end{figure}

A sample of a spectrum produced in the Monte Carlo is shown in Fig. \ref{fig_I}. This has a S/N equal to 3.9, which is a little below the ALFALFA detection criterion. As such it represents a very typical example of the quality of profile commonly seen in such surveys.

\subsubsection{Fitting to the data and analysing the results} \label{sss_coarse_fitting}

The simplex algorithm was used to fit each spectrum. Several fitting runs were done with increasing values of maximum permitted number of iterations, and decreasing values of the convergence criterion, in order to be sure that the fits were properly converging. The posterior described in Sect. \ref{sss_fitting_bayes} was the object function fitted, but this time some attention was paid to priors. Details of these and their justification can be found in appendix \ref{ss_app_B_justification}.

There are two ways one can analyse such data: one can compare the fitted values of the model parameters to those values obtained from fits to the original THINGS template spectra; and one can compare measurements derived from fitting a model to those obtained via more traditional, non-parametric methods. Both sorts of analysis have been done for the `coarse' data. The corresponding results are described respectively in Sects. \ref{sss_coarse_results_I} and \ref{sss_coarse_results_II}.

It is always going to be difficult to obtain reliable estimates of line parameters from data at such low S/N values as in Fig. \ref{fig_I}. There are three quantities which are of central importance: the total flux under the spectral line, its central or systemic velocity, and its width.

In the case of total flux, a typical non-parametric estimation method is that described in \citet{haynes_1984}: i.e. simply to `integrate over the observed signal'. Implicit in this however is a knowledge of where the `observed signal' begins and ends. For high S/N signals this is often done by visual inspection. In this case exactitude is not too important - one can afford to be generous, since the added noise per extra channel is small. Judging the line boundaries is of course much more problematic in the weak-signal limit. There are also practical difficulties involved in human inspection of large numbers of spectra.

Line width and line centre are often derived from the same source, namely two measurements of the low- and high-velocity edges of a line. Line width is taken from their difference and line centre from their average. \citet{bicay_1986} present several typical methods of estimating the edge velocities. These all start by deciding a flux density value, then interpolate linearly between adjacent channels which straddle this value. The flux density chosen is some percentage of a characteristic value for the line as a whole, which may be the mean flux density, or the maximum within the line profile, or some function of the peaks at the horns of the profile, in cases where these can be measured. In all cases such values for the characteristic flux density depend upon either the channel width or on implicit assumptions about where the line begins and ends.

The largest blind \ion{H}{i} survey (still on-going) is ALFALFA \citep{giovanelli_2005}; their most recent catalog release paper is \citet{haynes_2011}. The latter authors describe their edge-estimation procedure as similar to that of \citet{springob_2005}, in which polynomials (in practice nearly always `polynomials of order 1', i.e. straight lines) are fitted to several channels at the line rolloff on each side, the interpolation flux density being 50\% of the horn height on that side. In practice this is not much different to schemes already mentioned because, unlike the profiles shown in Fig. 2 of \citet{springob_2005}, in ALFALFA the effective channel width of 10 km s$^{-1}$ would not provide many channels in the rolloff.

To settle on a point of comparison therefore we decided to employ, as our `traditional, non-parametric method' of estimating the line edges, the simple scheme as follows:

\begin{enumerate}
  \item Determine $s_\mathrm{max}$, the highest flux density within the spectrum.
  \item Identify the number $j_\mathrm{lo}$ of the lowest-velocity channel whose $s$ value exceeds $0.5 s_\mathrm{max}$.
  \item Identify the number $j_\mathrm{hi}$ of the highest-velocity channel whose $s$ value exceeds $0.5 s_\mathrm{max}$.
  \item Interpolate between $j_\mathrm{lo}-1$ and $j_\mathrm{lo}$ to obtain $v_\mathrm{lo}$, and between $j_\mathrm{hi}$ and $j_\mathrm{hi}+1$ to obtain $v_\mathrm{hi}$. 
\end{enumerate}
It is thus a maximizing algorithm.

\subsubsection{Monte Carlo results I: comparison with high-S/N fits} \label{sss_coarse_results_I}

\begin{figure}
   \centering
      \includegraphics[width=\hsize]{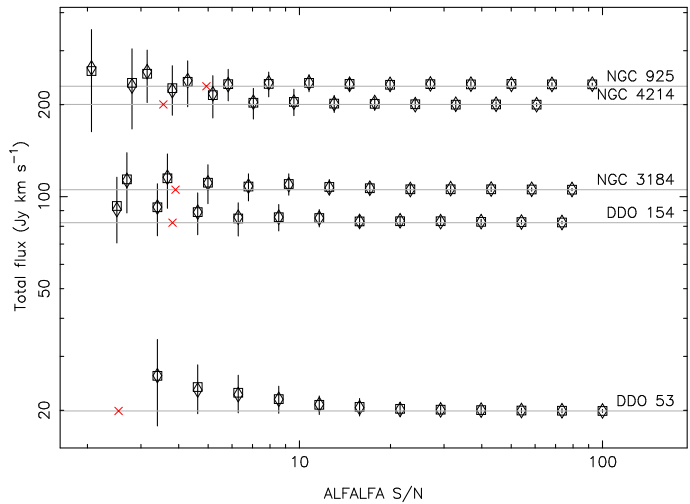}
      \caption{Mean total flux of an ensemble of fitted line profiles over a range of values of ALFALFA signal-to-noise ratio (S/N). Only 5 of the THINGS galaxies are shown, as labelled. Construction of these Monte Carlos is described in Sect. \ref{sss_coarse_mc}. The square symbols give the mean, the diamonds the median, and the vertical bars show the standard deviation for each ensemble. The half-tone lines give the values from the profile fits to the original THINGS data. Red Xs show the points at which the peak-to-RMS S/N equals 2.
              }
         \label{fig_K}
   \end{figure}

\begin{figure}
   \centering
      \includegraphics[width=\hsize]{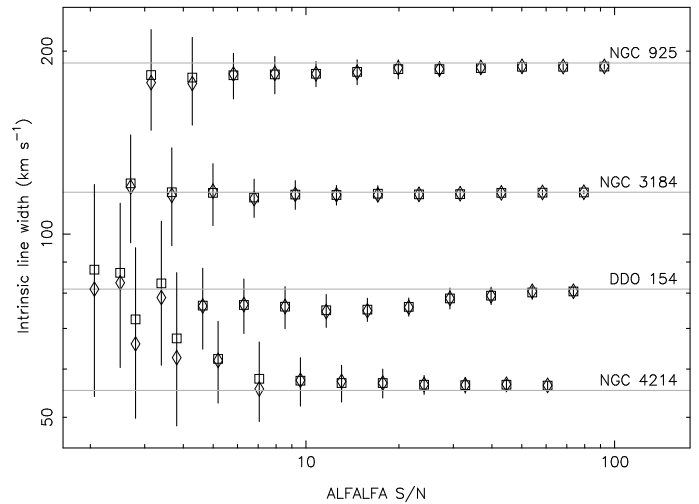}
      \caption{Similar to Fig. \ref{fig_K} except that the fitted value of intrinsic line width is shown instead of total flux. DDO 53 is not shown because the intrinsic line width for this galaxy is not significantly different from zero.
              }
         \label{fig_L}
   \end{figure}

\begin{figure}
   \centering
      \includegraphics[width=\hsize]{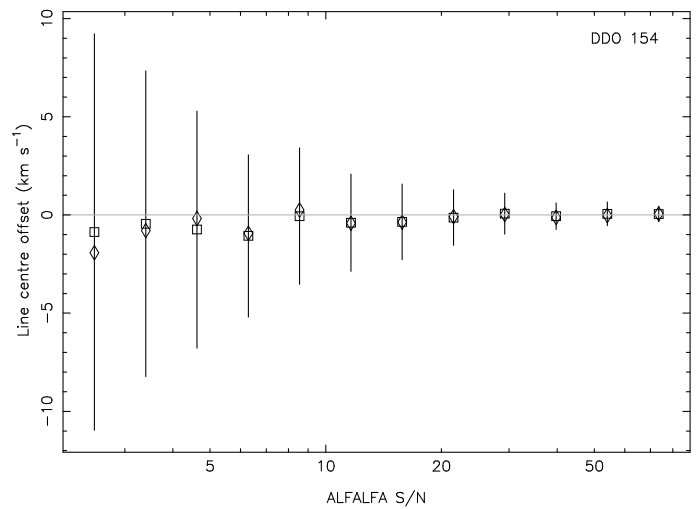}
      \caption{Mean fitted line centre offset is shown for DDO 154, at the same values of S/N as in Figs. \ref{fig_K} and \ref{fig_L}. No other galaxies in this subset are shown because all 5 show similar characteristics.
              }
         \label{fig_M}
   \end{figure}

Some of the more important results of the Monte Carlo described in Sects. \ref{sss_coarse_mc} to \ref{sss_coarse_fitting} are shown in Figs. \ref{fig_K} to \ref{fig_N3}.

In Fig. \ref{fig_K}, the variation of the mean value of total flux fitted to the spectra is shown against the S/N, for all five galaxies in the trial. Some upward bias in the values becomes visible at low S/N values, but this always remains within 1 standard deviation of nominal.

Figure \ref{fig_L} is a similar plot, this time for the intrinsic line width. DDO 53 is not shown, because line width is small and poorly constrained for this galaxy, even in fitting to the raw THINGS spectrum. Similar slight biases are observed in three of the four galaxies; it is not clear why NGC 4214 seems to be so badly affected at low S/N.

It is important to note that, at the highest S/N levels shown, the linewidth clearly asymptotes to the value measured in Sect. \ref{ss_tests_orig} and tabulated in Col. 3 of Table \ref{tab_O}, despite the fact that the spectral resolution of the present data is several times poorer than the original THINGS observations. This emphasises that no `instrumental broadening' correction to linewidth measures is necessary when fitting a line-profile model.

The curious under-estimation by up to $2 \sigma$ of the linewidth for DDO 154 at quite large values of S/N remains unexplained, although it is worth noting that it is never more than about 8 km s$^{-1}$, which is less than the channel width of these spectra. In the next section it is demonstrated however that for DDO 154 the $W_{50}$ linewidth remains correct in this range - so clearly the decrease in the fitted `intrinsic' value is compensated by an increase in the `turbulent broadening' value.

Note that the mean fitted linewidth for all five galaxies remains accurate down to the ALFALFA limit of detectability.

The mean and standard deviation of the line centre offset (that is, the difference between the fitted line centre and the input value) are displayed in Fig. \ref{fig_M} just for DDO 154. All the galaxies show similar results. No significant bias is seen, and the scatter remains within the channel width for ALFALFA S/N values greater than 5.

From these results we can conclude that, so far as the important parameters of the fitted profile model go, there is little bias evident when the spectral resolution and S/N are lowered to typical levels expected in blind surveys.

\subsubsection{Monte Carlo results II: comparison with non-parametric methods} \label{sss_coarse_results_II}

\begin{figure}
   \centering
      \includegraphics[width=\hsize]{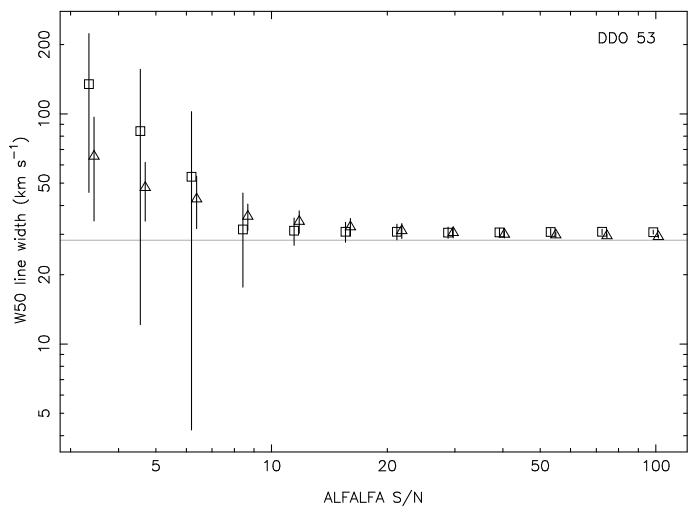}
      \caption{For the same range of S/N values as in Figs. \ref{fig_K} to \ref{fig_M}, line width at the 50\% height for the raw spectrum versus the same for the fitted profile is compared. DDO 53 is shown here. The square symbol indicates the mean value for the raw spectra, the triangle the mean for the fitted profiles. These respective points have been slightly offset horizontally for clarity. The halftone line indicates the value for the profile fitted to the original THINGS data.
              }
         \label{fig_N1}
   \end{figure}

\begin{figure}
   \centering
      \includegraphics[width=\hsize]{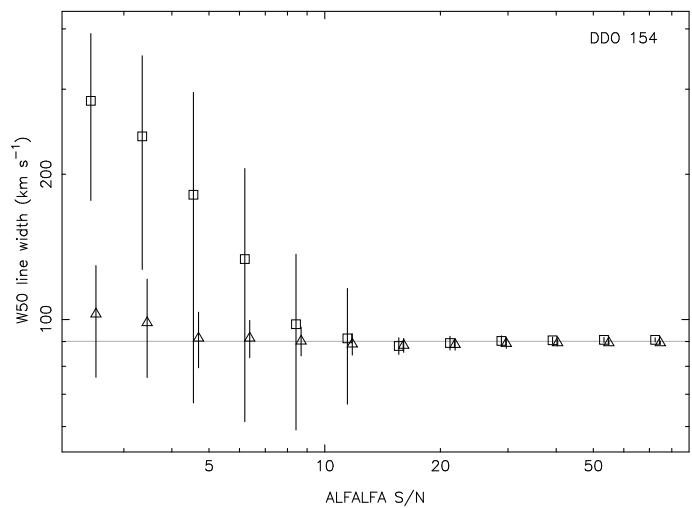}
      \caption{Same as for Fig. \ref{fig_N1} except DDO 154 is shown.
              }
         \label{fig_N2}
   \end{figure}

\begin{figure}
   \centering
      \includegraphics[width=\hsize]{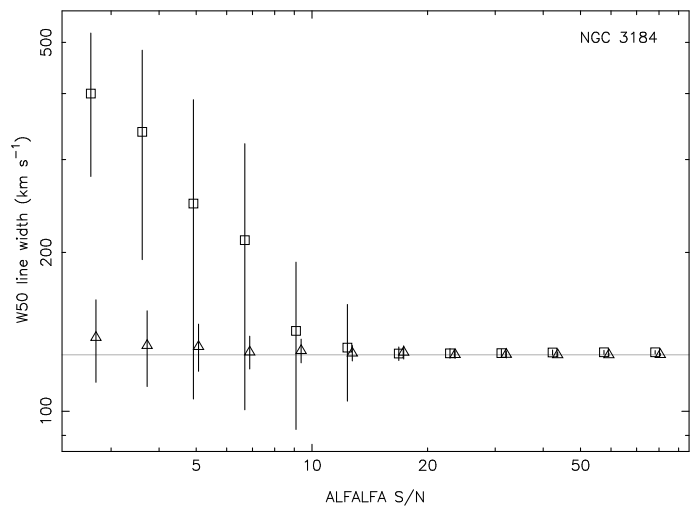}
      \caption{Same as for Fig. \ref{fig_N1} except NGC 3184 is shown.
              }
         \label{fig_N3}
   \end{figure}

Figures \ref{fig_N1} through \ref{fig_N3} all make the same comparison, but a single galaxy is plotted per figure for clarity. The other two galaxies have not been shown because their results are similar. For each value of S/N in each plot, two values of W50, the line width at 50\% of maximum height, are shown: the error bar centred on a triangle comes from measuring the fitted profile; that centred on the square, from the raw spectrum, using the simple algorithm discussed in Sect. \ref{sss_coarse_fitting}. It is obvious from all plots that finding W50 from the fitted profile yields a more accurate result. The effect is most stark for broad profiles, which for the same ALFALFA-formula S/N have a much lower peak-to-RMS value.

\subsubsection{Discussion} \label{sss_coarse_discussion}

\citet{lewis_1983} made a similar study of the effect of noise on measurements of \ion{H}{i} line position and width. Although Lewis's study was in many respects more comprehensive than the present one, this author applied traditional direct methods to measure the positions of line edges, and deprecated the use of functions fitted to the line shoulders over several channels. Lewis argued that the number of channels spanning the line shoulder was of the same order as the minimum number of parameters required by any useful fitting function, and therefore that fitting was equivalent merely to interpolation, and could therefore not be expected to yield any improvement in precision.

In opposition to this view is the Bayesian probability theory, which says that the most complete knowledge possible of the probability distribution of linewidth values is obtained by use of Bayes' theorem in conjunction with a model of the spectral line. A seeming oversupply of model parameters is no objection provided the unwanted ones are marginalized out. The only necessary criterion is that the model reproduces accurately the widths of real spectral lines. For the present model, this is well demonstrated in Sect. \ref{sss_orig_comparison}.

Nevertheless it seems reasonable that a model with 6 parameters will be poorly constrained if the spectral line extends over no more than about the same number of velocity channels. It is of interest therefore to consider the likely shape of the Bayesian posterior distribution in the extreme case that the spectral line has most of its flux in a single channel. In this case it seems clear that the line centre and total flux parameters will remain well-constrained. On the other hand, the fraction-solid and asymmetry parameters can be expected to be completely unconstrained, with the posterior density remaining about the same if they are varied through their respective ranges (as for example in Fig. \ref{fig_G}). Since the Gaussian spread of the spectral line is modelled by the $\Delta v_\mathrm{rand}$ parameter, its value will be most tightly constrained by the flux densities in the channels adjacent to the central one. If these are insignificant, then this parameter will be unconstrained within a range between zero and about the width of a channel. The final parameter, the intrinsic linewidth, will be expected to have the same character (this again is similar to what is seen in Fig. \ref{fig_G}).

In fact none of these indeterminacies and degeneracies matter, because in practice one integrates the posterior over all the dimensions (i.e. parameters) of no present interest. For example if one wants a probability distribution of $W_{50}$ for a given spectrum, this is very easy to obtain from the Bayesian formulation for the respective data. The number of parameters in the model should be viewed merely as a detail of the way the problem is worked out, having no importance for the final result.

Another issue which deserves some discussion concerns the biases in fit parameters observed in Figs. \ref{fig_K} to \ref{fig_N3}. If the Bayesian formulation is optimum, one might ask, why does it still lead to biased results? The answer is that the Monte Carlo procedure adopted in the present section is not the way one ought to proceed with real data. In no case is it optimum to fit separate spectra, then average the resulting parameter values. This is done here only because we have no option if we wish to make comparison with standard methods of estimating line parameters directly from the data. If one did not have the necessity of comparison with non-Bayesian methods, the correct approach depends on the nature of the quantity to be estimated. Either we have an ensemble of observations of the same object, or an ensemble of observations of different objects. In the former case, the correct Bayesian approach is to fit once to the entire data set. In this case we are back (given sufficient data) in the high S/N regime and any biases will be correspondingly insignificant. If the objects are all different, then the correct approach is hierarchical modelling. An interesting recent example of this treatment is \cite{brewer_2014}. Essentially one defines a hyper-model which describes the distribution of values of properties across the ensemble of objects, and tries to constrain its hyper-parameters. Although any individual spectrum is low S/N, again a sufficiently large data set will eventually reach the high-S/N, thus low-bias regime as regards the hyperparameters.

\subsection{The THINGS galaxies in EBHIS} \label{ss_tests_ebhis}
\subsubsection{Overview} \label{sss_ebhis_overview}

The semi-simulated spectra generated in Sect. \ref{ss_tests_coarse} still fall some way short of real life and therefore cannot be expected to present the full range of difficulties one encounters in trying to analyse real observations. We wanted also to test the profile model on single-dish data. The EBHIS survey (Effelsberg Bonn \ion{H}{i} Survey, \citealt{winkel_2010, kerp_2011}) is convenient for this, since its survey coverage has included most of the THINGS galaxies, yet as a single-dish survey, it necessarily has much reduced spatial resolution when compared to the interferometer used by \citet{walter_2008} to make their THINGS observations. An extensive comparison of the EBHIS observations of the THINGS galaxies might be interesting and useful, but is beyond the scope of the present paper. Here we just wish to show that profile-fitting works well also under less than ideal conditions. Specifically we are interested in the following questions:
\begin{itemize}
  \item Can we cope with non-flat baselines and interference?
  \item Can profile-fitting at separate pixels across the source return a map of the source flux which is better (less noisy) than one obtained simply by adding flux densities over a range of channels?
  \item Can the variation in the asymmetry parameter across the source give clues to the galaxy orientation and inclination, even with poorly-resolved sources?
\end{itemize}

\subsubsection{Baseline fitting} \label{sss_ebhis_baseline}

The baseline problem differs between single-dish vs. interferometer spectra. Single-dish measurements tend to be worse affected by Fabry-Perot type effects which can produce waves in spectral baselines. Spectra made with an interferometer on the other hand tend to have flat baselines, although a spatially superimposed continuum source can raise and otherwise perturb the baseline. Future, deep interferometric surveys for \ion{H}{i} may, however, begin to be affected by source confusion, which may necessitate wider use of baseline fitting.

Traditionally, non-flat baselines have been dealt with by fitting a function, often a simple polynomial, to stretches of channels on either side of the line of interest. The fit is interpolated across the line profile but channels within the profile make no contribution to the fit. This is less than ideal for three reasons: firstly because of the restricted number of channels which contribute information to the fit; secondly because it relies on a decision as to where the profile begins and ends; thirdly, selection of the degree of the fitting polynomial is usually done `by eye'.

The approach taken in the present paper is simply to add a baseline function to the existing profile model, and fit the combined model to all selected channels. This also lends itself to a Bayesian approach to determining the best order of the baseline function. This technique is more fully described in appendix \ref{ss_app_B_bkg}.

Chebyshev polynomials have been preferred for the baseline function in the present paper, because they are easy to generate, and more orthogonal and numerically tractable than simple polynomials. Note however that a Chebyshev polynomial is still just a polynomial, and an $n$th-order Chebyshev fitted to some data will result in exactly the same function as fitting a simple $n$th-order polynomial would (B. Winkel, private communication). The Chebyshev is just better behaved from a computational point of view.

The EBHIS cubes have 3000 spectral channels spanning nearly 3900 km/s, which is far wider than any of the lines fitted. It isn't necessary to fit to the full velocity range, so in all cases a small section only, about 3 or 4 times wider than the spectral line, was selected for fitting purposes.

\subsubsection{Results} \label{sss_ebhis_results}

Results are presented here for only a single galaxy, DDO 154.

\begin{figure}
   \centering
      \includegraphics[width=\hsize]{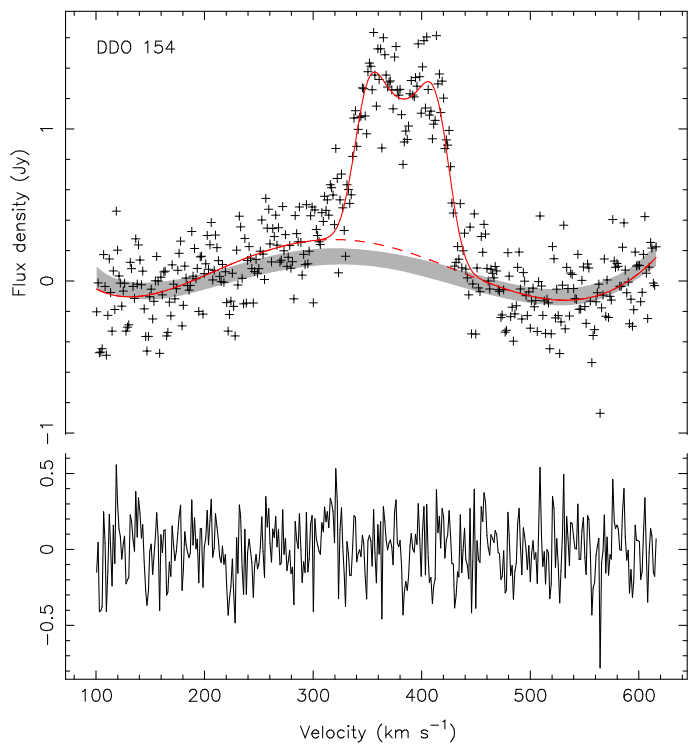}
      \caption{Constructed from the EBHIS cube for DDO 154. The upper plot shows a spectrum from a section of the cube, integrated over the same spatial dimensions as shown in Figs. \ref{fig_Q1} to \ref{fig_Q3}. The lower plot shows the residuals after subtraction of the fitted model. The red line shows the MCMC-mean profile model, with 6 orders of Chebyshev function. The solid line shows the whole profile, whereas the dashed line just shows the baseline. The $\pm 1$ sigma range of the baseline prior is indicated by the grey band.
              }
         \label{fig_R}
   \end{figure}

Figure \ref{fig_R} shows the spectrum of DDO 154 obtained by integrating a $9 \times 9$ pixel area of the relevant data cube. The profile shown is the mean of $10^4$ iterations of a converged MCMC, but it is almost indistinguishable to the eye from the best-fit profile from a Levenberg-Marquardt optimisation. The baseline is fitted by a sum of Chebyshev polynomials truncated at order 5 (i.e. giving 6 orders in total), which number gave the maximum Bayesian evidence (see appendix \ref{ss_app_B_bkg}).

A spectrum summed across the entire breadth of this cube showed no evidence for narrow-band RFI in any channel, so no channels were excised before fitting.

Because there is a significant baseline contribution in this spectrum, careful attention was paid to baseline priors, which were estimated from the non-source portions of the cube as described in Sect. \ref{ss_app_B_baseline}. As can be seen from the figure, the prior accounts for quite a lot of the baseline over the small spatial extent of the source. This gives one confidence that the baseline subtraction is accurate, and thus also the fitted total flux of the spectral line.

A Levenberg-Marquardt optimization returned a best-fit value of $102 \pm 4$ Jy km s$^{-1}$ for the total flux of DDO 154, whereas the MCMC returned a mean value of $100 \pm 5$ Jy km s$^{-1}$. These values are significantly higher than the value of 82 fitted to the data of \citet{walter_2008}. However, it is known that interferometric measurements can miss flux. Our value agrees very well with the value of $105 \pm 5$ Jy km s$^{-1}$ reported by \citet{carignan_1998} from a careful combination of interferometer and single-dish data.

\begin{figure}
   \centering
      \includegraphics[width=\hsize]{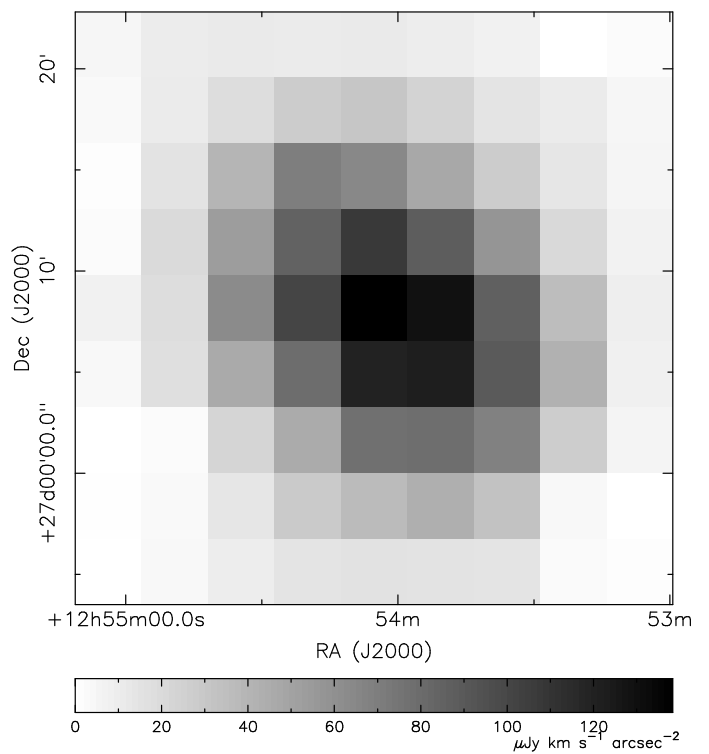}
      \caption{Constructed from the EBHIS cube for DDO 154. The plot shows the fitted total flux at each of a small range of spatial pixels.
              }
         \label{fig_Q1}
   \end{figure}

\begin{figure}
   \centering
      \includegraphics[width=\hsize]{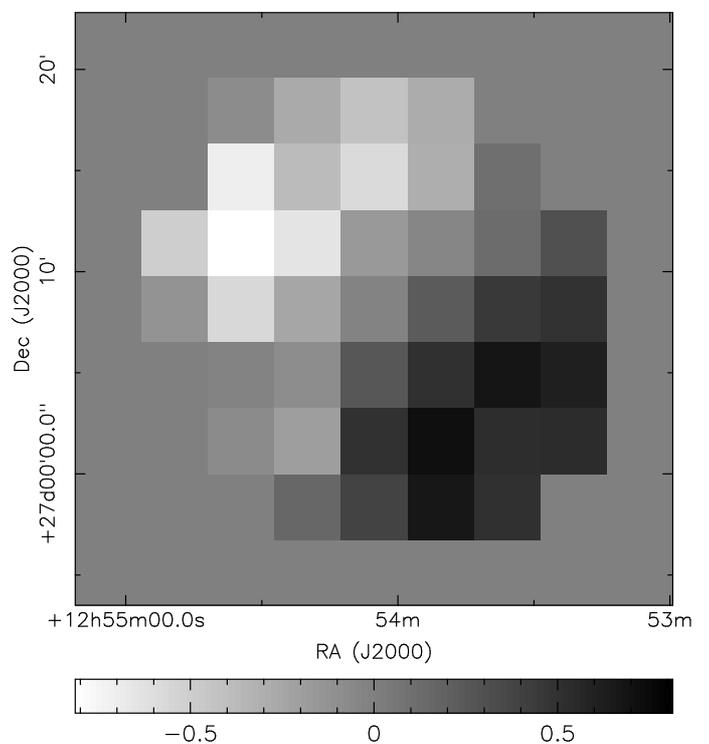}
      \caption{Same as for Fig. \ref{fig_Q1} except the asymmetry parameter is shown. Pixels are set to median grey where the total flux drops below 15\% of maximum.
              }
         \label{fig_Q2}
   \end{figure}

Next, the baseline-plus-line model was fitted to spectra extracted at single spatial pixels across the same $9 \times 9$ area integrated to produce the spectrum in Fig. \ref{fig_R}. $9 \times 9$ maps of the fitted parameter values were made. Figures \ref{fig_Q1} and \ref{fig_Q2} show two of these, respectively for the total-flux parameter and the asymmetry parameter. (The pixels of Fig. \ref{fig_Q2} were set to null where the total flux fell below 15\% of maximum.) It is of interest to compare these to the much higher-resolution versions in Fig. 53 in \citeauthor{walter_2008}. As one might expect, the single-dish EBHIS survey seems to detect more extended structure than Walter et al's interferometric measurements.

Clearly visible in the asymmetry map is the orientation of the galaxy's major and minor axes, even though DDO 154 is barely resolved by the 9 arcmin FWHM beam of the Effelsberg telescope at L band \citep{kerp_2011}.

The remaining parameter maps are of less interest and are not shown. The map of the fitted line centre also (as one might expect) shows indications of the major and minor axes, but the signal is not so unequivocal as with the asymmetry map.

\begin{figure}
   \centering
      \includegraphics[width=\hsize]{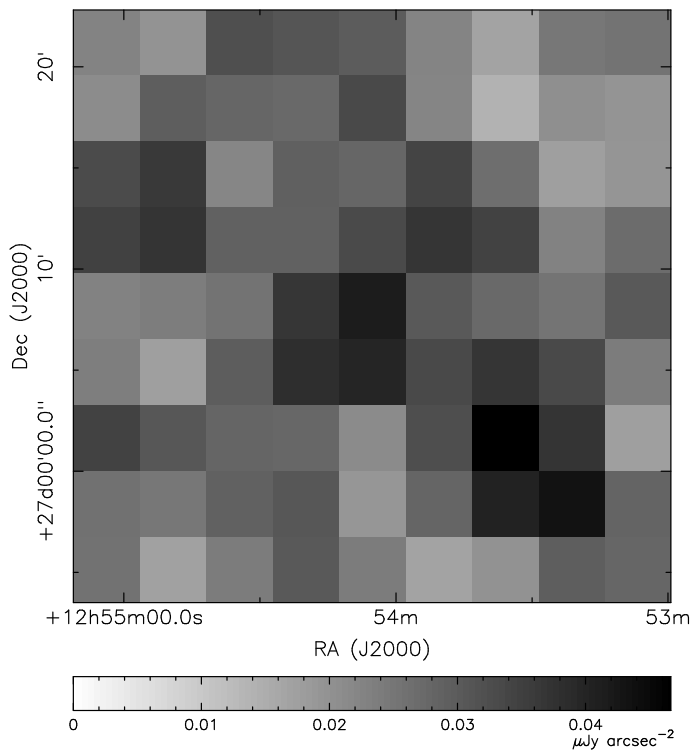}
      \caption{Same as for Fig. \ref{fig_Q1} except the RMS of the Chebyshev coefficients is shown.
              }
         \label{fig_Q3}
   \end{figure}

As far as the baseline fits go, no obvious trends are visible either in the 6 maps of the individual Chebyshev coefficients, or a map of their RMS value (Fig. \ref{fig_Q3}). There is no bright continuum radio source to disturb the baseline at the location of DDO 154: NVSS \citep{condon_1998} shows that there is no source with an average flux density at L band greater than 90 mJy/beam within the approximately 30" by 30" square shown in Figs. \ref{fig_Q1} to \ref{fig_Q3}. The first Chebyshev coefficient (which is just a DC offset) does appear slightly elevated at the centre of the source, but only at the 2-sigma level.

\section{Conclusion} \label{s_conclusion}

The two most desirable measurements of the global \ion{H}{i} spectral profile of a galaxy are its width and area. Combined with optical measurements of the inclination and brightness of the galaxy, these allow the galaxy's \ion{H}{i} mass and intrinsic optical luminosity, and thence its distance, to be estimated via the baryonic Tully-Fisher relation. The systemic velocity of the galaxy comes next in importance. This quantity, when compared to the Hubble velocity appropriate to that distance, additionally allows one to calculate the peculiar or local velocity of the galaxy in the line of sight to the observer. A profile model is described here which permits one to make accurate estimates of these quantities.

In Sect. \ref{ss_tests_orig} the model was fitted to a variety of \ion{H}{i} spectra from the THINGS survey \citep{walter_2008}. These galaxies were observed at high velocity resolution and under very low-noise conditions, and exhibited negligible baselines.

The fits are nearly always excellent. Local densities and voids in the \ion{H}{i} distribution give rise to noisiness in the profiles, but residuals from these wiggles typically have amplitudes less than about 10\% of the profile height. Such local deviations seem to have little effect on the bulk fitted properties.

Particularly important to note is the close fit to the slopes of the lines, which means that linewidth parameters of the profile such as $W_{50}$ or $W_{20}$ are accurately (within 2\%) reproduced by the best-fit model. There is a great deal of existing lore on how to correct such linewidth measurements so as to obtain estimates of the maximum rotation speed of the galaxy, which is the quantity desired for Tully-Fisher calculations. The availability of a good model proxy for the linewidth allows one to make use of such formulae without radical modification, even in cases of low spectral resolution or signal-to-noise ratio (S/N).

These correction formulae include a correction for instrumental broadening. Strictly speaking, this is no longer necessary if the linewidth is taken from the model profile, since this profile is what would actually be seen with no instrumental broadening (i.e., infinite resolution). Trials in Sect. \ref{ss_tests_coarse} in which the model is fitted to spectra with coarser velocity binning indeed show no significant `instrumental' bias in linewidth in the high-S/N limit. Biases do begin to show themselves as S/N decreases toward the limit of detectability, but they are always significantly less than those observed with linewidths measured directly from the noisy, coarsely-resolved profiles. As described in Sect. \ref{sss_coarse_discussion}, correct Bayesian treatment of ensemble observations can circumvent ill effects from such low-S/N biases.

Comparison of the total flux parameter of the fitted model to that obtained by summing the observed flux densities indicates again that the model value of flux is an accurate proxy to the real one (within 1\%). For about half the 34 galaxies fitted, the difference was less than the measurement uncertainty, which was nearly always significantly less than 1\%. There is however a significant minority of galaxies for which the total flux is slightly underestimated. It is thought likely that this is because of the frequent occurrence of extended tails in the profile wings, which are underfitted by the Gaussian wings of the model.

Systemic velocities are returned within measurement errors and without bias by the model fits, even for asymmetric profiles.

The use of the remaining three model parameters (turbulent broadening $\Delta v_\mathrm{rand}$, `solid rotating' fraction $f$ and asymmetry $\alpha$) has been little explored. The $\Delta v_\mathrm{rand}$ value fitted is typically about 12 km s$^{-1}$ (see appendix \ref{ss_app_B_justification}), which is 2 or 3 km s$^{-1}$ larger than typical values seen in spatially resolved maps of velocity distribution. As explained in Sect. \ref{sss_orig_comparison}, this is likely an artefact of the simplistic rotation curve implied by the model. Deviations of the rotation curves of real galaxies from this simple framework are readily absorbed into the $\Delta v_\mathrm{rand}$ parameter, tending therefore to inflate its value.

It might be of interest to see whether the $f$ and $\alpha$ parameters can be useful as proxies for features of galaxy morphology, in analogy with studies such as those of \citet{richter_1994} and \citet{andersen_2009}, but this question is beyond the scope of the present paper.

In Sect. \ref{ss_tests_coarse} it was shown that the model remains a useful way to extract parameters of interest even when the spectral resolution and S/N more nearly approximate those of the bulk of detected sources in surveys such as ALFALFA.

The final test performed on the model was to fit to a representative \ion{H}{i} profile from the EBHIS survey \citep{winkel_2010, kerp_2011}. In contrast to the THINGS observations, this survey was done using a single dish antenna, and thus comes with a significant amount of baseline. This was dealt with in a Bayesian fashion by using the non-source areas of the relevant data cube to constrain the baseline priors. The success of this procedure is shown by the accurate agreement between the resulting fitted value of total flux and previous careful measurements for this galaxy (DDO 154).

Fitting the profile to individual pixels instead of to a spatial sum across several pixels allowed something of the spatial distribution of neutral hydrogen in this galaxy to be mapped in a low-noise fashion, despite the intrinsically poor spatial resolution of the survey. The orientation of the galaxy's rotation axis could also be approximately determined via mapping the fitted value of the asymmetry parameter. The success of these fits shows that the model is not only suitable for \ion{H}{i} profiles which result from a complete spatial integration across the extent of the galaxy, but also for partially integrated profiles.

\begin{acknowledgements}
IMS wishes to thank F Walter and J Kerp for generous provisions of non-public data, and is grateful to L Fl\"{o}er, A Schr\"{o}der and B Winkel for help and advice. We are grateful to the anonymous referee for helpful and informative comments.
\end{acknowledgements}


\bibliographystyle{./aa} 
\bibliography{./things} 

\appendix

\section{The model in Fourier space} \label{s_app_A}
\subsection{The transform components} \label{ss_app_A_1}

As mentioned in Sect. \ref{ss_theory_method}, the model is generated in Fourier space partly to avoid singularities, partly to facilitate convolution, and partly to mimic the process by which real radio spectra are generated in an XF-type correlator. To follow the latter process requires the accurate construction of a synthetic discrete autocorrelation function $Z_j$ for $j=-N$ to $N-1$. This is done by sampling the Fourier transform $Z(\tau)$ of the profile model at appropriate intervals in $\tau$, which is proportional to the autocorrelation lag time. The correct autocorrelation values result, provided only that the spectral line to be modelled has no significant power outside the chosen velocity interval $[v_\mathrm{lo},v_\mathrm{hi}]$. This interval needs to be decided on before calculating the transforms. Together with the number of channels $N$, it defines the channel width $\Delta v_\mathrm{chan}$.

The Fourier transform $Z(\tau)$ of the entire profile model is obtained by multiplying together the separate transforms $Z_\mathrm{intrinsic}(\tau)$ and $G(\tau)$ of the expressions respectively for $s_\mathrm{intrinsic}(v)$ and the dispersive convolver $g(v)$ given in equations \ref{equ_model_a} and \ref{equ_dispersion}. The transform of the intrinsic or undispersed profile evaluates to
{\setlength\arraycolsep{2pt}
  \begin{eqnarray} \label{equ_Z_intrinsic}
	Z_\mathrm{intrinsic}(\tau) & = & 2 S \exp (\mathrm{i} \psi \tau) \left[ \left\{ (1-f) J_0(\tau) + \frac{2 f}{\tau}J_1(\tau) \right\}\right. +\nonumber\\
	                           & + & \left. \mathrm{i} \alpha \left\{ (1-f) J_1(\tau) +  \frac{2 f}{\tau} \left[ \frac{2}{\tau} J_1(\tau) - J_0(\tau) \right] \right\} \right]
  \end{eqnarray}
}
where $J_0$ and $J_1$ are respectively the zeroth and first order Bessel functions of the first kind, and 
\begin{displaymath}
	\psi = \frac{2 (v_\mathrm{hi} - v_\mathrm{ctr})}{\Delta v}.
\end{displaymath}
The transform of the dispersive-motion convolver gives
\begin{displaymath}
	G(\tau) = \exp \left[ -2 \left( \frac{\Delta v_\mathrm{rand}}{\Delta v} \tau \right)^2 \right].
\end{displaymath}
The correct sampling is at $\tau = j \Delta \tau$ for integer $j \in [-N,N-1]$, where
\begin{displaymath}
	\Delta \tau = \frac{\pi \Delta v}{2 N \Delta v_\mathrm{chan}}.
\end{displaymath}

Recall that $\Delta v$ in all these expressions is the linewidth parameter of the model, as described in Sect. \ref{ss_theory_model}.

\subsection{Computing close to singularities} \label{ss_app_A_2}

Note that we did not group all the $J_0$ and $J_1$ terms together in equation \ref{equ_Z_intrinsic}. This is because terms with $\tau$ in the denominator need special treatment as $\tau \to 0$.

The expression $J_1(\tau)/\tau$ in equation \ref{equ_Z_intrinsic} is analogous to the sinc function. For values of $\tau$ close to zero it should be approximated by its power series
\begin{displaymath}
	\frac{J_1(\tau)}{\tau} = \frac{1}{2} \left( 1 - \frac{\tau^2}{8} + \frac{\tau^4}{192} -  \dots \right).
\end{displaymath}

Let us denote the last set of terms in equation \ref{equ_Z_intrinsic} as
\begin{displaymath}
	E(\tau) = \frac{1}{\tau} \left[ \frac{2}{\tau} J_1(\tau) - J_0(\tau) \right].
\end{displaymath}
The components of this function also have singularities at $\tau=0$ which cause computational difficulties as $\tau \to 0$. The power series for $E$ evaluates to
\begin{displaymath}
	E(\tau) = \frac{\tau}{8} \left( 1 - \frac{\tau^2}{12} + \frac{\tau^4}{384} - \dots \right).
\end{displaymath}

\subsection{Derivatives} \label{ss_app_A_3}

It can be convenient (for example when doing Levenberg-Marquardt optimization) to have expressions for the derivatives of the model with respect to each of the six parameters. We give the Fourier transforms of these here.

For convenience we define $C(\tau)$ such that the whole-model transform can be expressed as
\begin{displaymath}
	Z(\tau) = Z_\mathrm{intrinsic}(\tau) \times G(\tau) = 2 S G(\tau) \exp (\mathrm{i} \psi \tau) C(\tau).
\end{displaymath}
$C$ thus contains all terms within the outer square bracket of equation \ref{equ_Z_intrinsic}. The derivatives can then be given as
{\setlength\arraycolsep{2pt}
\begin{eqnarray*}
  \frac{\partial Z}{\partial v_\mathrm{ctr}}      & = & \frac{-2 \mathrm{i} \tau}{\Delta v} Z \\
  \frac{\partial Z}{\partial \Delta v}            & = & 2 S G \exp(\mathrm{i} \psi \tau) \frac{\partial C}{\partial \Delta v} \\
  \frac{\partial Z}{\partial S}                   & = & Z / S\\
  \frac{\partial Z}{\partial \Delta v_\mathrm{rand}} & = & -4 \Delta v_\mathrm{rand} \left( \frac{\tau}{\Delta v} \right)^2 Z \\
  \frac{\partial Z}{\partial f}                   & = & 2 S G \exp(\mathrm{i} \psi \tau) \frac{\partial C}{\partial f} \\
  \frac{\partial Z}{\partial \alpha}              & = & 2 S G \exp(\mathrm{i} \psi \tau) \frac{\partial C}{\partial \alpha}.
\end{eqnarray*}
}
The $C$ derivatives are as follows:
{\setlength\arraycolsep{2pt}
\begin{eqnarray*}
	\frac{\partial C}{\partial \Delta v} & = & \frac{-\tau}{\Delta v} \left\{ (1-f) J_1(\tau) + 2 f E(\tau) - \phantom{\frac{1}{1}}\right.\\
	                                     & - & \left. \mathrm{i} \alpha \left[ (1-f) \left( J_0(\tau) - \frac{J_1(\tau)}{\tau} \right) + \frac{2 f}{\tau} J_1(\tau) - \frac{6 f}{\tau} E(\tau) \right] \right\}
\end{eqnarray*}
}
\begin{displaymath}
	\frac{\partial C}{\partial f} = \left[ -J_0(\tau) + \frac{2}{\tau}J_1(\tau) \right] +\mathrm{i} \alpha \left[ -J_1(\tau) + 2 E(\tau) \right]
\end{displaymath}
\begin{displaymath}
	\frac{\partial C}{\partial \alpha} = \mathrm{i} \left[ (1-f) J_1(\tau) + 2 f E(\tau) \right].
\end{displaymath}

\section{Model fitting - technical issues} \label{s_app_B}

\subsection{Priors in general} \label{ss_app_B_general}

Bayes' theorem (equation \ref{equ_bayes}) describes the posterior probability distribution of the model parameter values for a given set of data. This represents all that can be known about the parameters. However usually it suffices to settle for less than this. Typically one wishes to approximate the posterior via a single set of optimum parameter values coupled with widths of the posterior function in each dimension of the parameter space. This represents a full description of the posterior only if this function is an unskewed Gaussian; but it is often a good enough approximation. `Optimum' can mean the values of the parameters at the maximum of the posterior, but mean values are often an acceptable proxy for these. Only if the posterior is very asymmetric will these two sets of values be significantly different.

The posterior is a product of the likelihood and a function which represents prior knowledge of the parameter distribution. It is often easier to specify the likelihood than the priors. A typical example is where nothing definite at all is known about the likely distribution of the parameter values. What form of prior best encodes such a state of ignorance? There has been considerable debate about the correct forms for `ignorance' priors \citep[see for example references in][]{dagostini_2003}. The question depends however to some degree on what one wants to do with the posterior. If a value for the Bayesian evidence is what is needed, then priors have to be more carefully chosen: in particular, they must be integrable. Priors which are not integrable are named `improper'. However in most cases discussed in the present paper, the location of the maximum in the posterior is all that is desired. If in addition the S/N of the data is high, the shape of the posterior will be dominated by a sharply-peaked likelihood, and the exact form of the priors becomes unimportant. One may even choose an improper prior without disadvantage.

Unbounded, thus improper, priors were used for most of the parameters in fitting the profile model to the original, high-S/N THINGS data. The fraction-solid and asymmetry parameters $f$ and $\alpha$ have natural high and low bounds, but all the other parameters are unbounded on one or both sides.

\subsection{Priors chosen for the low S/N spectra} \label{ss_app_B_justification}

The distributions of fitted parameters from the high-S/N fits were used to generate some priors for the low-S/N fits. Identical priors were applied in fitting both the semi-simulated spectra in Sect. \ref{ss_tests_coarse} and the EBHIS spectra in Sect. \ref{ss_tests_ebhis}.

\begin{figure}
   \centering
      \includegraphics{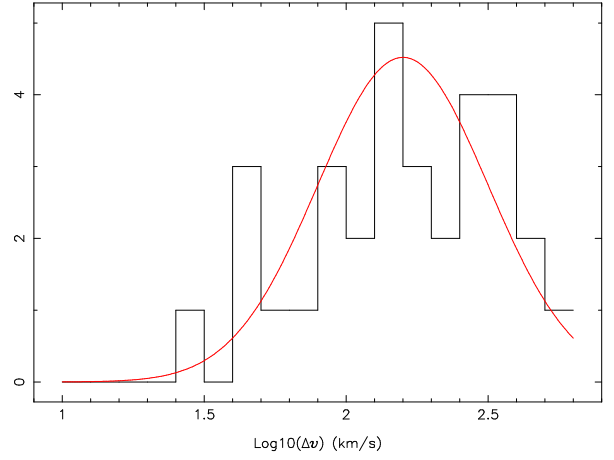}
      \caption{Distribution among the THINGS galaxies of $\Delta v$, the linewidth parameter of the fitted profiles. The red curve shows the prior which was chosen approximately to represent this distribution.
              }
         \label{fig_J3}
   \end{figure}

\begin{figure}
   \centering
      \includegraphics[width=\hsize]{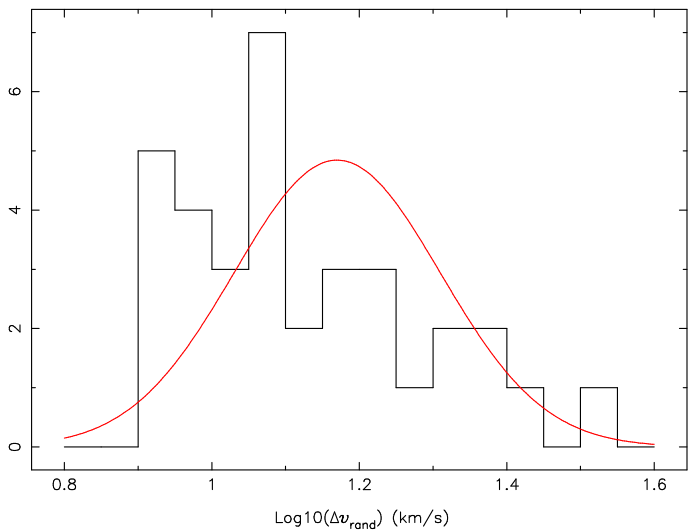}
      \caption{Distribution among the THINGS galaxies of $\Delta v_\mathrm{rand}$, the velocity dispersion parameter of the fitted profiles. As in Fig. \ref{fig_J3}, the red curve shows the prior.
              }
         \label{fig_J0}
   \end{figure}

\begin{figure}
   \centering
      \includegraphics[width=\hsize]{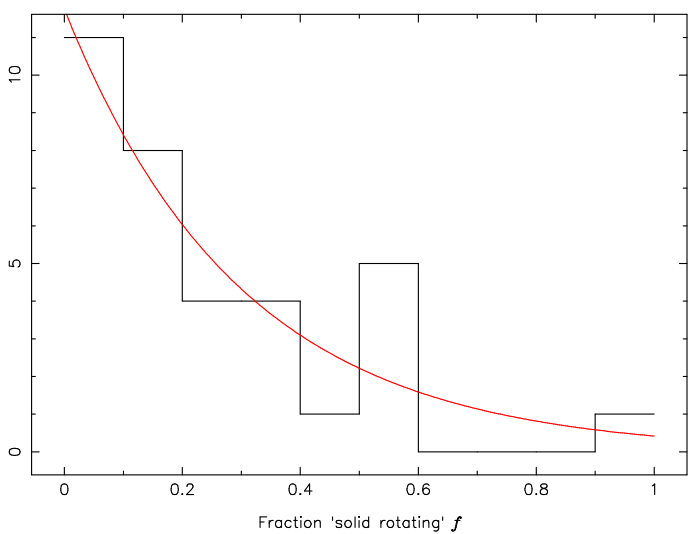}
      \caption{Distribution among the THINGS galaxies of $f$, the `fraction solid rotating' parameter of the fitted profiles. As in Fig. \ref{fig_J3}, the red curve shows the prior.
              }
         \label{fig_J1}
   \end{figure}

\begin{figure}
   \centering
      \includegraphics[width=\hsize]{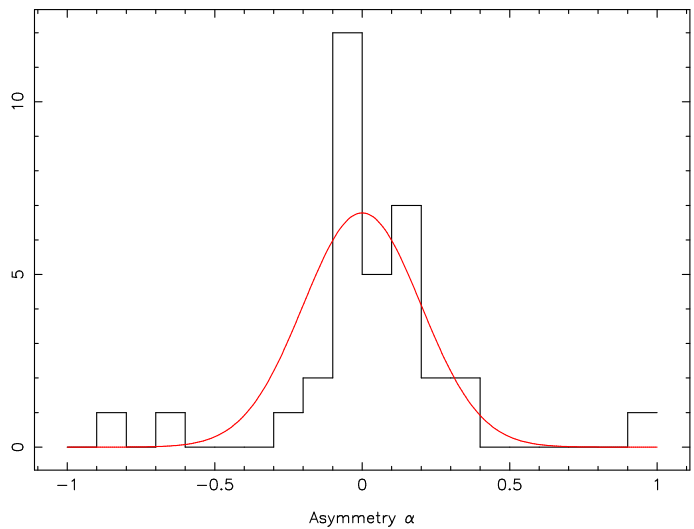}
      \caption{Distribution among the THINGS galaxies of $\alpha$, the asymmetry parameter of the fitted profiles. As in Fig. \ref{fig_J3}, the red curve shows the prior.
              }
         \label{fig_J2}
   \end{figure}

The spectral-line fits performed on the 34 THINGS galaxies in Sect. \ref{ss_tests_orig} don't provide a very numerous sample, but some trends do become apparent. Using this information, non-flat priors were decided for four of the six model parameters, namely the linewidth $\Delta v$, the dispersion velocity width $\Delta v_\mathrm{rand}$, the \ion{H}{i} mass fraction in the core $f$, and the asymmetry parameter $\alpha$. Histograms of the distribution of these four across the 34 THINGS fits are shown in Figs. \ref{fig_J3}, \ref{fig_J0}, \ref{fig_J1} and \ref{fig_J2}.

A logarithmic histogram of the fitted linewidth parameter $\Delta v$ is shown in Fig. \ref{fig_J3}. Despite the single galaxy (DDO 53) with linewidth close to 10 km/s, a log-normal prior centred on 2.2 (i.e., at $10^{2.2}$ km/s) and of width 0.3 seemed a reasonable fit to the THINGS values.

Figure \ref{fig_J0} shows the dispersion width. As already discussed, this is typically wider than would be expected from turbulence and thermal velocities alone. For this parameter we chose a log-normal prior function with a center in log space of 1.17 and a width of 0.14. It isn't a particularly good fit to the histogram however: something with a sharper low-end cutoff and a longer tail would be better.

In Fig. \ref{fig_J1} a similar histogram for the core-fraction parameter is presented. The decrease in the incidence of this parameter as it increases in value is consistent with an exponentially decreasing prior, although a value of 0.3 was chosen for the characteristic constant so as to be consistent with the presence of the outlier in the histogram at the high end of the scale.

The final histogram in Fig. \ref{fig_J2} shows the asymmetry parameter. Values which depart from the midpoint by more than about 0.2 are seen to be unusual. The outliers here occur for spectra for which the model is not a good fit to the profile. A Gaussian with sigma equal to 0.2 was chosen for this prior.

In addition to these four priors, and purely in order to exclude nonsense results due to fitting the profile to noise away from the true spectral line, we also chose a prior for the line centre parameter, this being a Gaussian centred on the true line centre (as determined from the high-S/N fits), and with a width somewhat arbitrarily set equal to 20 km s$^{-1}$. This is considered acceptable because we assume that in a real survey, the rough location of the line would already been found by other means. We remind the reader that we are not treating source detection here, but source characterisation.

For the remaining total-flux parameter, a flat prior was retained, with non-physical, negative-valued results excluded as before.

To summarize, the priors adopted in the present section were:
\begin{displaymath}
  \begin{array}{cl}
	x                      & p(x) \\
	\hline
	v_\mathrm{ctr}         & \exp(-0.5 [\{x-v_\mathrm{true}\} / 20]^2 ) / (20 \sqrt{2 \pi}) \\
	\Delta v               & \exp(-0.5 [\{\log10(x)-2.2\} / 0.3]^2 ) / (0.3 \ln 10 \sqrt{2 \pi}) \\
	S                      & x>0 \\
	\Delta v_\mathrm{rand} & \exp(-0.5 [\{\log10(x)-1.17\} / 0.14]^2 ) / (0.14 \ln 10 \sqrt{2 \pi}) \\
	f                      & 0.3 \exp(-0.3 x ) \\
	\alpha                 & \exp(-0.5 [x / 0.2]^2 ) / (0.2 \sqrt{2 \pi})\textrm{ for }|x|\le 1, =0\textrm{ else.}\\
	\hline
  \end{array}
\end{displaymath}

\subsection{Possible fitting problems} \label{ss_app_B_problems}
\subsubsection{Degeneracy} \label{sss_app_B_degeneracy}

\begin{figure}
   \centering
      \includegraphics[width=\hsize]{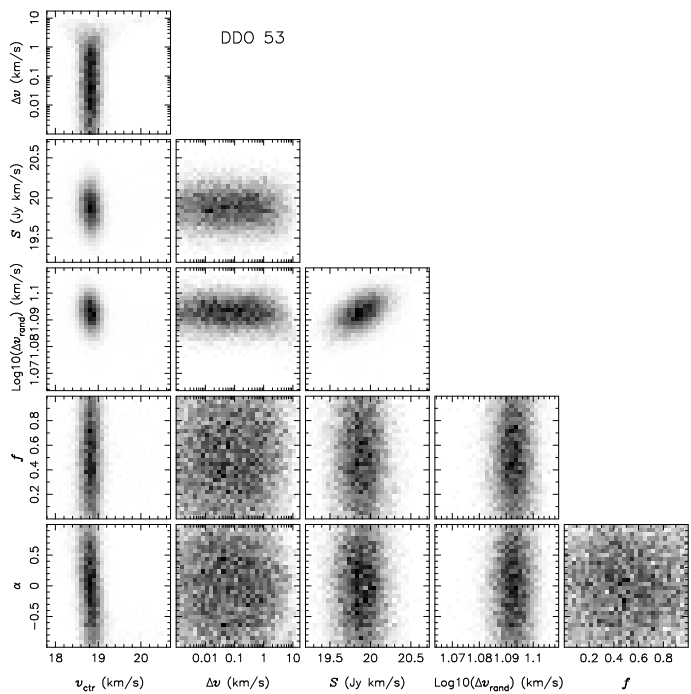}
      \caption{The shape of the posterior distribution for the six-parameter line profile model applied to DDO 53, as explored by the Markov-chain Monte Carlo (MCMC) procedure. Each pixel is shaded to represent the density of points in the MCMC.
              }
         \label{fig_G}
   \end{figure}

The chosen set of parameters are convenient for many reasons but have the problem of not being orthogonal under all circumstances. For example, if the intrinsic linewidth $\Delta v$ is small compared to the dispersion width $\Delta v_\mathrm{rand}$, the $f$ and $\alpha$ parameters begin to be poorly constrained. Figure \ref{fig_G} serves to illustrate this point. And for at least one of the fits in Sect. \ref{ss_tests_orig}, the best-fit value of $\alpha$ is close to unity, which causes nearly all the other parameters to be poorly constrained. These degeneracies can occur for both scenarios described in Sect. \ref{sss_fitting_bayes} and can cause convergence problems when trying to fit the profile. An MCMC is again the preferred way to deal with such cases.

\subsubsection{Lag-space problems} \label{sss_app_B_lag}

Filtering in lag space in general causes the noise values in adjacent velocity channels to be correlated. The only common filtering scenario which avoids this is Hanning filtering followed by discarding every second channel. This was the scheme employed by \citet{walter_2008}, so we have not had to address this problem. Other data sets are not so favourable though: for example, HIPASS made use of a 25\% Tukey filter \citep{barnes_2001} which introduces correlations between adjacent channels. The way to deal with this is called generalized least squares and involves reformulating the expression for $\chi^2$ to have the form
\begin{equation} \label{equ_GLSQ}
	\chi^2 = (\mathbf{s} - \mathbf{m})^\mathrm{T} \mathbf{C}^{-1} (\mathbf{s} - \mathbf{m}).
\end{equation}
Here the T superscript indicates transpose and $\mathbf{C}$ is the covariance matrix of the data channels. For uncorrelated data $\mathbf{C}$ is of course diagonal and equation \ref{equ_GLSQ} reverts to the standard expression. Equation \ref{equ_GLSQ} seems straightforward but in fact $\mathbf{C}$ is often singular for correlated data. This is effectively because there are now fewer than $N$ degrees of freedom. Hanning smoothing for example reduces the number of degrees of freedom by a factor of 2, which is why one usually discards half the channels afterwards - they don't contain any new information. In general the solution to this is not to try to invert $\mathbf{C}$ but to calculate instead its pseudo-inverse \citep{penrose_1955}, which reduces the rank of the problem to the proper number. But this is not further elaborated here.

An alternative approach to the case of correlated noise would be to back-transform the data and perform the fitting in lag space, where there is no correlation between lag samples. This is a little complicated if the velocity channels have different standard deviations, which is the case in Sect. \ref{ss_tests_orig}, so it has not been attempted here.

When fitting to a real spectral line, in principle both the velocity range and the number of channels should match that in the original observation. This is because the end points where the autocorrelation function is truncated (particularly the low-frequency, i.e. high-velocity end) generate some ringing at places in the spectrum where its slope is steep \citep[see e.g.][]{willis_1993}. If there is such ringing in the data, it's obviously advantageous if the model reproduces this artefact.

There were two reasons why, for many of the THINGS galaxies, this was difficult to do. Firstly, the cubes published by \citet{walter_2008} don't include all the observed channels. The starting channels had to be estimated from the tabulated description of the observations. Secondly, for some of the targets, Walter et al spliced together several observations taken with different velocity ranges. These cannot be accurately represented via a single simulated autocorrelation. These inaccuracies were however not regarded as significant.

\subsection{Markov-chain Monte Carlo} \label{ss_app_B_mcmc}

A Markov-chain Monte Carlo (MCMC) is a sequence of steps within the parameter space. According to the Metropolis-Hastings algorithm \citep{metropolis_1953}, at each step a random displacement from the last good position is generated, and a test is applied to the value of the posterior at the resulting trial position. If the test is passed, the trial position is accepted as the next position in the chain; if not, a new trial is generated. It can be shown that the distribution of accepted positions will in time converge to that of the posterior.

This procedure is simple and robust to apply, but practical issues can arise, principally two: the speed of convergence, and how to know when the MCMC has converged. If the MCMC starts a long way from the centre of the posterior, and if the dimensions of the step size distribution are very different from those of the posterior, the rate of acceptances can be very low. This is the reason for using a LM fit in Sect. \ref{ss_tests_orig} to find the location and dimensions of the posterior maximum before starting the MCMC. After starting the MCMC, convergence speed was further improved via an iterative procedure, described as follows. The MCMC was divided into a sequence of separate chains. To begin with, the chains were short - only tens to hundreds of accepted steps were required. After finishing each chain, the covariance of the parameter points was calculated and used to specify the shape of an $N$-dimensional Gaussian step-size function for use in the next chain ($N$ being the number of model parameters). The chains thus gradually converged at a faster rate, and more acceptances could be demanded. The final sequence of chains used had
\begin{displaymath}
	N_\mathrm{A} = \{30, 100, 200, 1000, 1000, 10000\},
\end{displaymath}
where $N_\mathrm{A}$ is the number of acceptances required.

\begin{figure}
   \centering
      \includegraphics[width=\hsize]{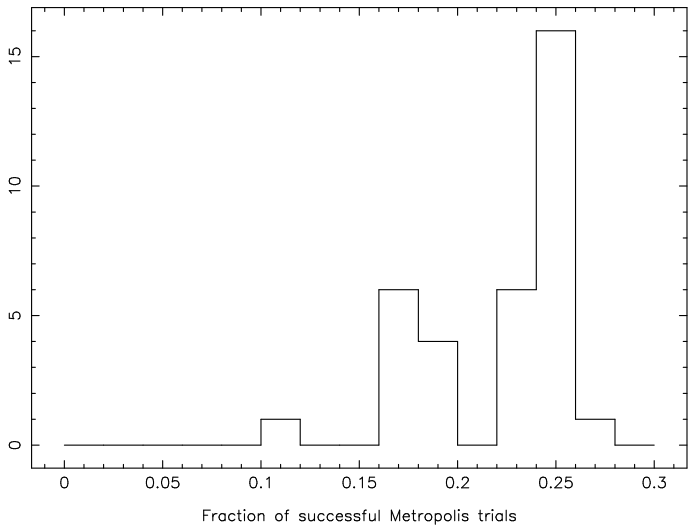}
      \caption{The distribution among the 34 THINGS galaxies of the fraction of successful trials returned by the Metropolis algorithm in the Markov-chain Monte Carlo profile-fitting procedure. The fraction was calculated from a chain of $10^4$ successful trials after a burn-in chain of 2330 successes.
              }
         \label{fig_F}
   \end{figure}

For a Gaussian posterior in a 6-dimensional parameter space, the fraction of accepted steps is expected to be a little less than 0.3 (estimated from Figs. 3 and 4 of \citealt{hanson_1998}). This figure may be compared with the distribution of this fraction among the THINGS galaxies shown in Fig. \ref{fig_F}.

The parameter values and uncertainties listed in Table \ref{tab_O} were obtained by calculating the means and covariances of the final chain of $10^4$ acceptances.

Others who employ the MCMC method commonly use far more points in the final chain than $10^4$; typically also convergence tests are applied; sometimes multiple chains are used, with sparse sampling of the final chains, in order to avoid the obvious correlation between successive points. We have not felt any of this to be necessary. Convergence was judged by monitoring the acceptance rate from chain to chain in the sequence, also by examining some plots of the successive parameter values. When no drift in the mean could be detected, it was judged that these chains have converged enough. The sequence given above was found to be adequate for all 34 profiles. Correlation was thought not to be a problem, since it is the ensemble of points which matters to us, and if there are enough points, then any sparse subset of them ought to have the same distribution as any other, and thus of the whole.

\subsection{Calculation of the spectrum uncertainties} \label{ss_app_B_uncerts}

The THINGS spectra were formed by summing, for each velocity channel, the values of a subset of pixels in the corresponding plane of the data cube. The subset was chosen by \citet{walter_2008} to include only those pixels where some \ion{H}{i} emission was detectable above the background. The uncertainty in each summed value can be expected to scale with the square root of the number of pixels summed, but calculating an exact value has some complications, because the noise values in adjacent spatial pixels are already correlated via convolution by the interferometer beam.

It is easiest to examine this in the Fourier domain. Given input uncorrelated noise of standard deviation $\sigma$ which is then convolved by a beam $b_{j,k}$, the standard deviation or RMS $\sigma_I$ of the output is given by
{\setlength\arraycolsep{2pt}
\begin{eqnarray} \label{equ_rms}
  \sigma_I^2 & = & \sigma^2 \langle \mathbf{B} \mathbf{B}^\ast \rangle\nonumber\\
             & = & \sigma^2 \Sigma b^2 \textrm{ or } \sigma^2 \mathbf{b}\cdot\mathbf{b}
\end{eqnarray}
}
where $\mathbf{B}$ is the discrete Fourier transform of the beam $\mathbf{b}$, $\ast$ signifies complex conjugation, and the angle brackets represent the mean value.

This is the state of the data in each image plane of the cube as we receive it. Note that in interferometry practice it is standard to normalize the beam such that $b_{0,0}=1$. This allows one to read the flux density of point sources from the point image value $I_{j,k}$ at their locations, and gives rise to the expression `janskys per beam' for the units of the image values. For an extended source, the total flux density $S$ can be approximated by
\begin{equation} \label{equ_spectrum}
  S \sim \frac{1}{A_\mathrm{beam}} \sum_{j,k}^{A_\mathrm{patch}} I_{j,k}
\end{equation}
where $A_\mathrm{patch}$ represents the subset of image pixels summed over, and $A_\mathrm{beam}$ is the `area' in pixels of the beam, defined by
\begin{displaymath}
  A_\mathrm{beam} = \frac{1}{b_{0,0}} \sum b_{j,k}.
\end{displaymath}

The problem we have is to calculate the uncertainty in the $S$ values given by equation \ref{equ_spectrum}, starting with a knowledge of $\mathbf{b}$ and the value of $\sigma_I$. This becomes easy once we realize that equation \ref{equ_spectrum} can be thought of as a further convolution
\begin{displaymath}
  \mathbf{S} = \frac{\mathbf{I} \star \mathbf{p}}{A_\mathrm{beam}} ,
\end{displaymath}
where the convolver $\mathbf{p}$ equals unity for pixels within the patch or subset contributing to the sum, and zero otherwise. Bold font has been used to emphasize that all these quantities are images; the desired scalar value of spectrum flux density is in this notation given by $S_{0,0}$. The uncertainty in $S$ can thus be related directly to the original $\sigma$ by
\begin{displaymath}
  \sigma_S^2 = \frac{\sigma^2}{A^2_\mathrm{beam}} \langle (\mathbf{B} \mathbf{B}^\ast) (\mathbf{P} \mathbf{P}^\ast) \rangle,
\end{displaymath}
where $\mathbf{P}$ is the discrete Fourier transform of $\mathbf{p}$. More directly useful to us is to make use of equation \ref{equ_rms} to express $\sigma_S$ in relation to the RMS $\sigma_I$ of the cube planes:
\begin{displaymath}
  \sigma_S^2 = \frac{\sigma_I^2}{A^2_\mathrm{beam}} \frac{\langle (\mathbf{B} \mathbf{B}^{\ast}) (\mathbf{P} \mathbf{P}^{\ast}) \rangle}{\Sigma b^2}.
\end{displaymath}
This can be calculated for any given $\mathbf{p}$, but it is of interest to look at its behaviour when $A_\mathrm{patch} \gg A_\mathrm{beam}$. In this case, we expect $\mathbf{P}$ to be much more compact than $\mathbf{B}$, thus we can approximate
{\setlength\arraycolsep{2pt}
\begin{eqnarray*}
  \sigma_S^2 & \sim & \frac{\sigma_I^2}{A^2_\mathrm{beam}} \frac{\left( B_{0,0} B_{0,0}^\ast \right) \langle \mathbf{P} \mathbf{P}^\ast \rangle}{\Sigma b^2}\\
             & \sim & \frac{\sigma_I^2}{A^2_\mathrm{beam}} \frac{\left( \Sigma b \right)^2 A_\mathrm{patch}}{\Sigma b^2}.
\end{eqnarray*}
}
For a Gaussian $\mathbf{b}$,
\begin{displaymath}
  \Sigma b^2 = \frac{b_{0,0}}{2} \Sigma b.
\end{displaymath}
so we arrive at a final approximation
\begin{displaymath}
  \sigma_S \sim \frac{\sigma_I}{b_{0,0}} \sqrt{\frac{2 A_\mathrm{patch}}{A_\mathrm{beam}}}.
\end{displaymath}
Thus, with $b_{0,0}$ set to 1 according to usual interferometer practice, the ratio between the output and input uncertainties $\sigma_S$ and $\sigma_I$ is given, not by the square root of the `number of beams' $A_\mathrm{patch}/A_\mathrm{beam}$, but by this quantity times root 2.

\subsection{Deciding the order of the baseline model} \label{ss_app_B_bkg}

In fitting to the EBHIS profiles we had to deal with spectra with non-flat baselines. For this purpose a sequence of Chebyshev polynomials $T_j$ was added to the model, as described in Sect. \ref{ss_tests_ebhis}:
\begin{displaymath}
  s_\mathrm{base}(v) = \sum_{j=1}^n a_j \; T_j(u)
\end{displaymath}
where
\begin{displaymath}
  u = 2 \left( \frac{v - v_\mathrm{lo}}{v_\mathrm{hi} - v_\mathrm{lo}} \right) - 1.
\end{displaymath}
One then has to decide at what value of $n$ to truncate the sequence. The Bayesian methodology offers a way to decide this.

The posterior function $p(\mathbf{q}|\mathbf{y})$ giving the probability that the data $\mathbf{y}$ is fitted by model parameter values $\mathbf{q}$ is given in equations \ref{equ_bayes} and \ref{equ_like}. So far we have not concerned ourselves with the evidence $E$, the denominator of equation \ref{equ_bayes}, which is the integral over the parameter space of the numerator. If we change the number of parameters, however, the value of $E$ will vary, and can be used to decide the optimum number.

In the present case the evidence integral has $6+n$ dimensions, and considering for example Fig. \ref{fig_G}, we can see that the integrand, at least in the directions of the 6 profile-only parameters, is unlikely to be well behaved. However, the model is linear in the Chebyshev functions, which means that in these $n$ directions, given Gaussian-distributed noise as we have, the likelihood function has also a Gaussian shape in these dimensions. In addition, the correlation between the $n$ Chebyshev parameters and the 6 profile parameters may be expected to be small, particularly if we include rather more baseline than there are channels under the line profile. Therefore we can approximate the posterior probability function for the whole model by a product between the profile and baseline contributions:
{\setlength\arraycolsep{2pt}
\begin{eqnarray} \label{equ_separate}
  p(\mathbf{q}|\mathbf{y},n) & = & p(\mathbf{q}_\mathrm{profile}+\mathbf{q}^\prime|\mathbf{y},n)\nonumber\\
                             & \sim & p(\mathbf{q}_\mathrm{profile}|\mathbf{y}) \times [ p(\mathbf{q}^\prime|n) \, p(\mathbf{y}|\mathbf{q}^\prime,n) ].
\end{eqnarray}
}
Here $\mathbf{q}^\prime$ represents the $n$ baseline parameters, and the dependence of the probabilities on $n$ has been made explicit.

Deciding the priors $p(\mathbf{q}^\prime|n)$ presents its usual difficulties. Since the Chebyshev polynomials have a significant degree of orthogonality, we can approximate $p(\mathbf{q}^\prime|n)$ by a product of $n$ separate priors. We choose the same very broad prior for each Chebyshev term. `Broad' of course implies slowly-varying compared to the likelihood function; thus each baseline prior may be replaced by its value at the posterior maximum, which we designate $P_\mathrm{base}$. Since all $n$ of them are separate and equal, $p(\mathbf{q}^\prime|n) = P^n_\mathrm{base}$.

The Gaussian likelihood for the $n$ Chebyshev parameters has, following equation \ref{equ_like}, the form
\begin{equation} \label{equ_bkg_like}
  p(\mathbf{y}|\mathbf{q}^\prime,n) = p(\mathbf{y}|\mathbf{q}^\prime_0,n) \exp \left( -[ \mathbf{q}^\prime  - \mathbf{q}^\prime_0]^T \frac{\mathbf{C}^{\prime -1}}{2} [ \mathbf{q}^\prime  - \mathbf{q}^\prime_0] \right).
\end{equation}
Here $\mathbf{q}^\prime_0$ represents the values of the baseline parameters at the posterior maximum, and $\mathbf{C}^\prime$ is the covariance matrix of the posterior with respect to those parameters. The T superscript indicates transpose.

From equations \ref{equ_evidence}, \ref{equ_separate} and \ref{equ_bkg_like} the evidence evaluates to
\begin{displaymath}
  E_n = \kappa_n \; P^n_\mathrm{base} \; \sqrt{(2\pi)^n \; \mathrm{det}(\mathbf{C}^\prime)}
\end{displaymath}
where
\begin{displaymath}
  \kappa_n = p(\mathbf{y}|\mathbf{q}^\prime_0,n) \int d\mathbf{q}_\mathrm{profile} \, p(\mathbf{q}_\mathrm{profile}) \, p(\mathbf{y}|\mathbf{q}_\mathrm{profile}).
\end{displaymath}
Because of the weak coupling between the Chebyshev contribution and the profile model proper, we assume that $\kappa_n$ can be decomposed into
\begin{displaymath}
  \kappa_n \sim K \; p(\mathbf{q}_0|\mathbf{y},n)
\end{displaymath}
where $K$ is a constant and $p(\mathbf{q}_0|\mathbf{y},n)$ is the maximum value of the posterior, which can be measured at the end of the fit process. We cannot easily calculate $K$, but since it will be the same for any $n$, for purposes of comparing different $n$ values we can simply evaluate
\begin{equation} \label{equ_En}
  E^\prime_n = \frac{E_n}{K} = p(\mathbf{q}_0|\mathbf{y},n) \; P^n_\mathrm{base} \; \sqrt{(2\pi)^n \; \mathrm{det}(\mathbf{C}^\prime)}.
\end{equation}

Intuitively, as $n$ increases, we would expect the fit to get better and better, reflected in an increased value of the posterior maximum, and thus of $\kappa_n$ in equation \ref{equ_En}; but since $P_\mathrm{base}$ for a broad baseline prior is significantly less than 1, $P^n_\mathrm{base}$ will of course become geometrically smaller, and will eventually offset the rise in $\kappa_n$. This is what is known as the `Ockham's Razor' feature of the Bayesian methodology, in that it weights against a model with too many parameters \citep{dagostini_2003,gregory_1992}.

\begin{figure}
   \centering
      \includegraphics[width=\hsize]{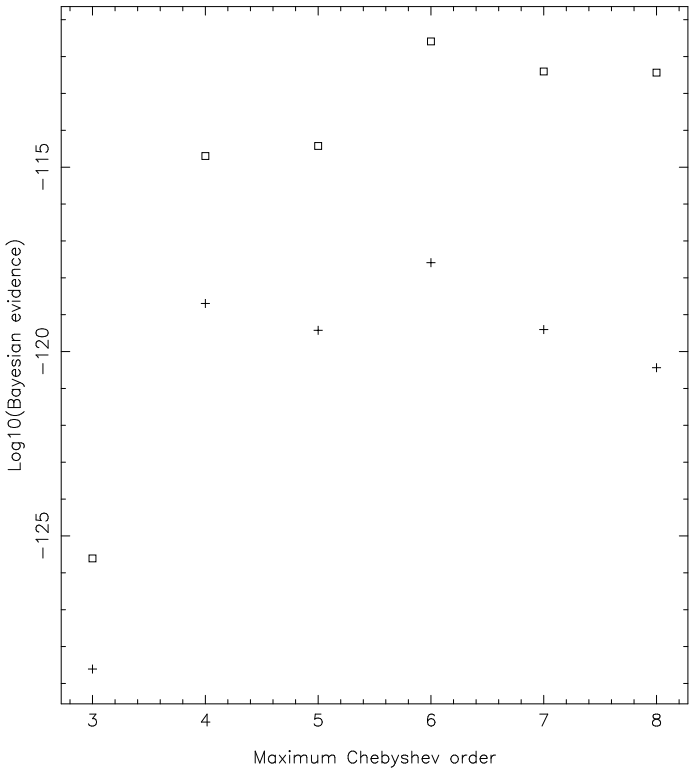}
      \caption{The Bayesian evidence for different maximum order of Chebyshev function. The fit was performed on an EBHIS spectrum for DDO 154.
              }
         \label{fig_P}
   \end{figure}

Shown in Fig. \ref{fig_P} is the value of $E^\prime_n$ for different values of $n$. Two Gaussian priors for the Chebyshevs are tested, one having a width equal to 10 times the noise standard deviation (square points), the other being 10 times wider again (crosses). It can be seen that the width of the prior has small effect, provided only it is wide. In both cases the evidence shows a peak at $n=6$. This was therefore the value chosen for the baseline fits.

\subsection{Baseline priors for the EBHIS profiles} \label{ss_app_B_baseline}

Many of the EBHIS cubes show significant baseline ripple. This must be fitted to obtain the most accurate estimates of the total flux (and to a lesser extent, the other parameters) of spectral lines. Of course the baseline variations continue smoothly through the spectral line itself, but if one includes the spectral line in an unconstrained baseline fit, one risks that the line itself will distort the baseline estimation. Traditionally this is accommodated by manually excluding the line channels from a separate baseline fit. Formally speaking however it is better to fit the same model to every channel, but to constrain the baseline parameters by priors. This has been our approach.

Gaussian priors were chosen for each of the Chebyshev amplitudes in the baseline model. The centre and spread for each were obtained in the following way. Firstly, the baseline model was fitted over the same channel range to the spectrum at every spatial pixel of the data cube, excluding only those pixels either where a significant source contribution was obvious, or at the edges of the cube where the noise was higher than average. Then, for each Chebyshev order, the centre of the prior was taken from the mean of the whole-cube fitted amplitudes, multiplied by the number of spatial pixels summed over the source at issue; and the spread (sigma value) of the prior was assigned from the standard deviation of these amplitudes, suitably corrected for the existing correlation between spatial pixels which is caused by the breadth of the telescope beam.

\section{Graphs of the fits to all spectra} \label{s_app_C}

   \begin{figure}[htbp]
   \centering
      \includegraphics[width=\hsize]{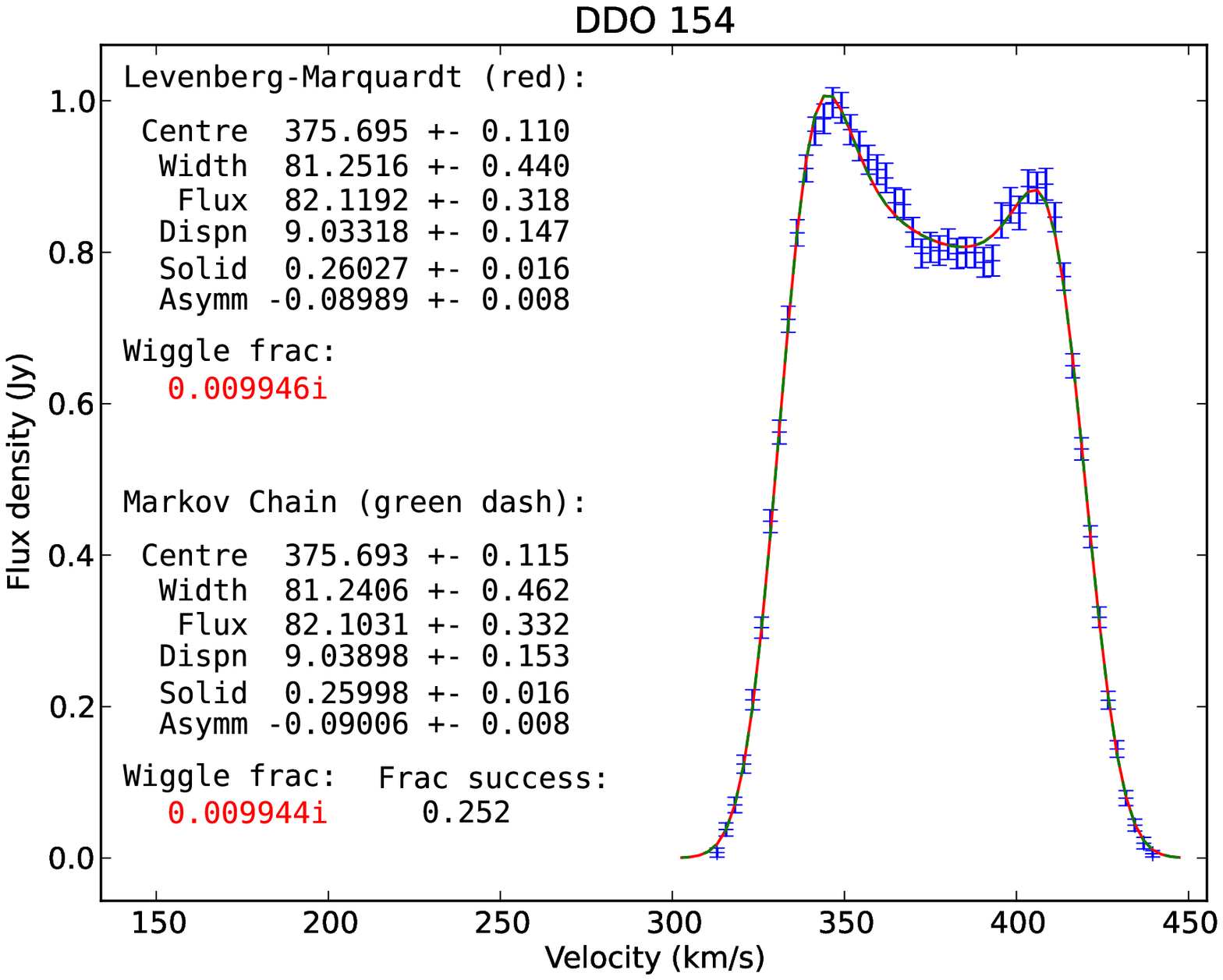}
      \caption{
              }
         \label{fig_gal1}
   \end{figure}

   \begin{figure}[htbp]
   \centering
      \includegraphics[width=\hsize]{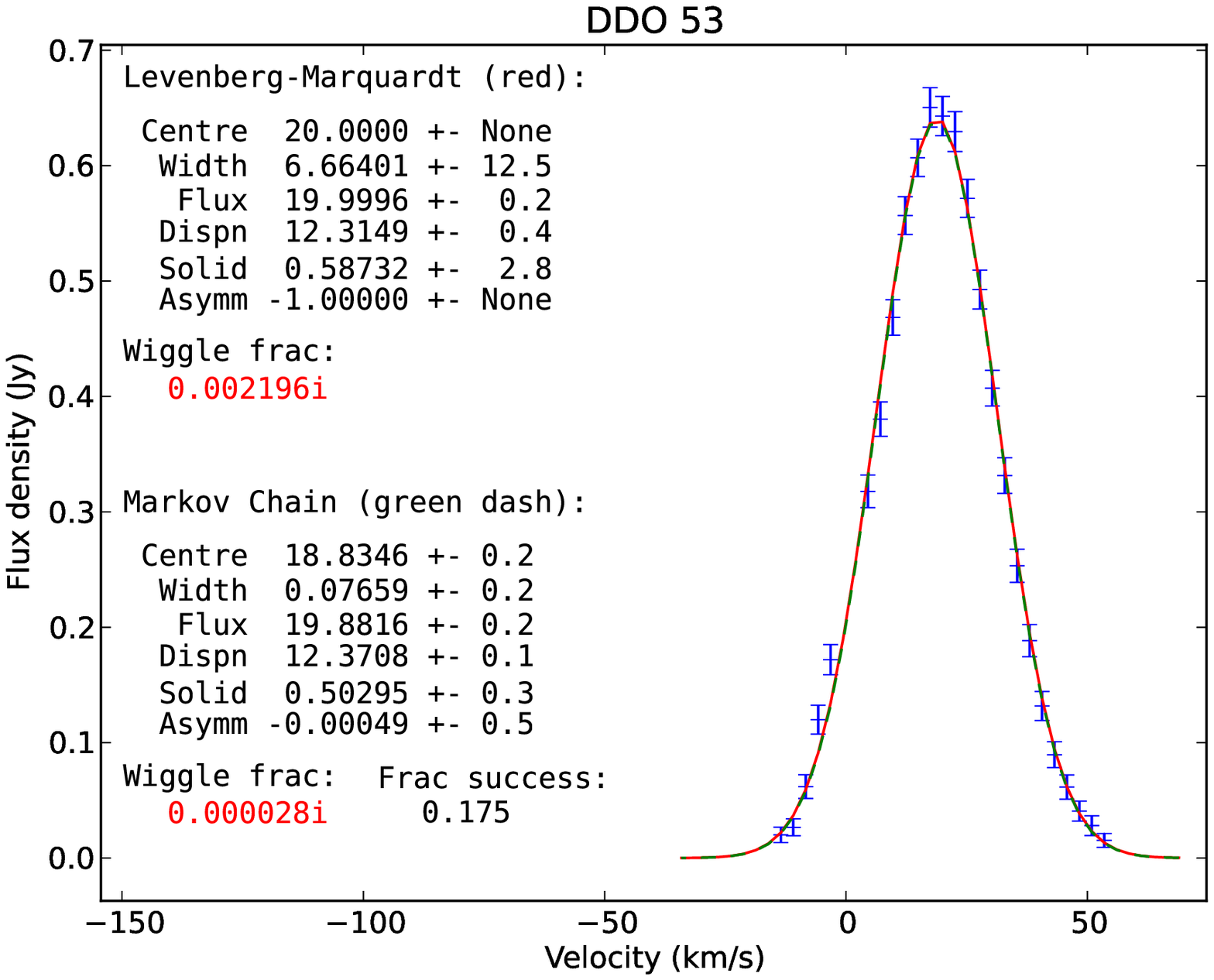}
      \caption{
              }
   \end{figure}

   \begin{figure}
   \centering
      \includegraphics[width=\hsize]{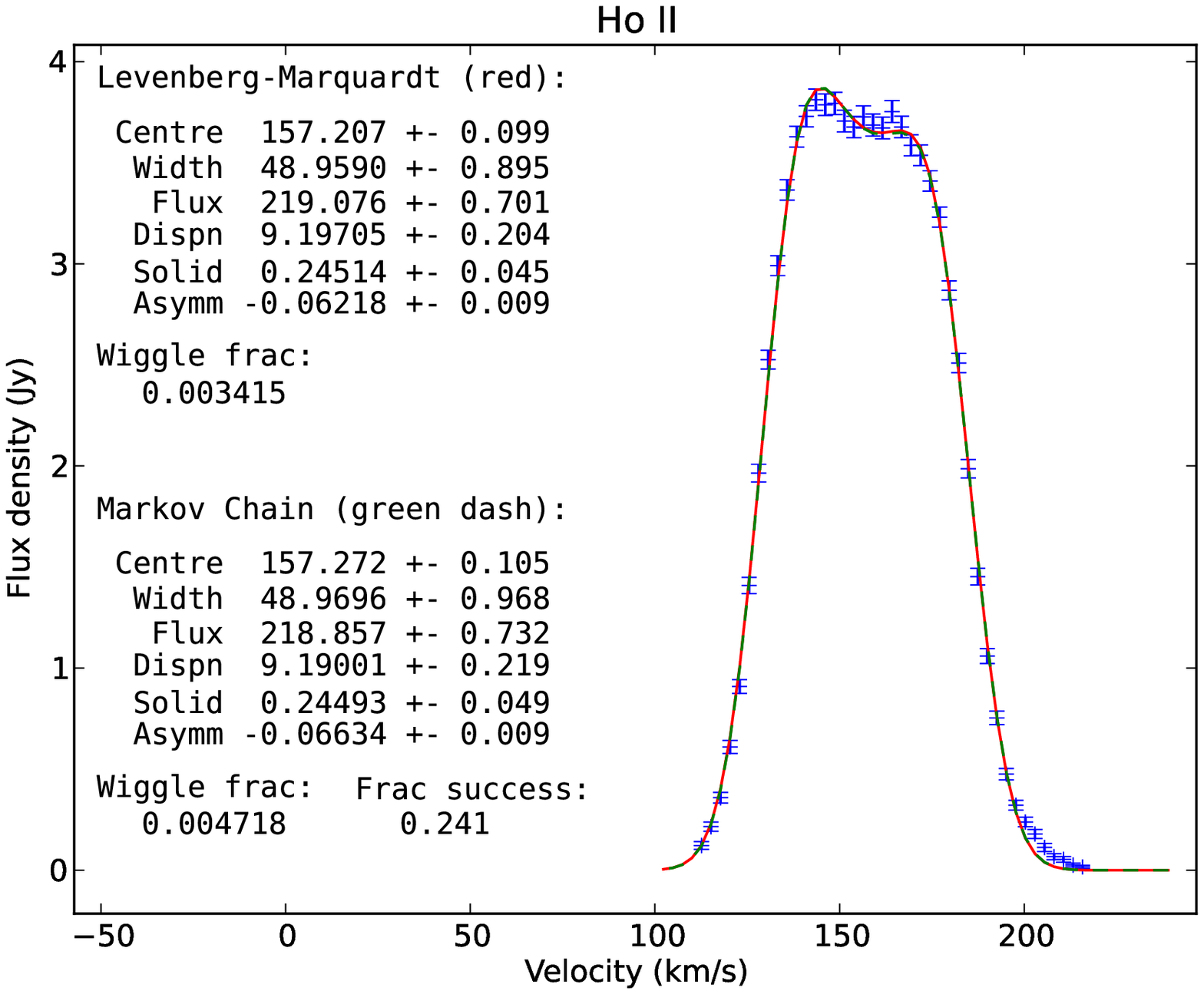}
      \caption{
              }
   \end{figure}

   \begin{figure}
   \centering
      \includegraphics[width=\hsize]{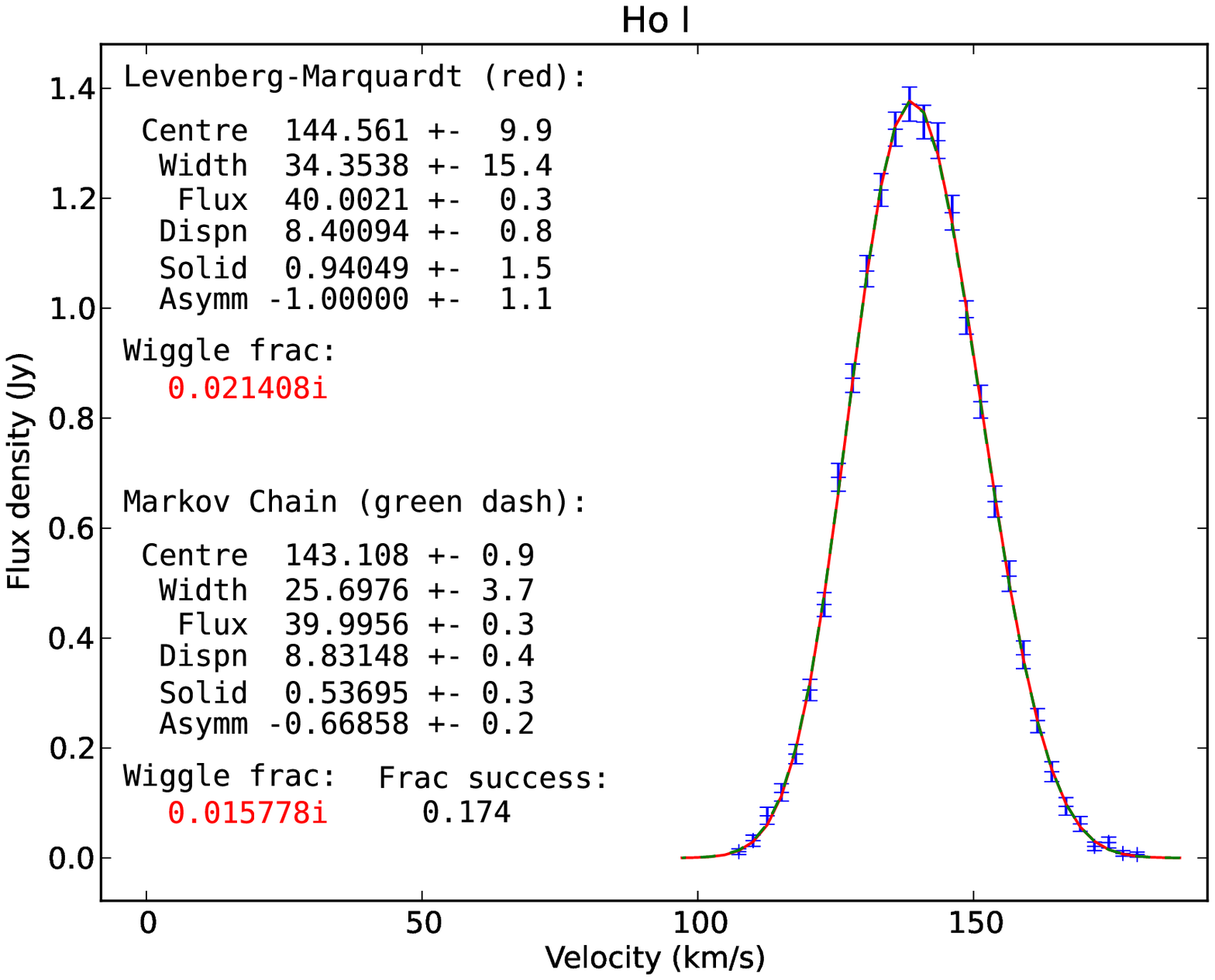}
      \caption{
              }
   \end{figure}

\clearpage

   \begin{figure}
   \centering
      \includegraphics[width=\hsize]{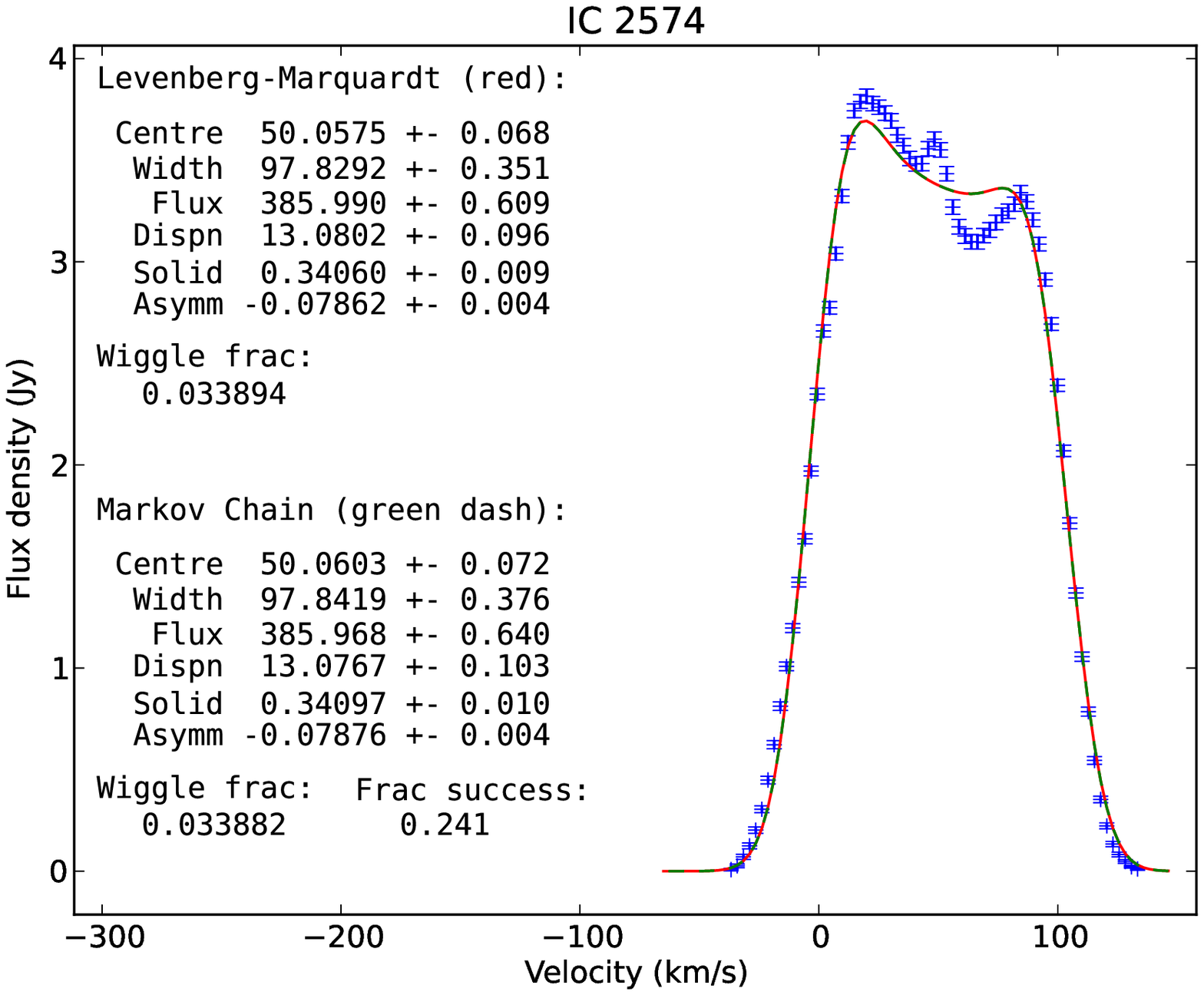}
      \caption{
              }
   \end{figure}

   \begin{figure}
   \centering
      \includegraphics[width=\hsize]{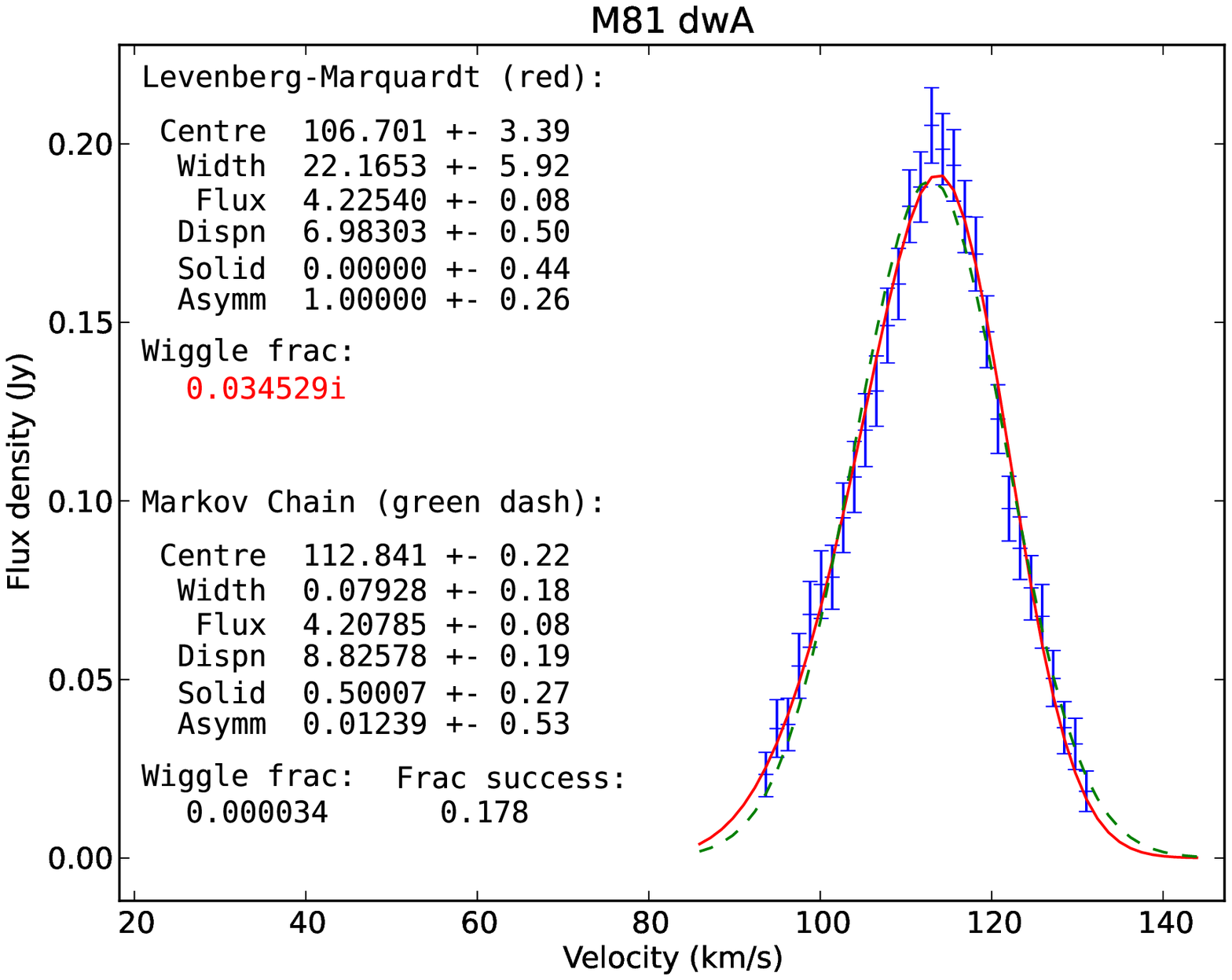}
      \caption{
              }
   \end{figure}

   \begin{figure}
   \centering
      \includegraphics[width=\hsize]{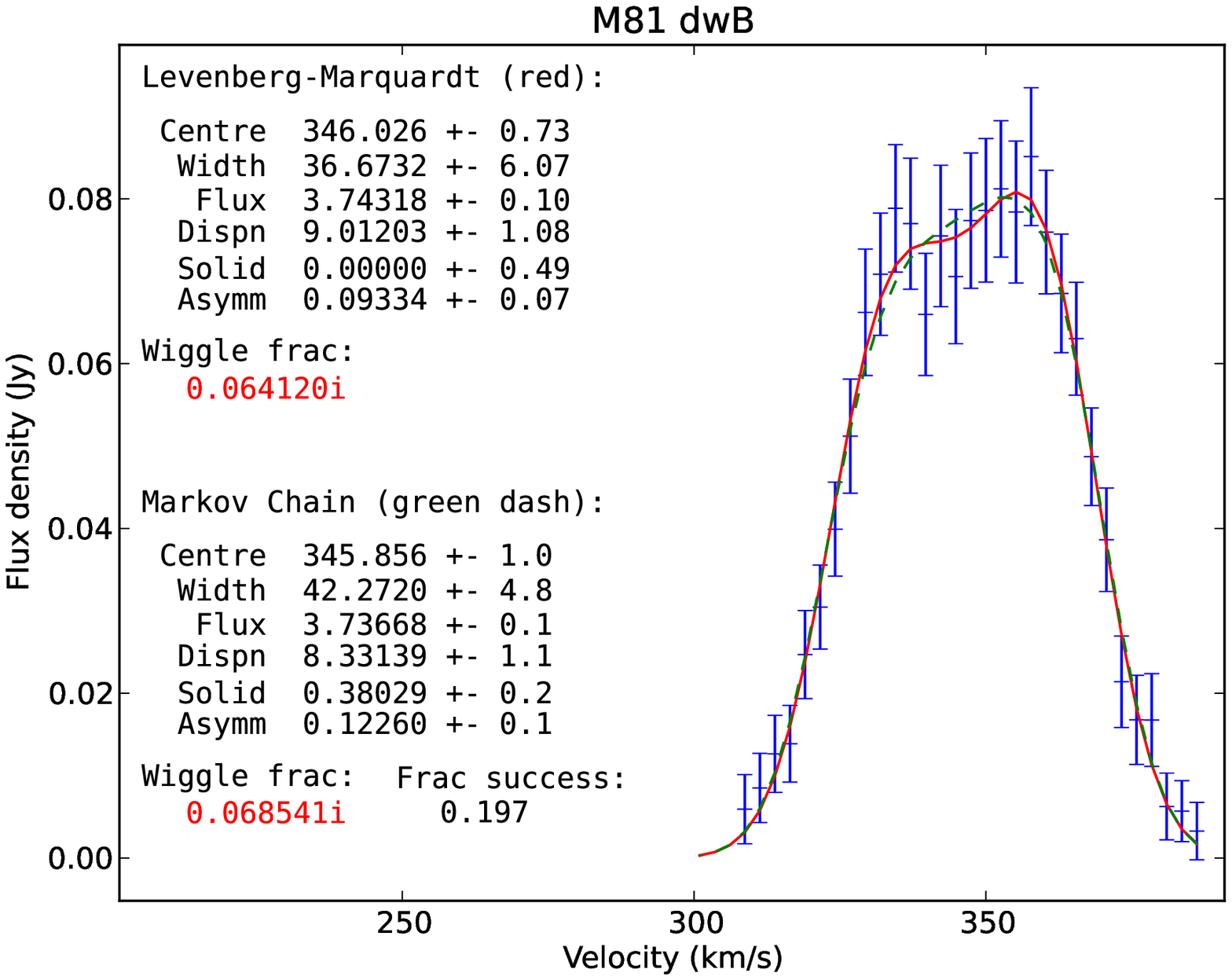}
      \caption{
              }
   \end{figure}

   \begin{figure}
   \centering
      \includegraphics[width=\hsize]{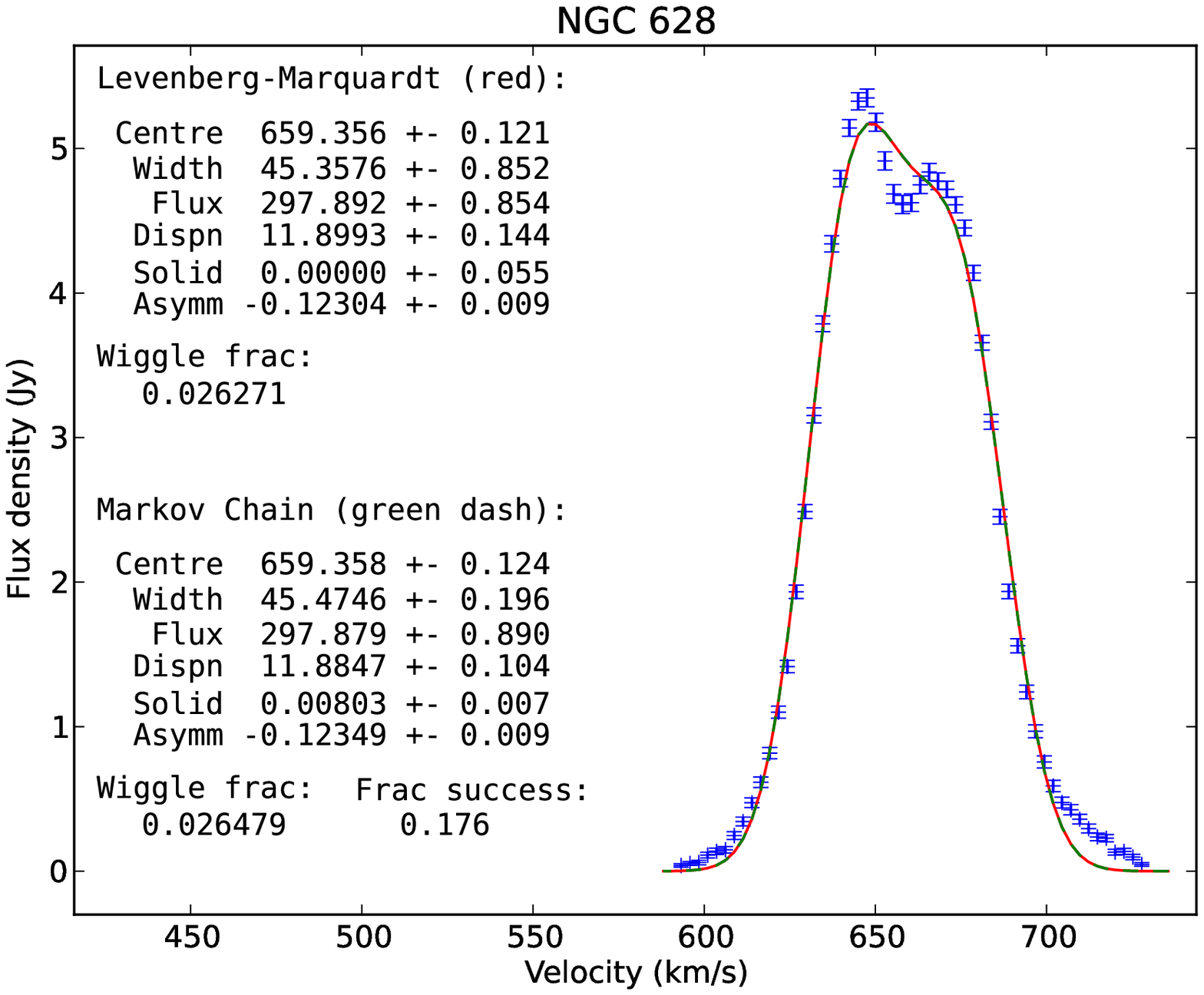}
      \caption{
              }
   \end{figure}

   \begin{figure}
   \centering
      \includegraphics[width=\hsize]{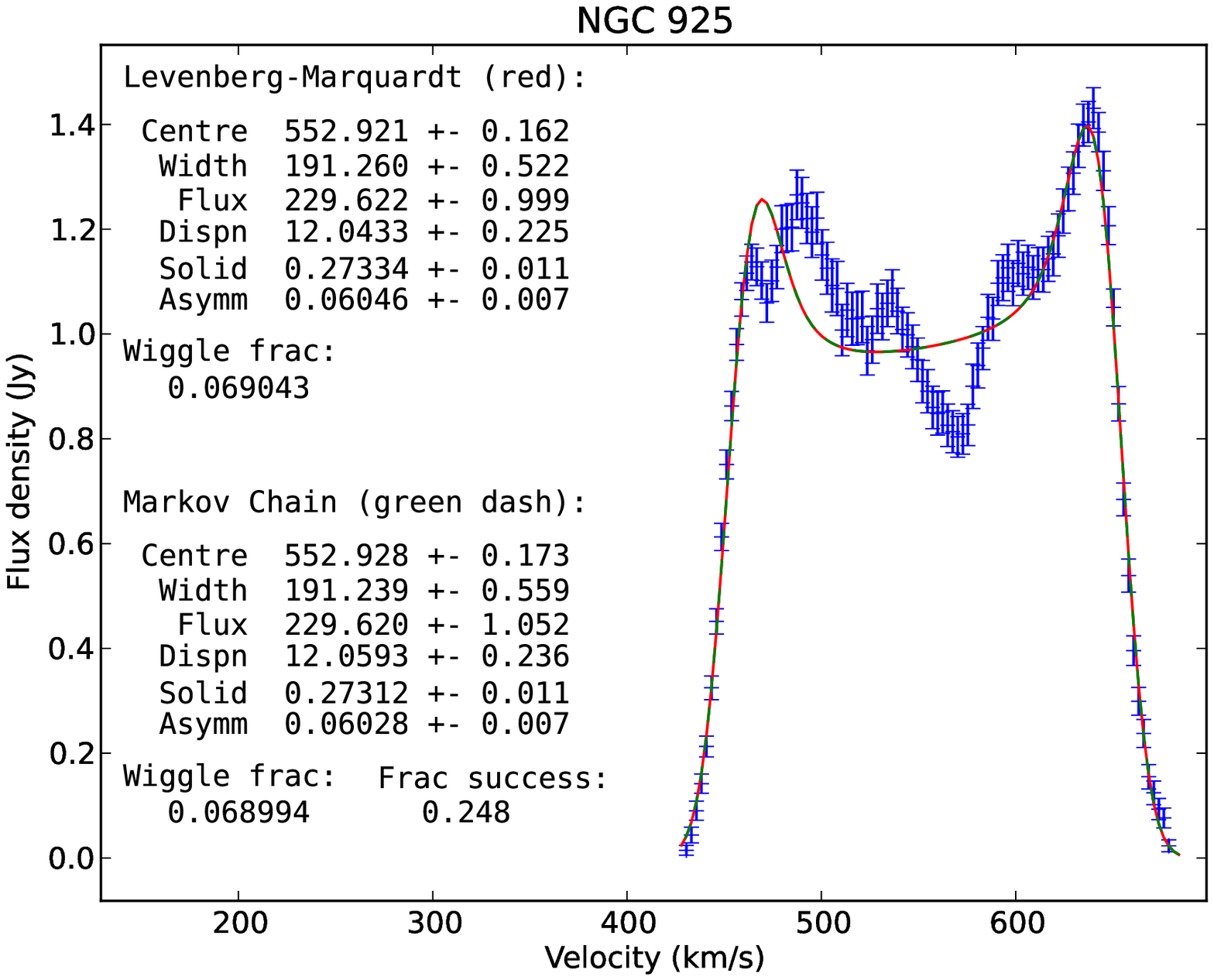}
      \caption{
              }
   \end{figure}

   \begin{figure}
   \centering
      \includegraphics[width=\hsize]{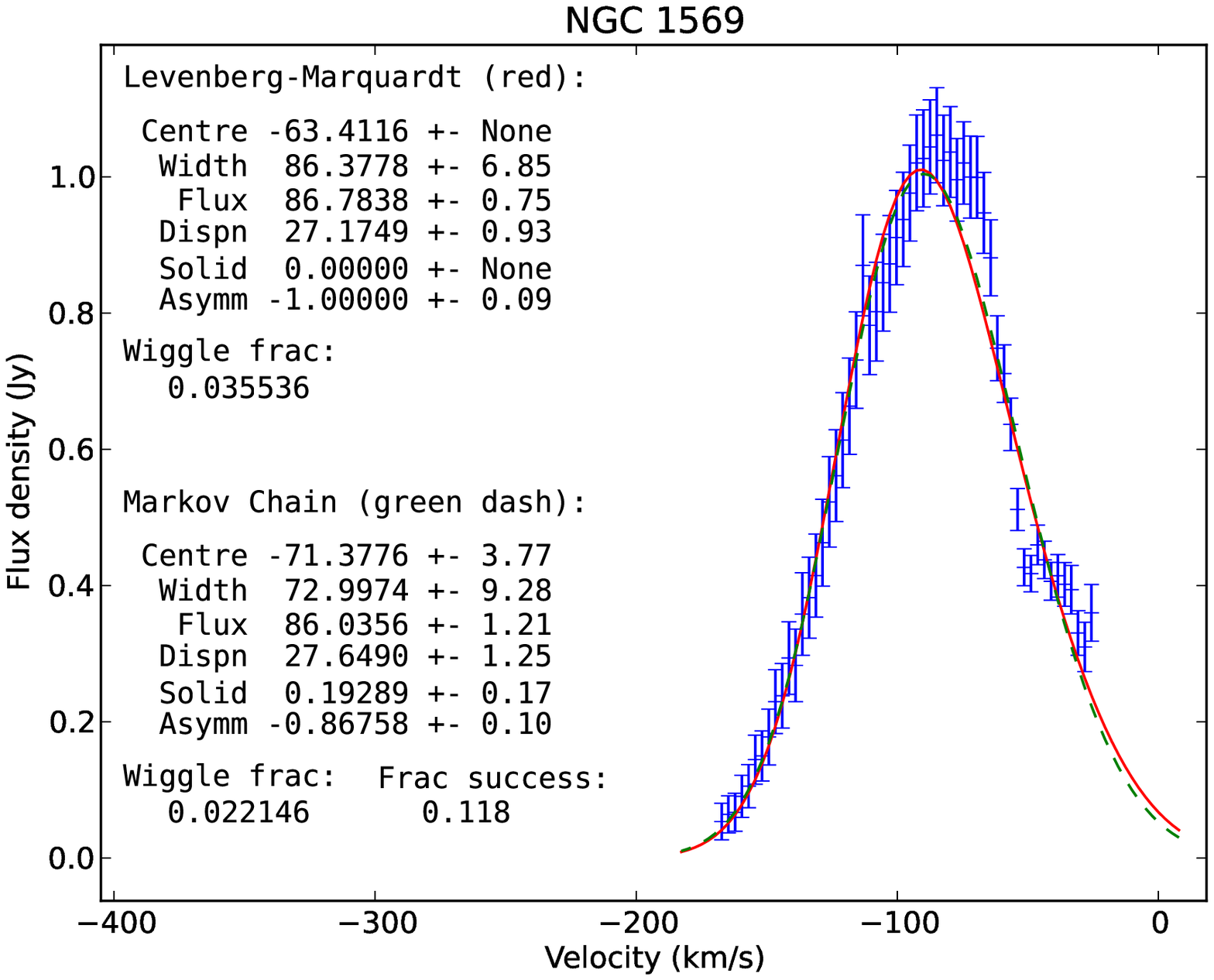}
      \caption{
              }
   \end{figure}

\clearpage

   \begin{figure}
   \centering
      \includegraphics[width=\hsize]{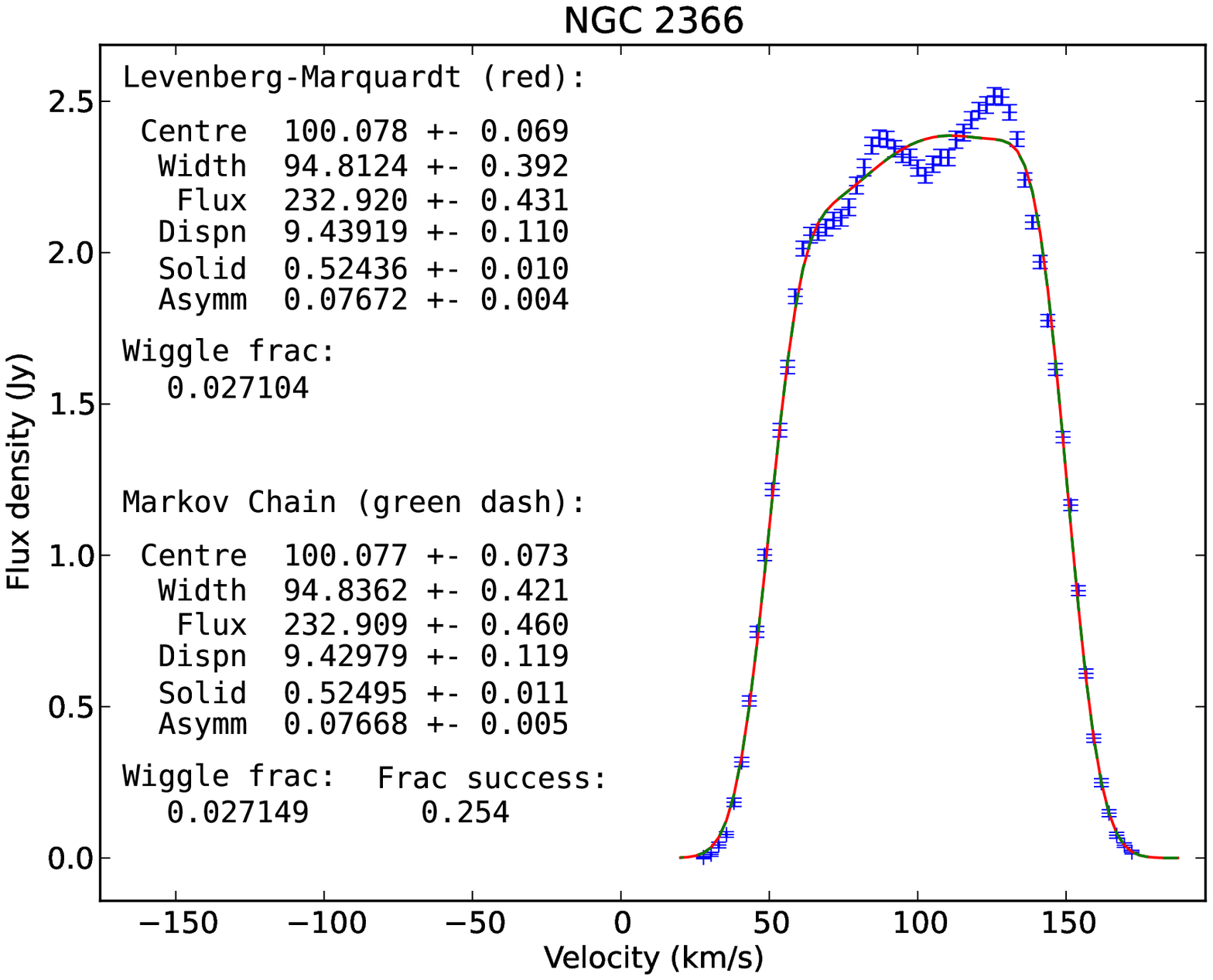}
      \caption{
              }
   \end{figure}

   \begin{figure}
   \centering
      \includegraphics[width=\hsize]{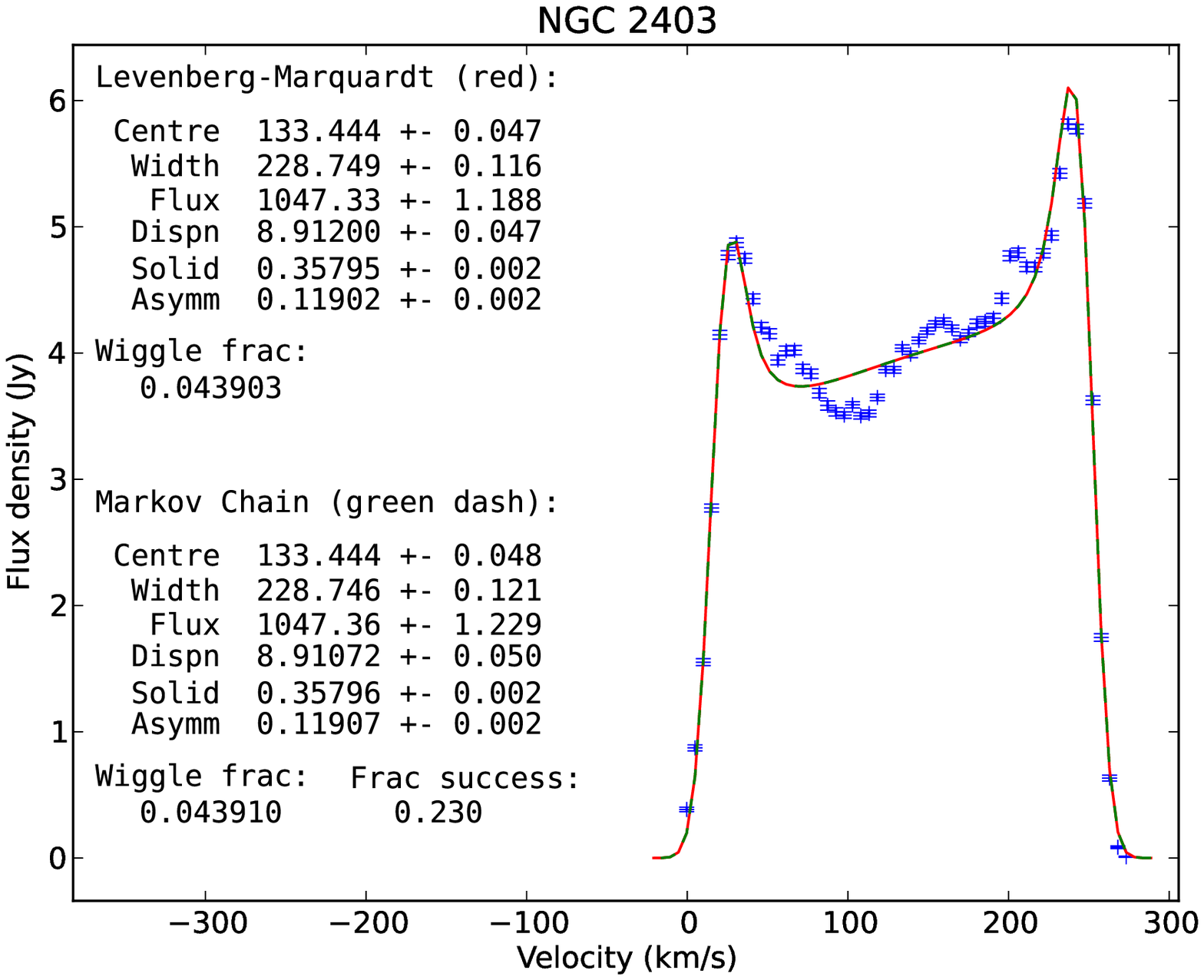}
      \caption{
              }
   \end{figure}

   \begin{figure}
   \centering
      \includegraphics[width=\hsize]{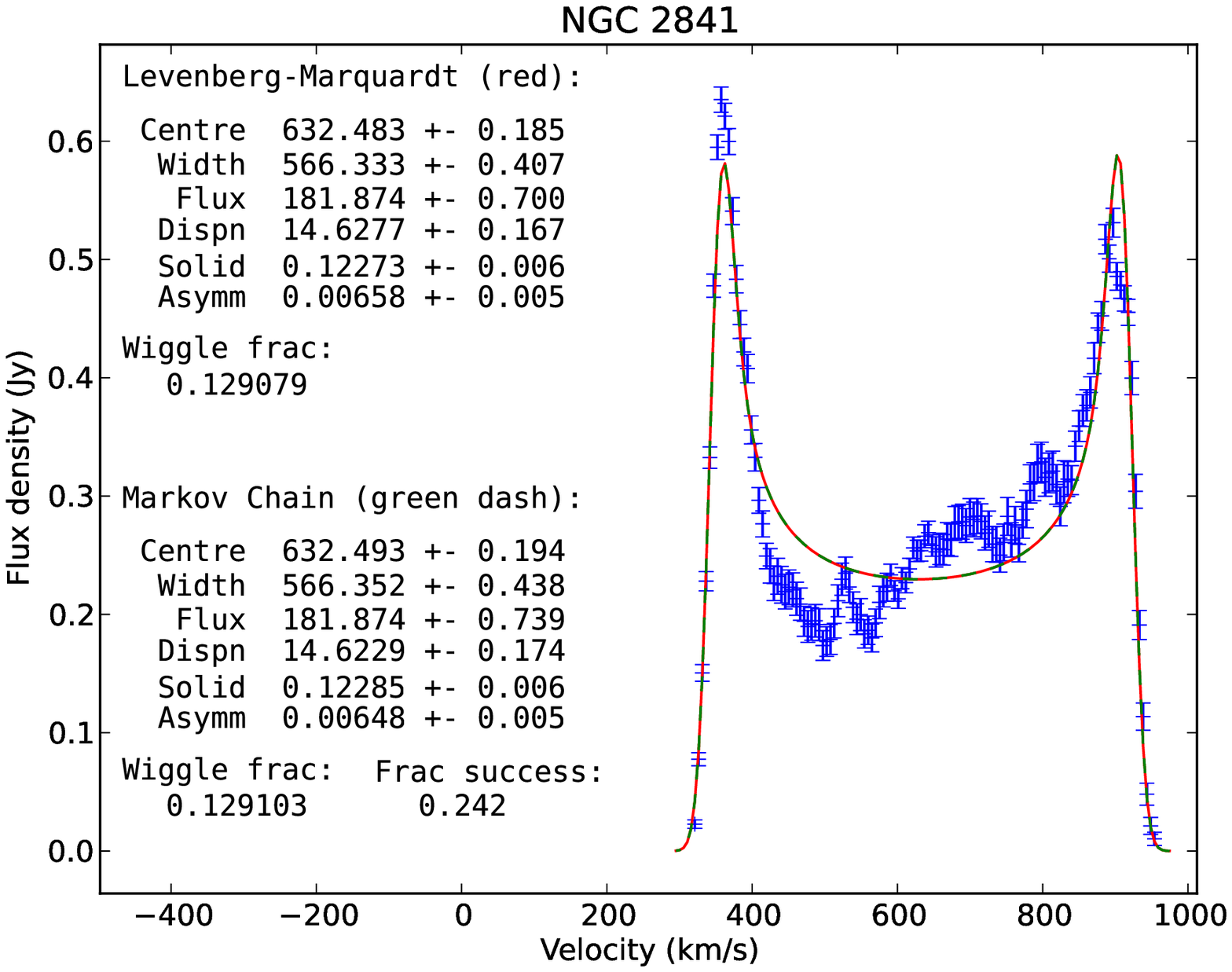}
      \caption{
              }
   \end{figure}

   \begin{figure}
   \centering
      \includegraphics[width=\hsize]{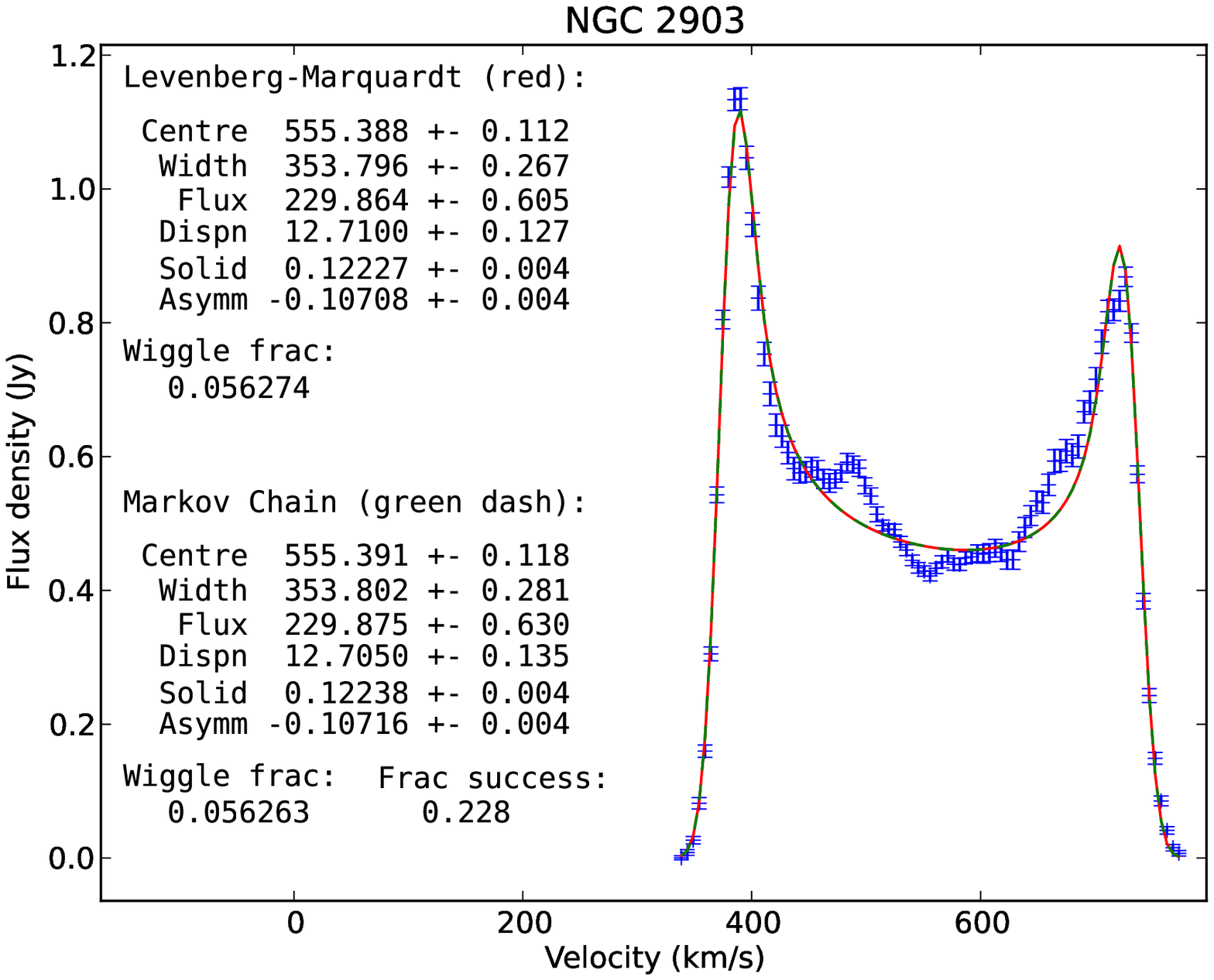}
      \caption{
              }
   \end{figure}

   \begin{figure}
   \centering
      \includegraphics[width=\hsize]{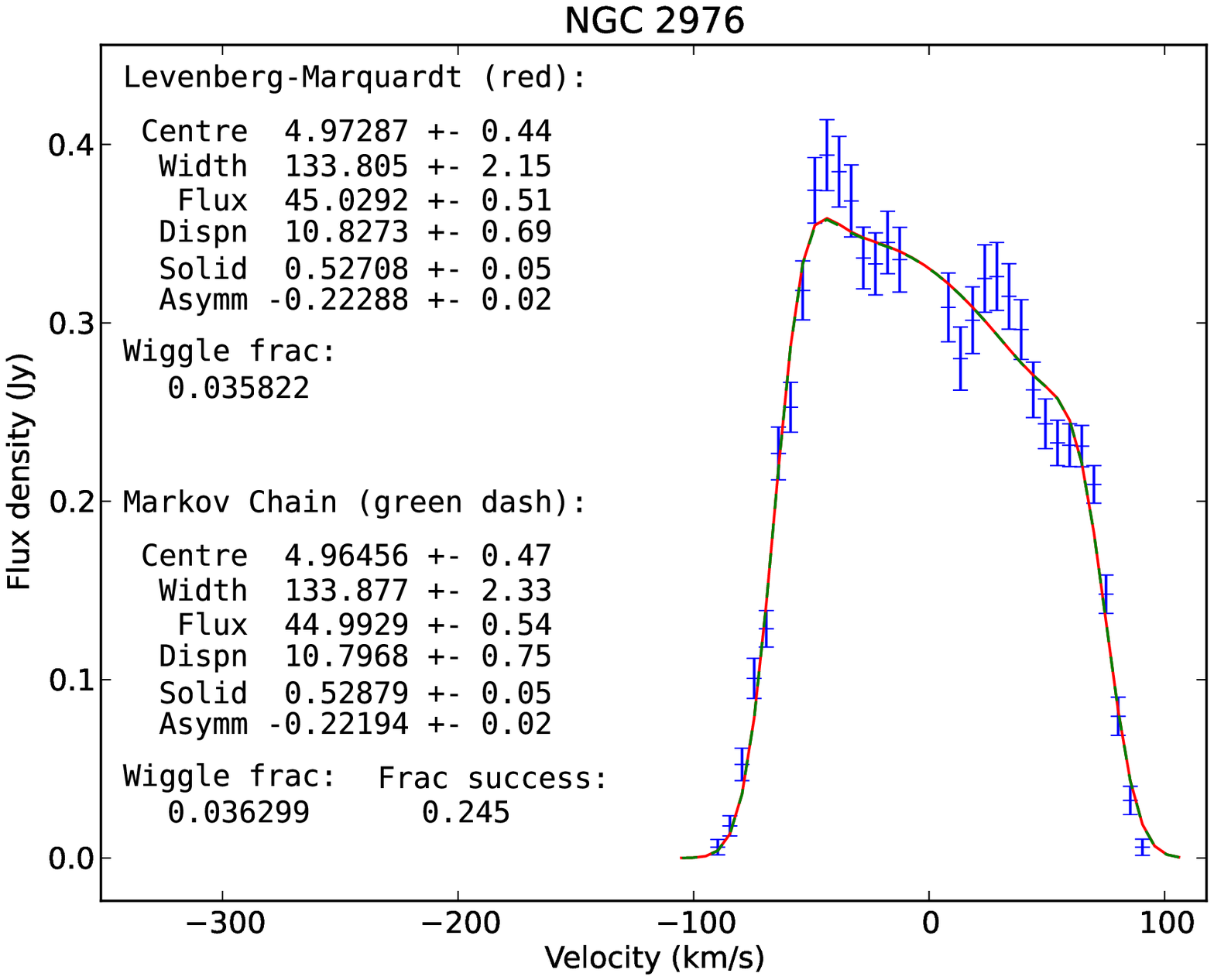}
      \caption{
              }
   \end{figure}

   \begin{figure}
   \centering
      \includegraphics[width=\hsize]{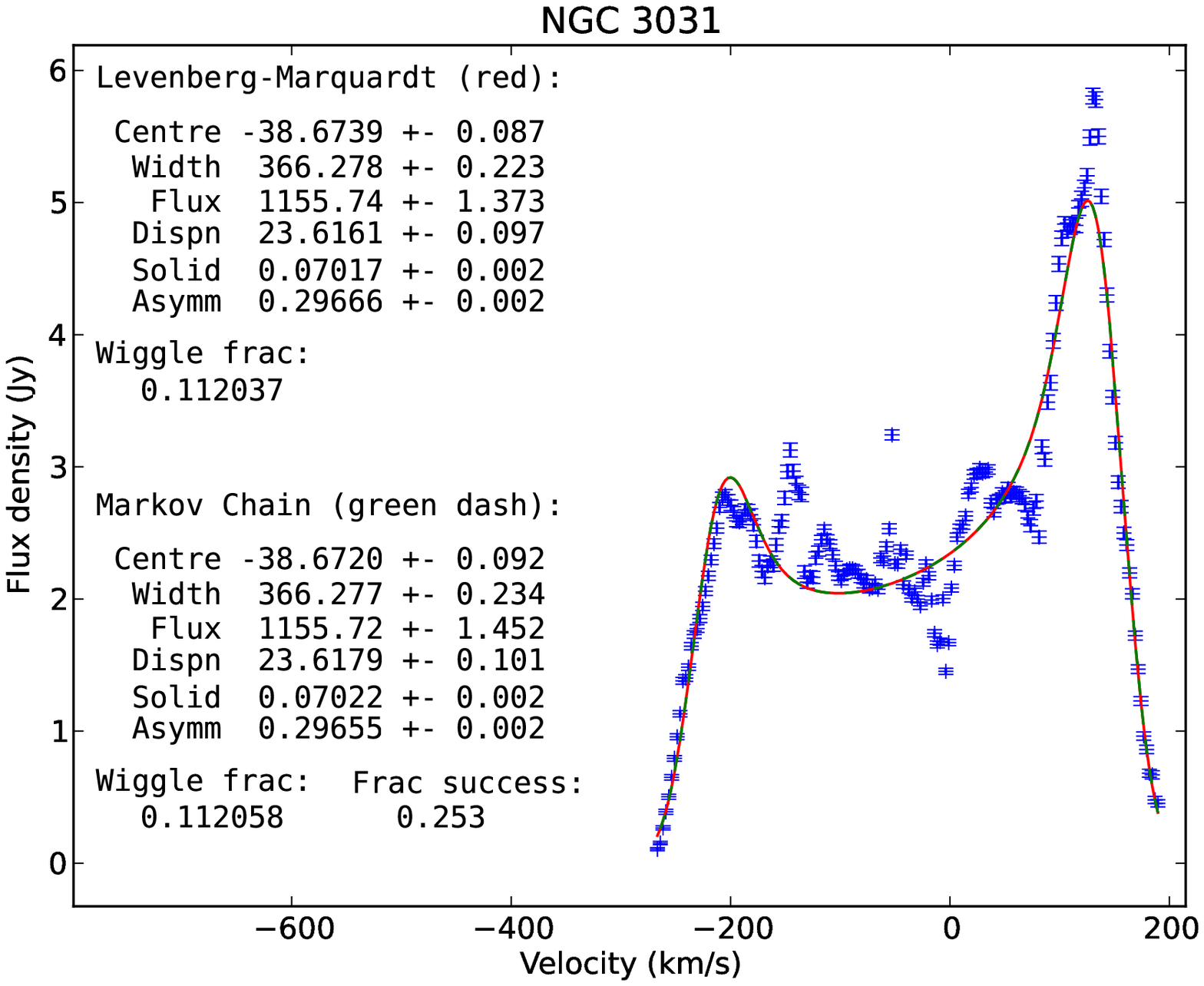}
      \caption{
              }
   \end{figure}

\clearpage

   \begin{figure}
   \centering
      \includegraphics[width=\hsize]{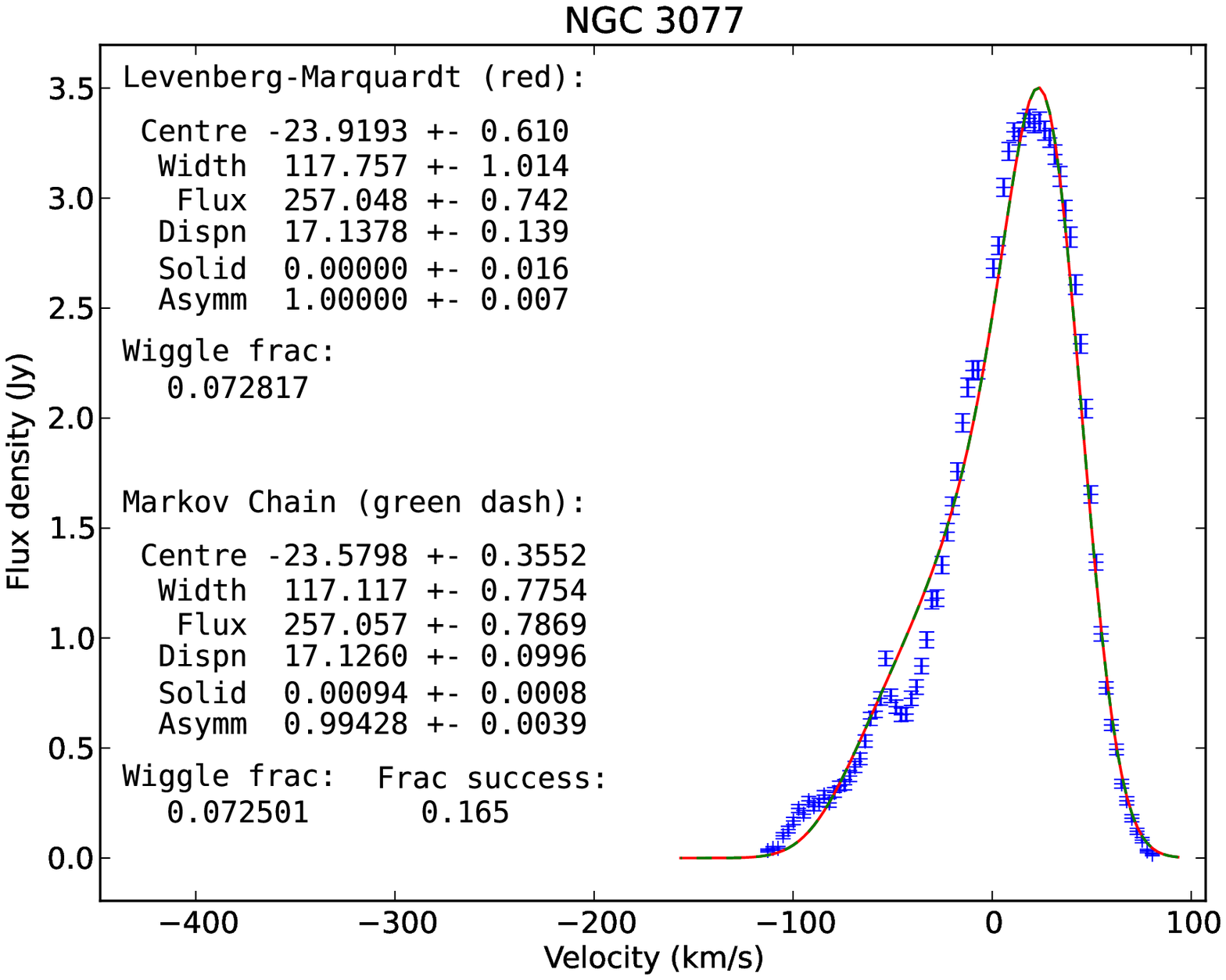}
      \caption{
              }
   \end{figure}

   \begin{figure}
   \centering
      \includegraphics[width=\hsize]{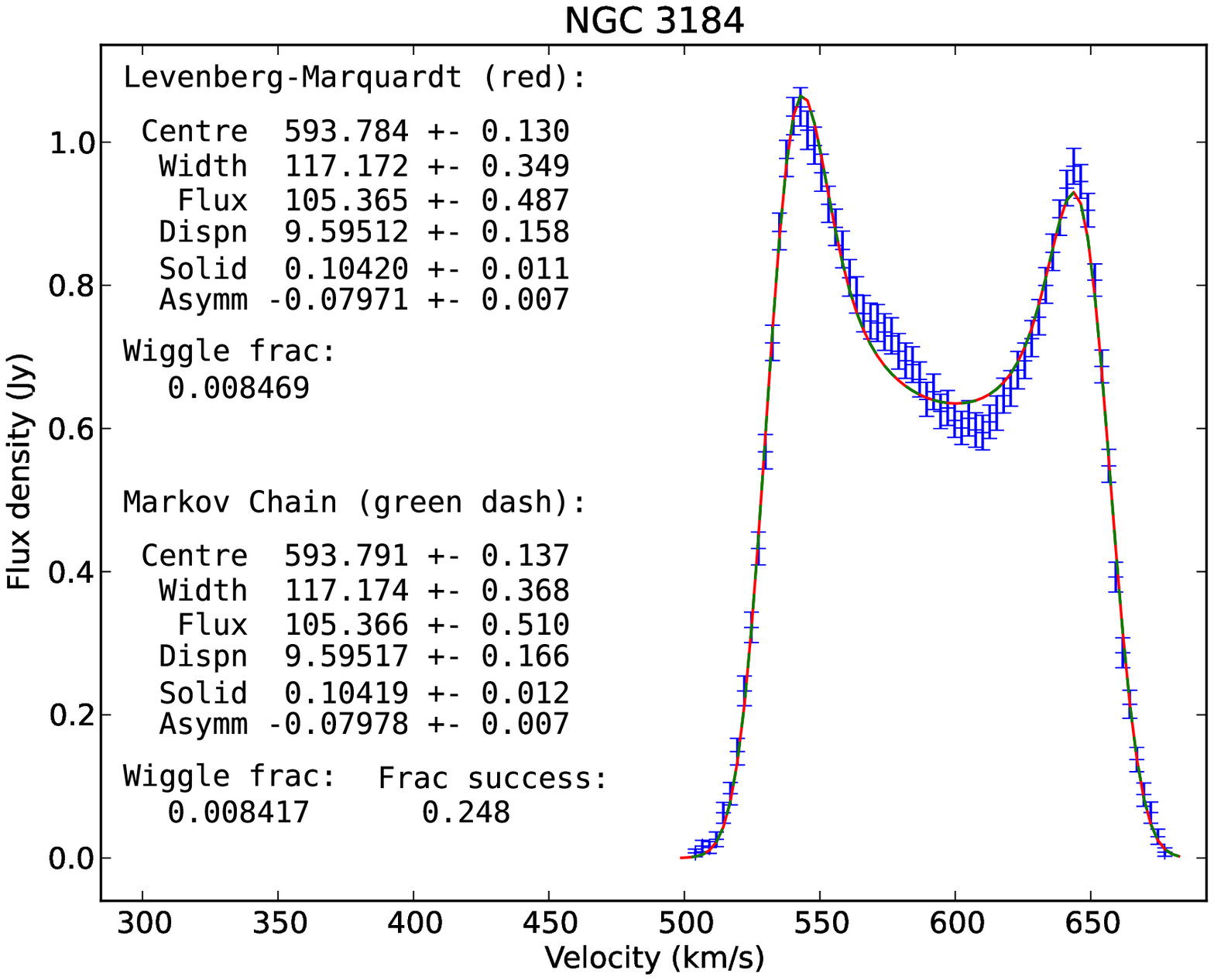}
      \caption{
              }
   \end{figure}

   \begin{figure}
   \centering
      \includegraphics[width=\hsize]{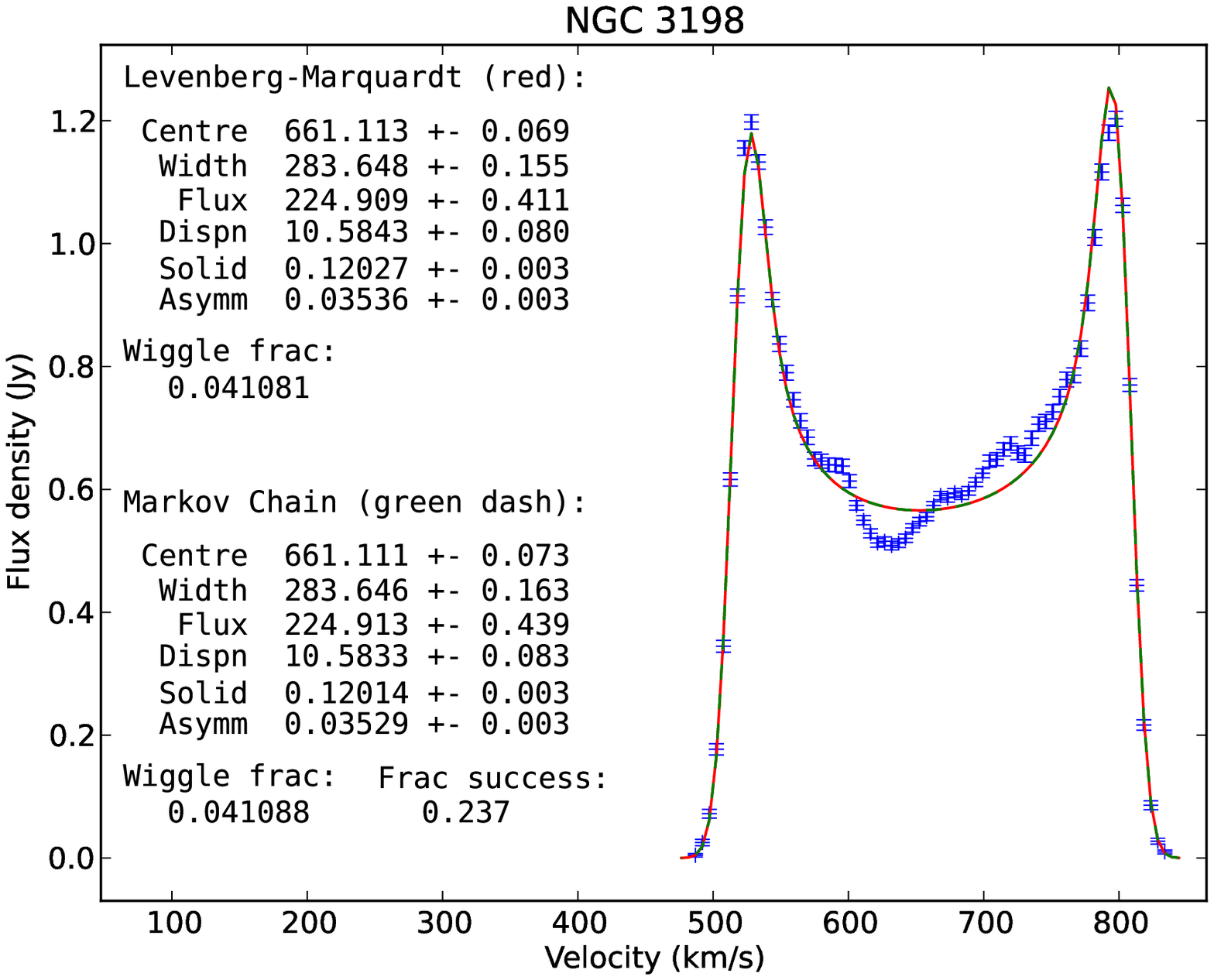}
      \caption{
              }
   \end{figure}

   \begin{figure}
   \centering
      \includegraphics[width=\hsize]{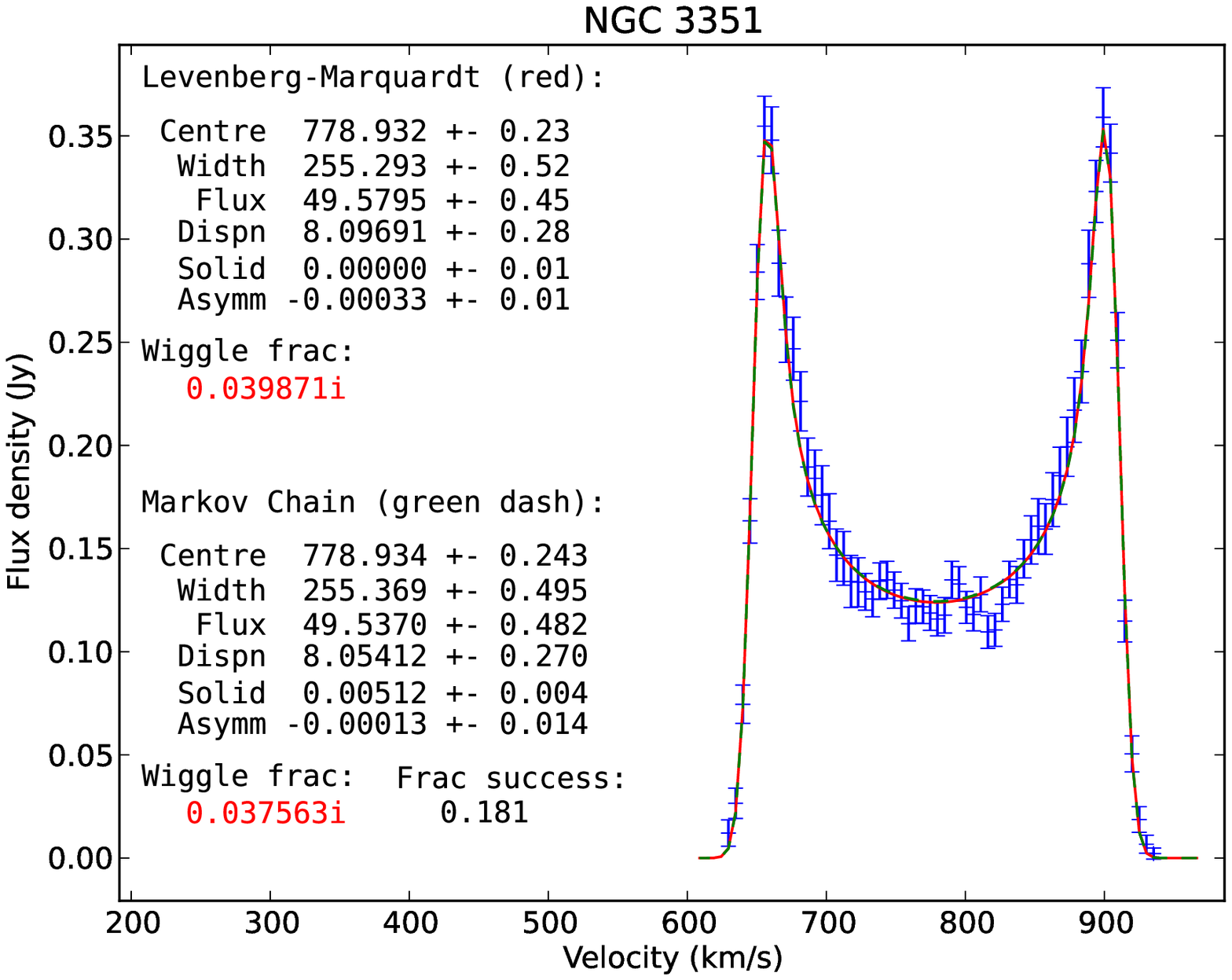}
      \caption{
              }
   \end{figure}

   \begin{figure}
   \centering
      \includegraphics[width=\hsize]{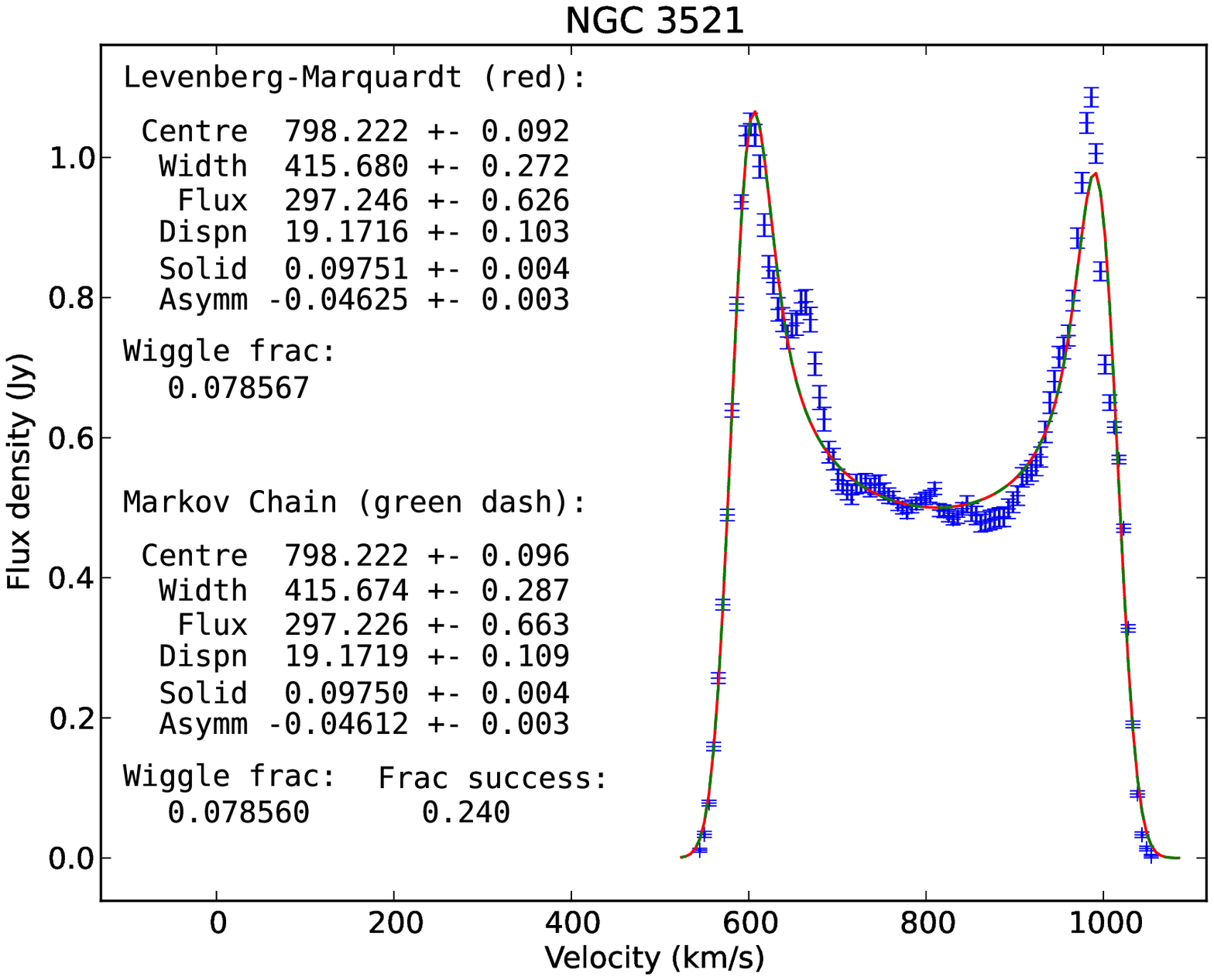}
      \caption{
              }
   \end{figure}

   \begin{figure}
   \centering
      \includegraphics[width=\hsize]{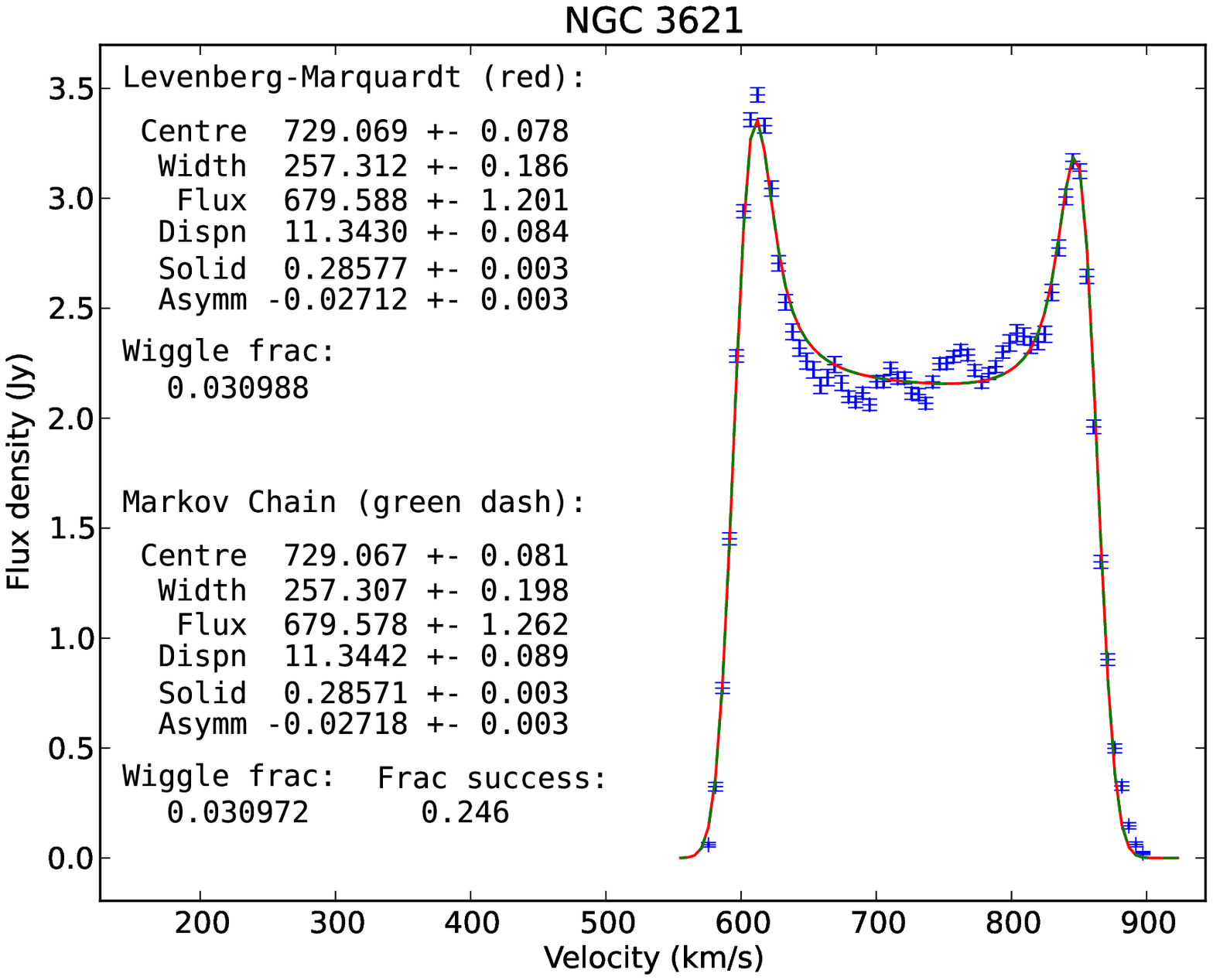}
      \caption{
              }
   \end{figure}

\clearpage

   \begin{figure}
   \centering
      \includegraphics[width=\hsize]{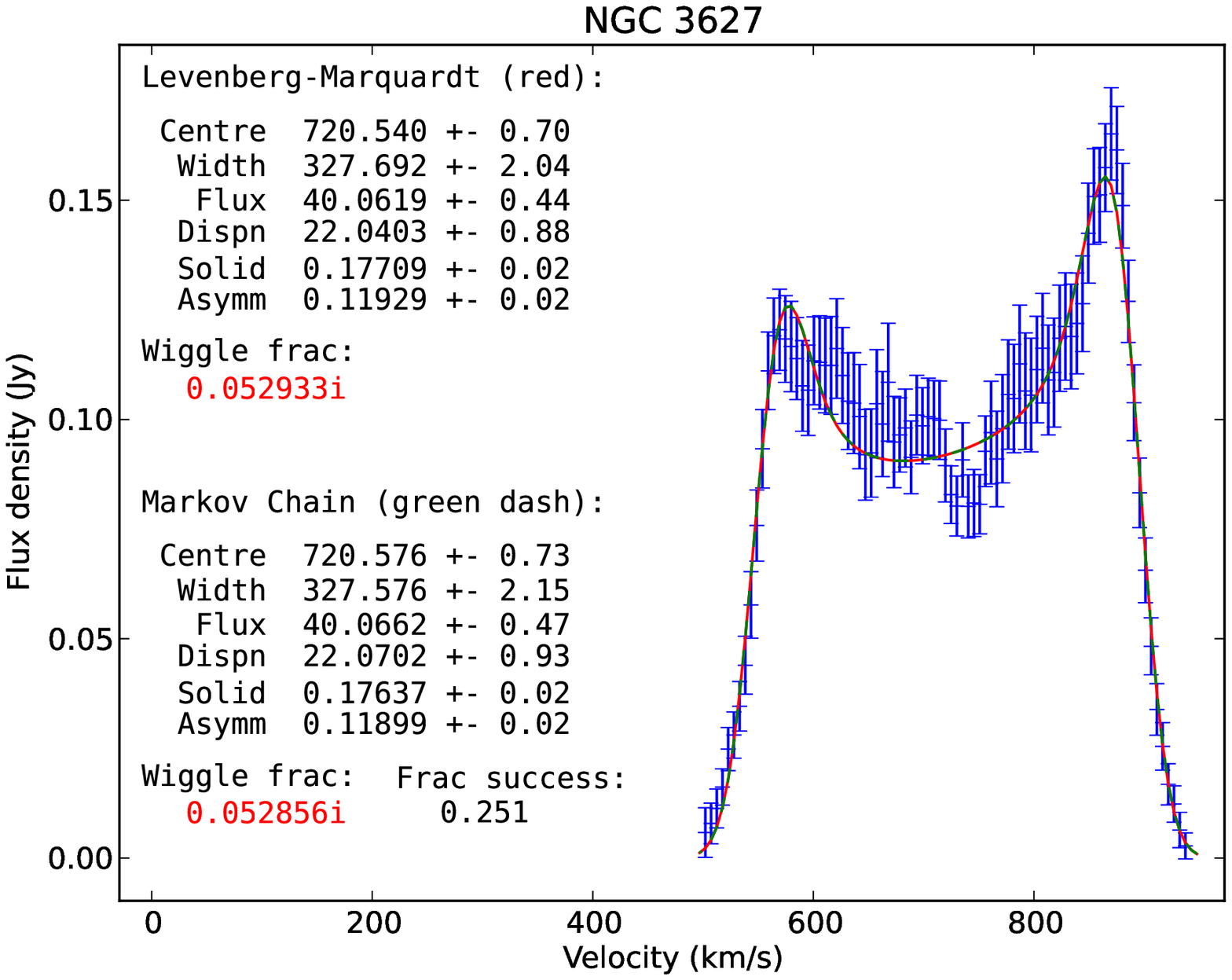}
      \caption{
              }
   \end{figure}

   \begin{figure}
   \centering
      \includegraphics[width=\hsize]{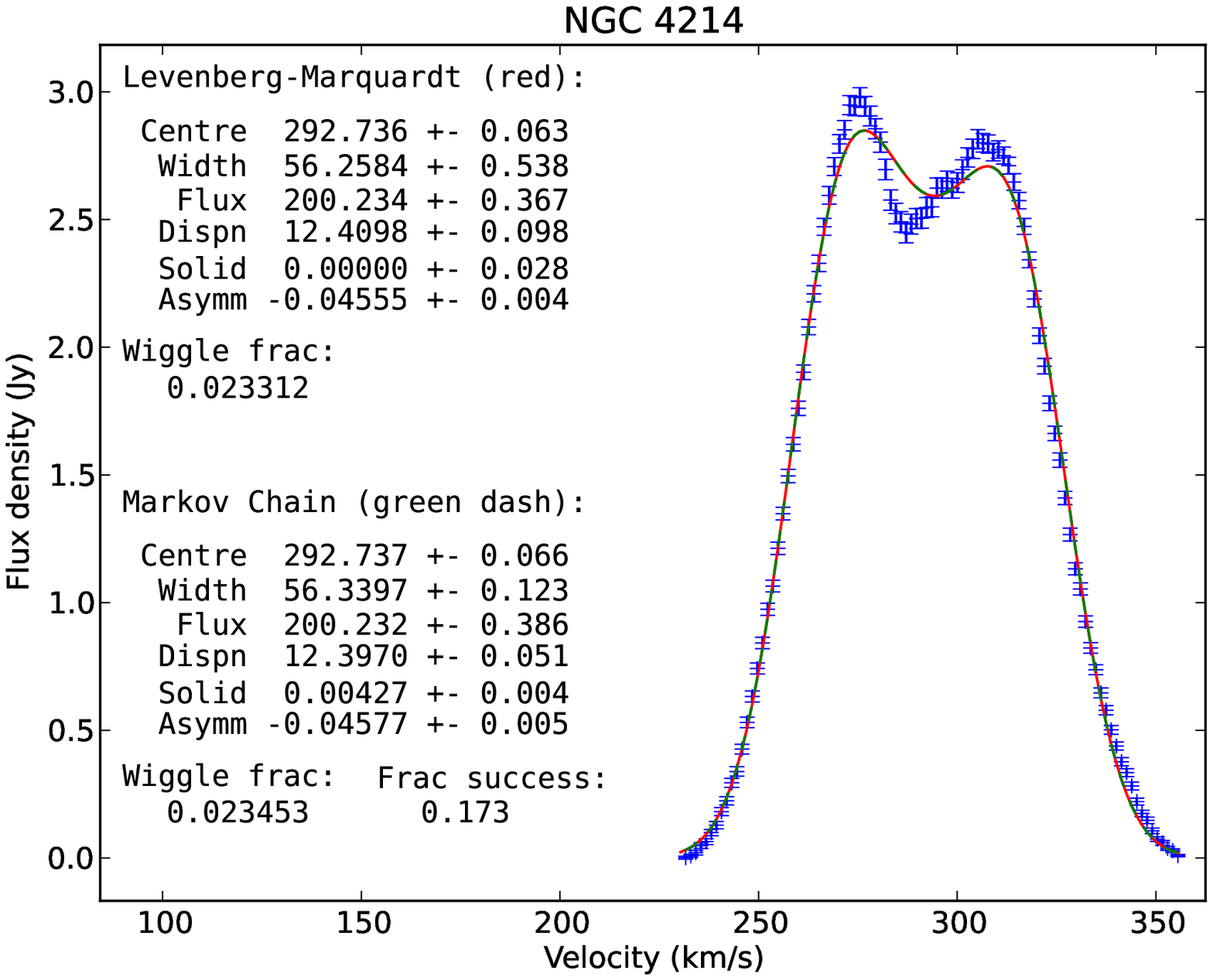}
      \caption{
              }
   \end{figure}

   \begin{figure}
   \centering
      \includegraphics[width=\hsize]{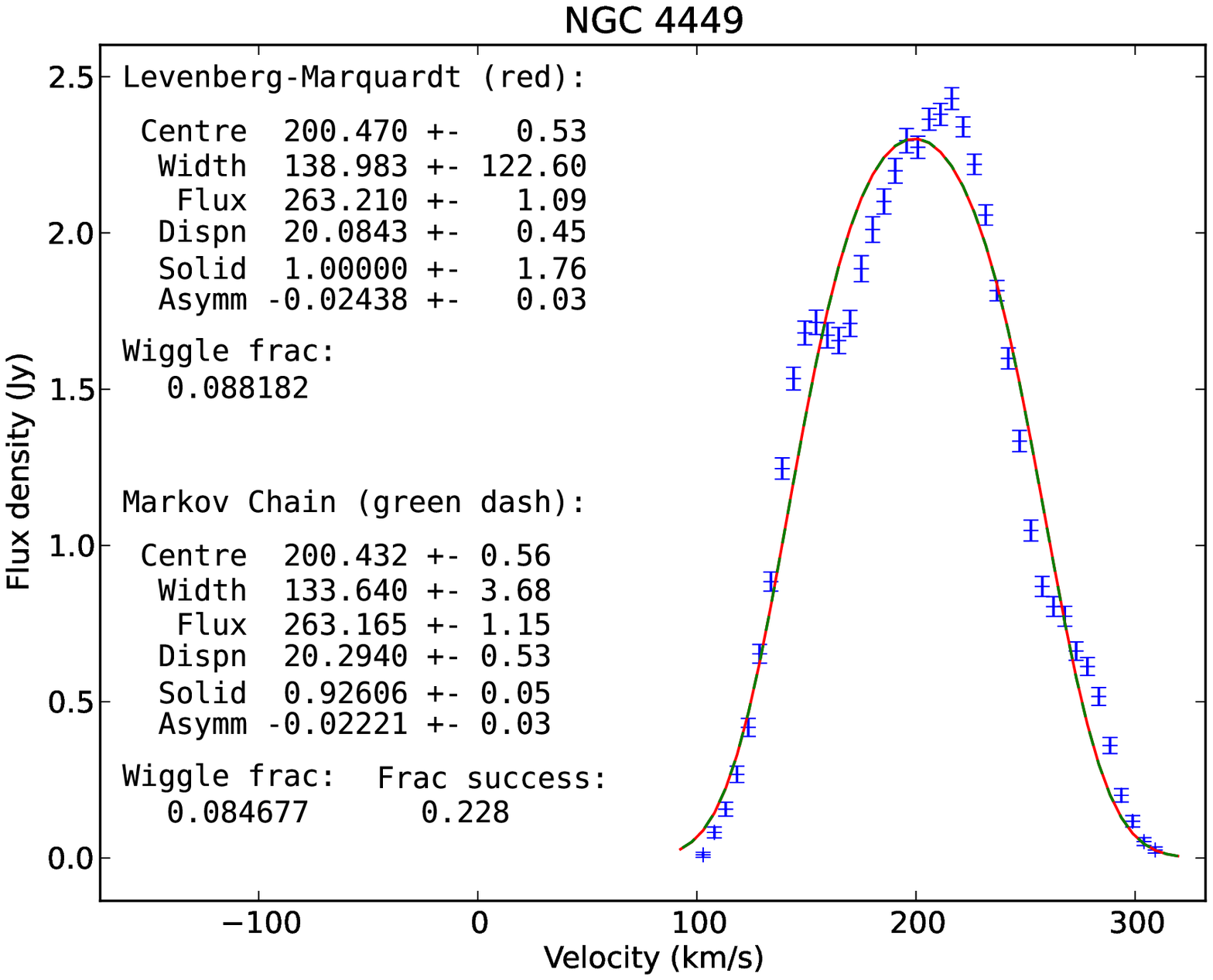}
      \caption{
              }
   \end{figure}

   \begin{figure}
   \centering
      \includegraphics[width=\hsize]{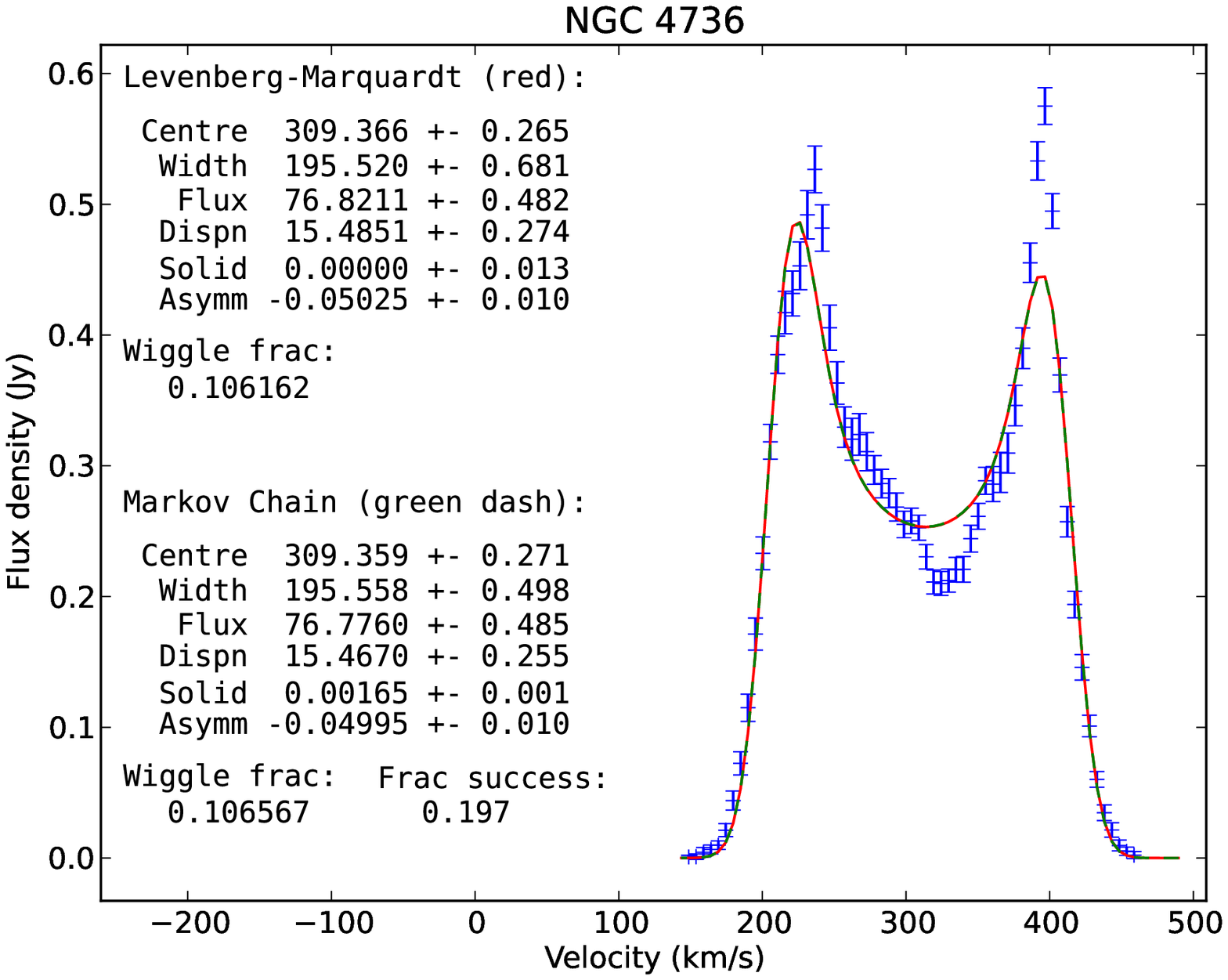}
      \caption{
              }
   \end{figure}

   \begin{figure}
   \centering
      \includegraphics[width=\hsize]{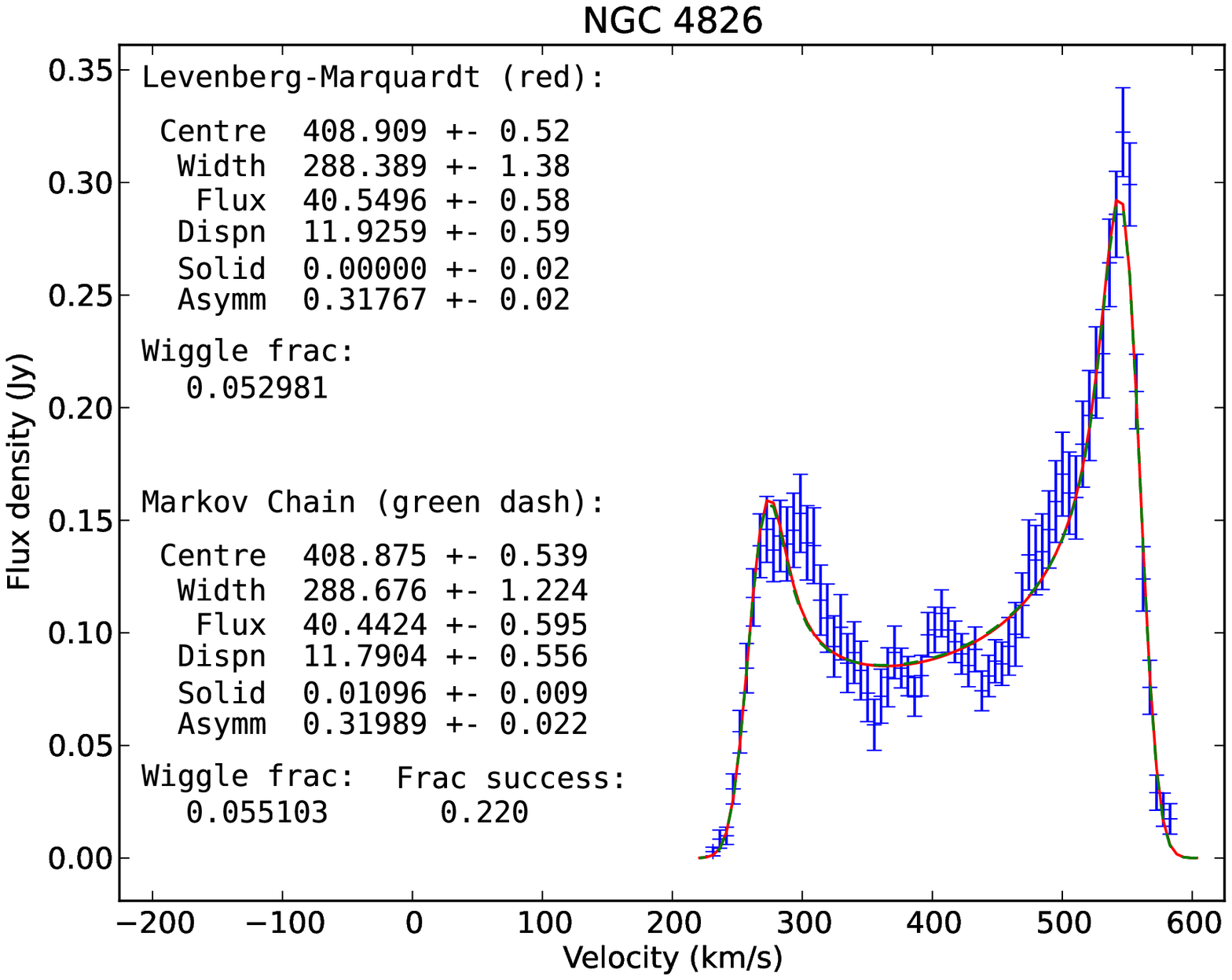}
      \caption{
              }
   \end{figure}

   \begin{figure}
   \centering
      \includegraphics[width=\hsize]{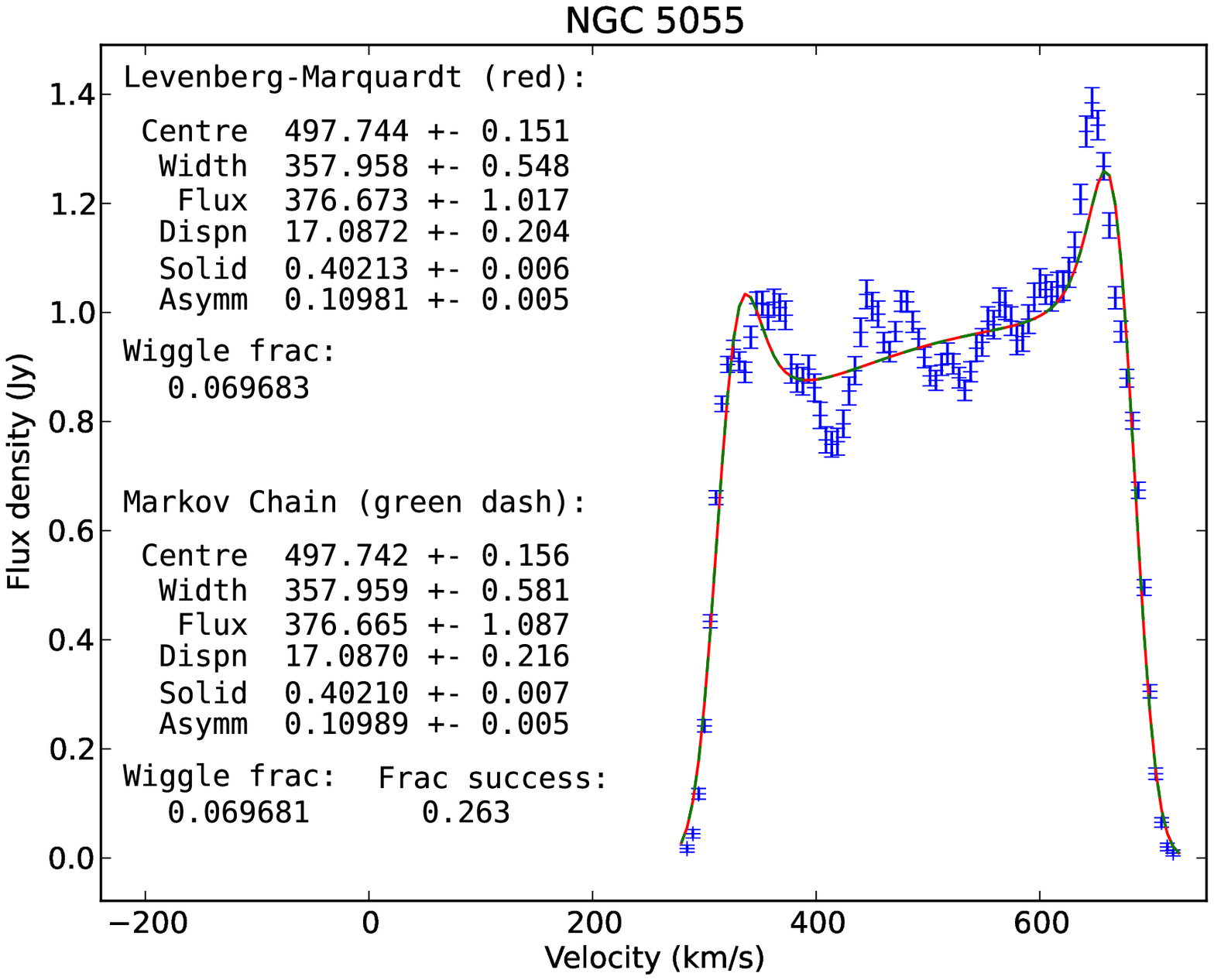}
      \caption{
              }
   \end{figure}

\clearpage

   \begin{figure}
   \centering
      \includegraphics[width=\hsize]{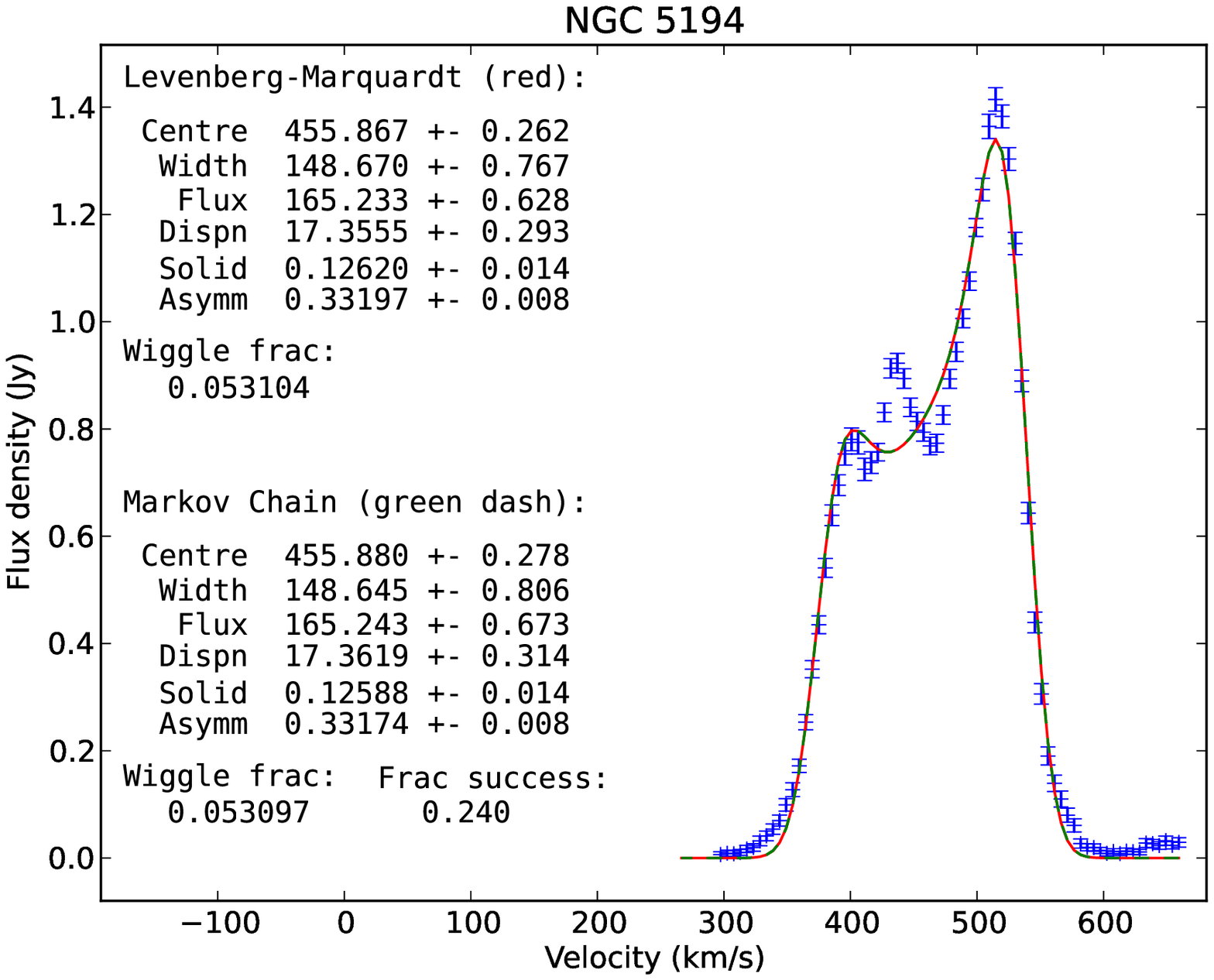}
      \caption{
              }
   \end{figure}

   \begin{figure}
   \centering
      \includegraphics[width=\hsize]{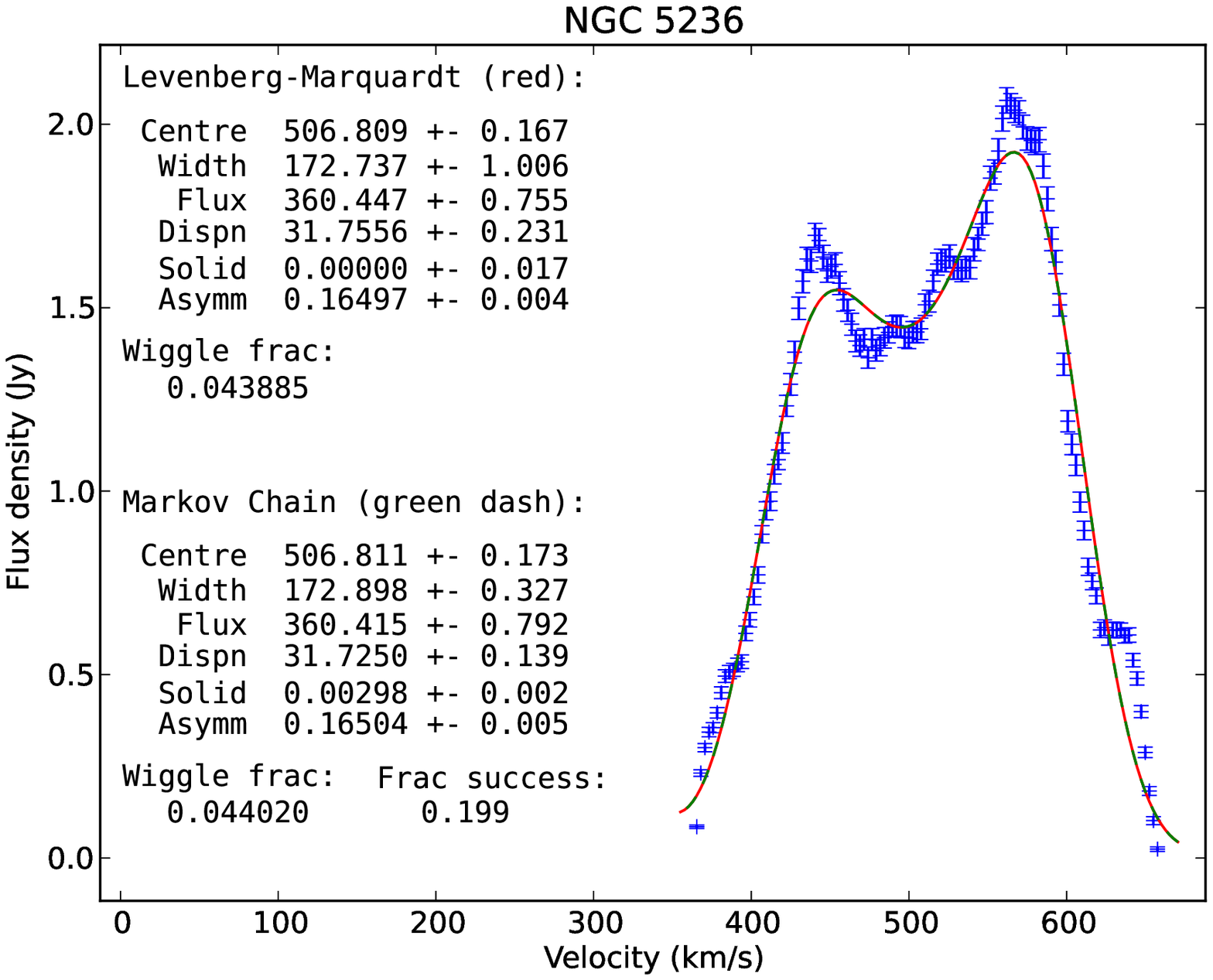}
      \caption{
              }
   \end{figure}

   \begin{figure}
   \centering
      \includegraphics[width=\hsize]{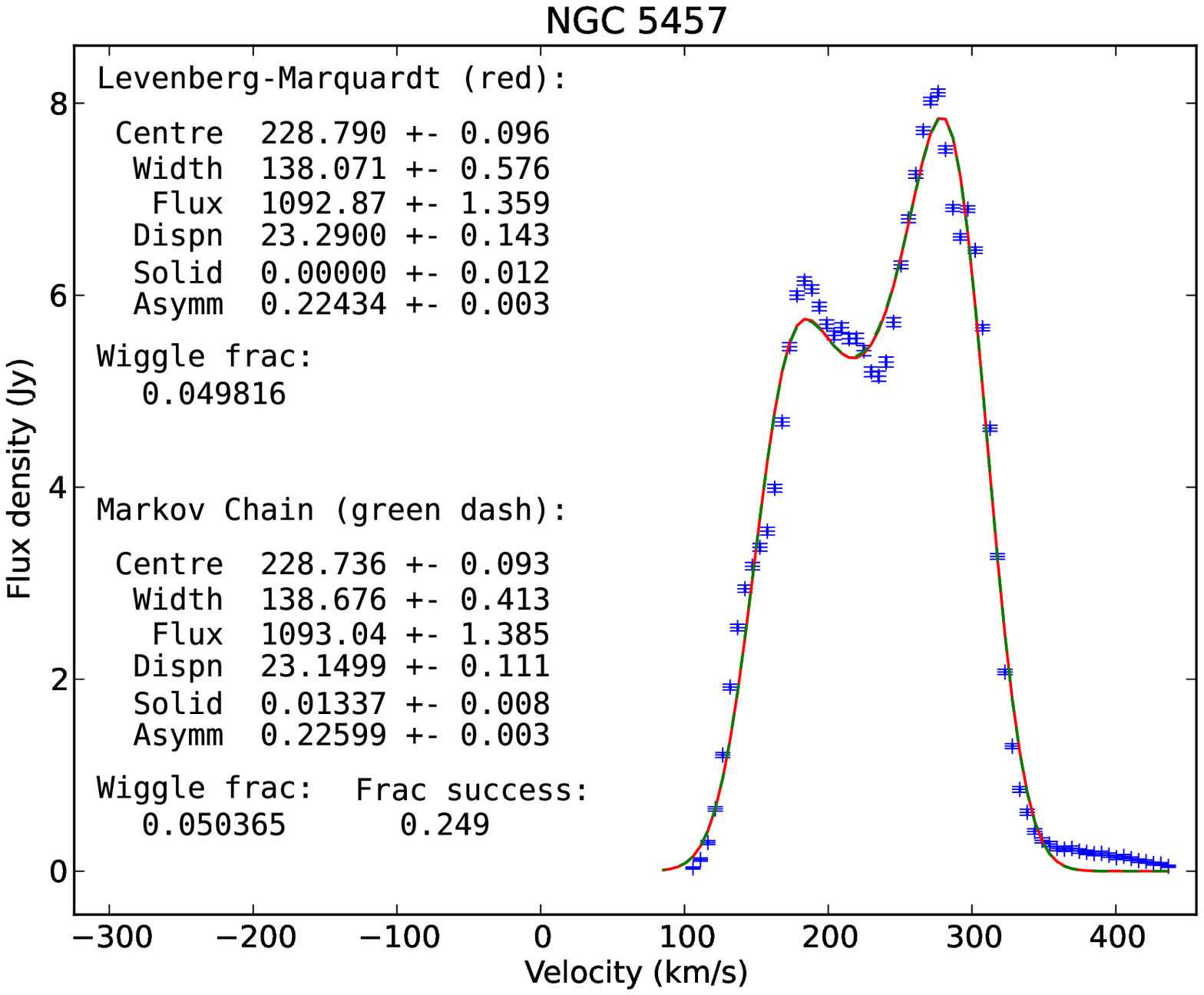}
      \caption{
              }
   \end{figure}

   \begin{figure}
   \centering
      \includegraphics[width=\hsize]{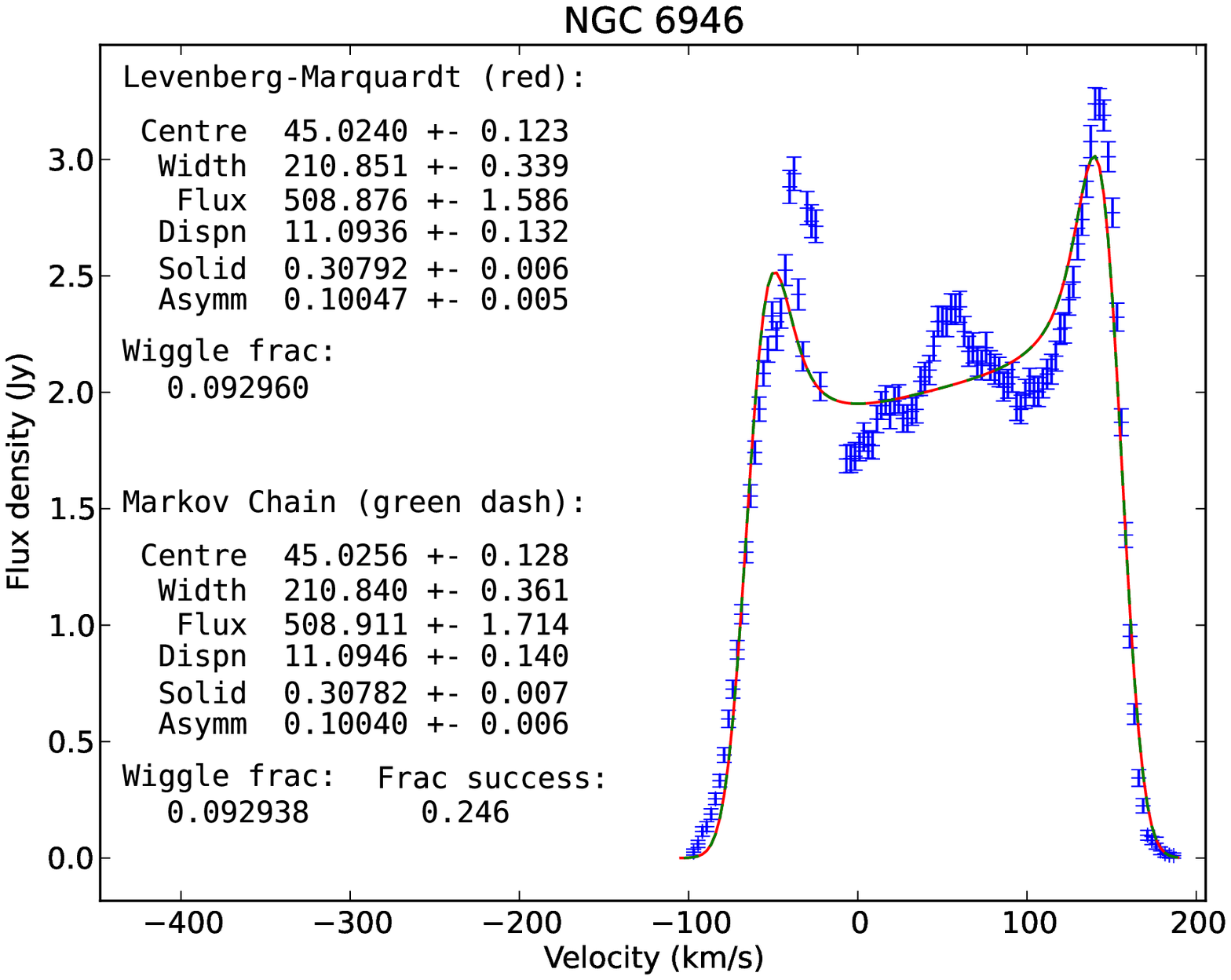}
      \caption{
              }
   \end{figure}

   \begin{figure}
   \centering
      \includegraphics[width=\hsize]{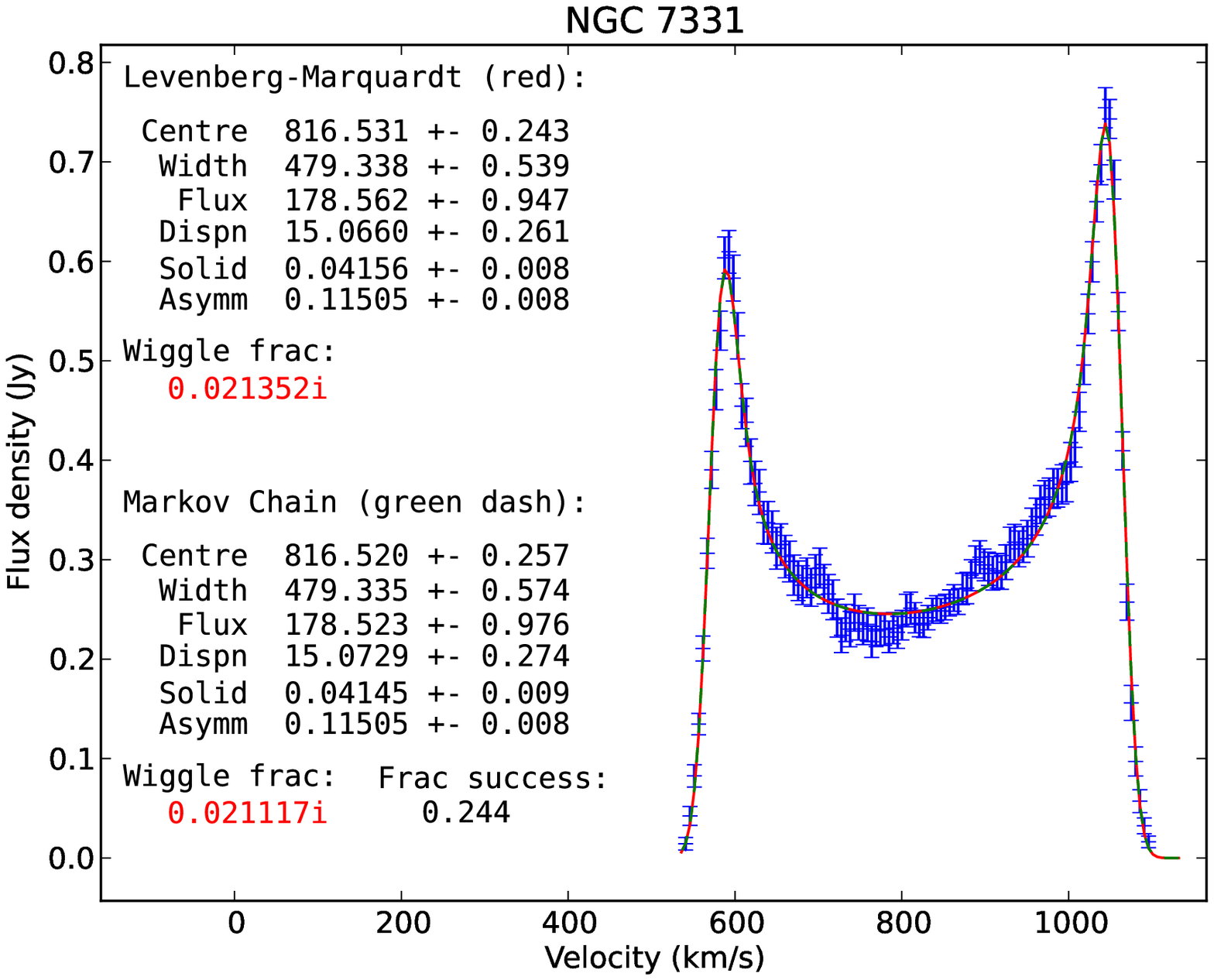}
      \caption{
              }
   \end{figure}

   \begin{figure}
   \centering
      \includegraphics[width=\hsize]{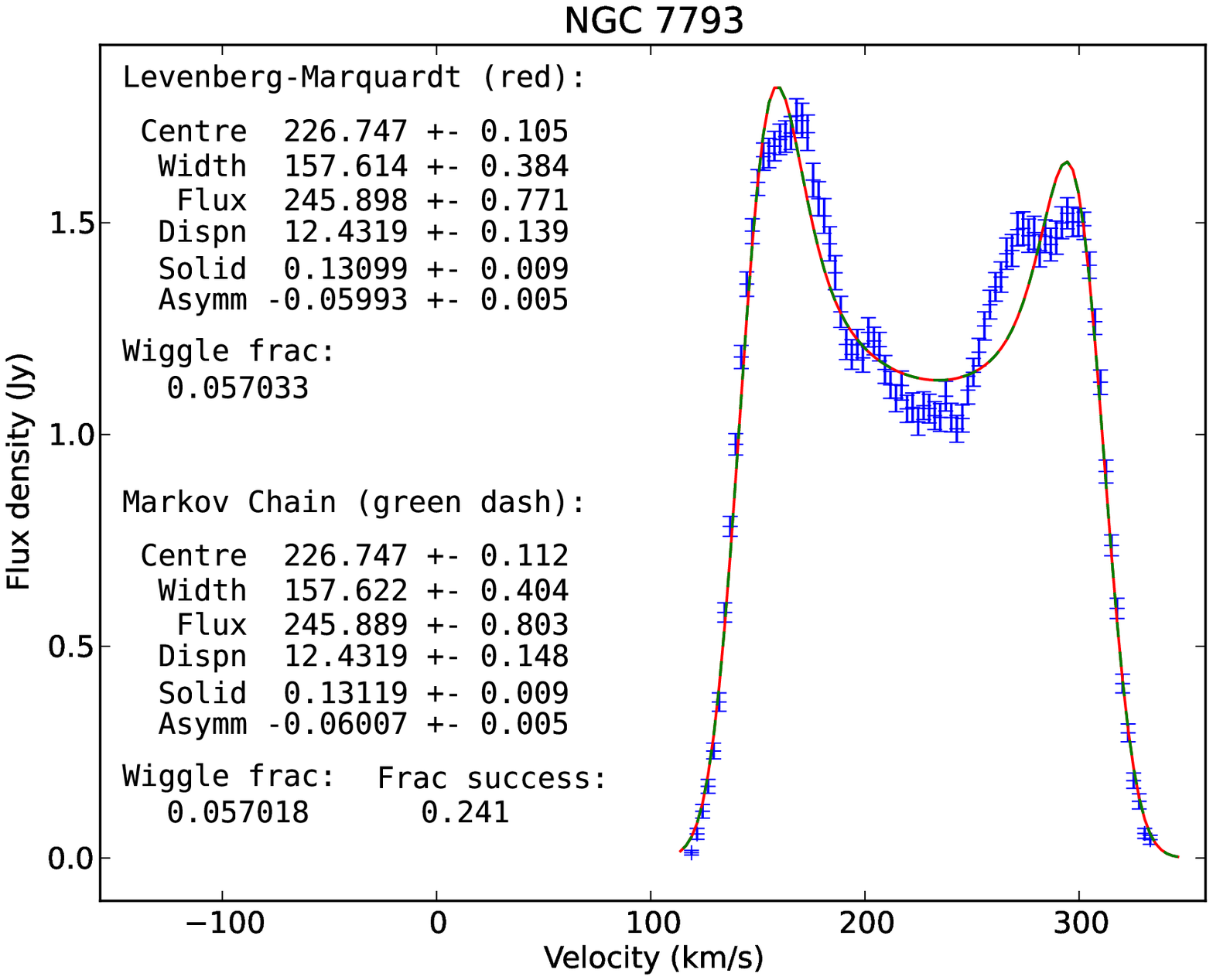}
      \caption{
              }
         \label{fig_gal34}
   \end{figure}

\end{document}